\pdfoutput=1

\documentclass[aps,prd,twocolumn,superscriptaddress,preprintnumbers,floatfix,nofootinbib,longbibliography]{revtex4-2}

\usepackage{subfigure}
\usepackage{graphicx}
\usepackage[dvipsnames, usenames]{xcolor}  %
\usepackage{tikz}
\usetikzlibrary{arrows.meta, fit, backgrounds, calc,
  decorations.pathmorphing}
\usepackage{amsmath, amssymb, mathrsfs}
\usepackage{bm}
\usepackage[T1]{fontenc}
\usepackage[utf8]{inputenc}
\usepackage{microtype}
\usepackage{booktabs}
\usepackage[colorlinks,urlcolor=NavyBlue,citecolor=NavyBlue,linkcolor=NavyBlue,pdfusetitle]{hyperref}
\usepackage[all]{hypcap}
\usepackage{orcidlink}

\newcommand{\Igroup}{\mathscr{I}^{+}}

\graphicspath{{Figures}}

\definecolor{cauchygreen}{RGB}{195, 218, 158}
\definecolor{asyncyan}{RGB}{145, 210, 220}
\definecolor{bondiyellow}{RGB}{252, 237, 125}
\definecolor{numpeach}{RGB}{248, 205, 168}
\definecolor{darkteal}{RGB}{22, 76, 100}
\definecolor{panelbrown}{RGB}{105, 62, 18}
\definecolor{annotgreen}{RGB}{0, 110, 55}

\newsavebox{\prlSchematicBox}

\newcommand{\figJunkOverview}{%
\begin{figure*}[t]
    \begin{center}
    \includegraphics[width=\textwidth]{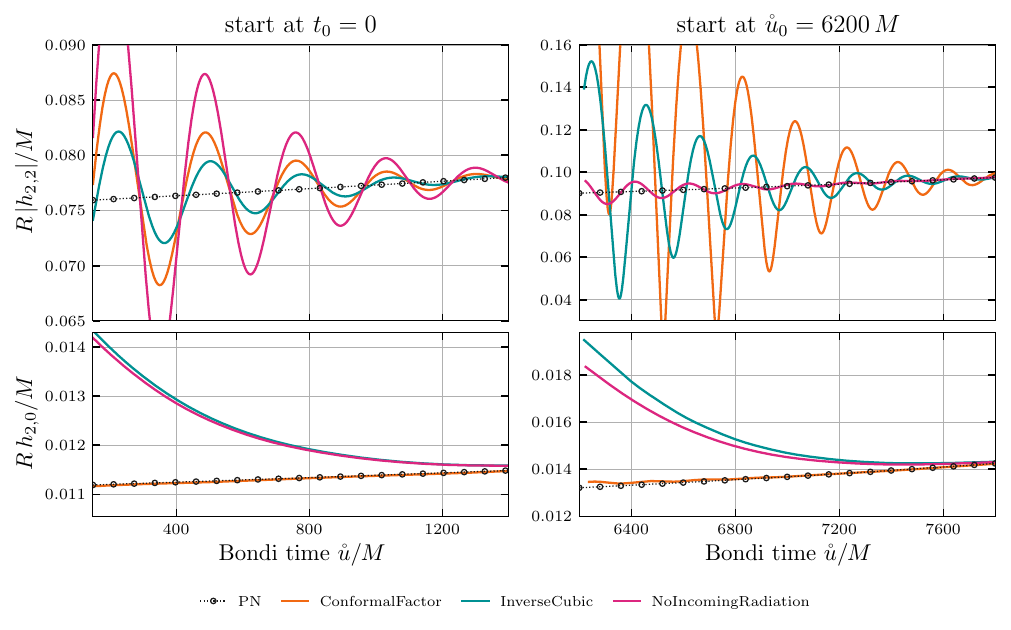}
    \end{center}
    \caption{Strain at $\Igroup$ for an equal-mass, non-spinning BBH for
    $R=300\,M$ and $R_{\mathrm{max}}=2000\,M$, with the three pre-existing
    initial-data schemes. The rows show
    $R|h_{2,2}|/M$ and $R\,h_{2,0}/M$; the latter is purely real for this
    system. Hollow circles show the post-Newtonian prediction.
    \emph{Left}: started on the Cauchy initial slice, $t_{0}=0$, i.e.\
    $\mathring{u}_{0}=-R$.
    \emph{Right}: restarted at $\mathring{u}_{0}=6200\,M$. All curves are
    mapped to the post-Newtonian BMS frame using windows outside the plotted range.}
    \label{fig:JunkOverview}
\end{figure*}
}
\newcommand{\figBondiConstraints}{%
\begin{figure*}[t]
    \begin{center}
    \includegraphics[width=\textwidth]{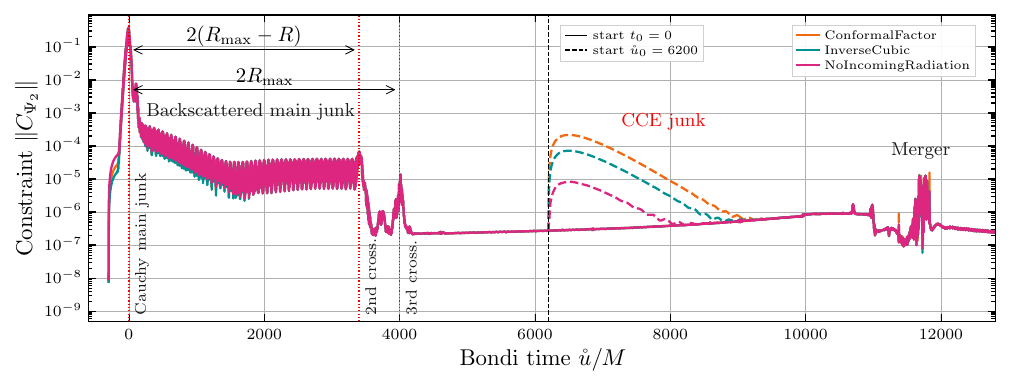}
    \end{center}
    \caption{Norm $\lVert C_{\Psi_{2}}\rVert$ of the Bondi--Sachs constraint
    associated with $\Psi_{2}$ at $\Igroup$, for $R=300\,M$ and
    $R_{\mathrm{max}}=2000\,M$. Red dotted lines mark the three crossings of the
    worldtube by the Cauchy junk: its first passage at $\mathring u = 0$, the
    return of its first outer-boundary reflection at $2(R_{\mathrm{max}}-R)$,
    and that same reflection crossing outward again at $2R_{\mathrm{max}}$,
    after passing through the center of the domain. The black dashed line is the
    restart time. Solid curves
    start at $t_{0}=0$, dashed curves at $\mathring u_{0}=6200\,M$.}
    \label{fig:BondiConstraints}
\end{figure*}
}

\newcommand{\figAngularMap}{%

\begin{figure*}[t]
    \begin{center}
    \includegraphics[width=\textwidth]{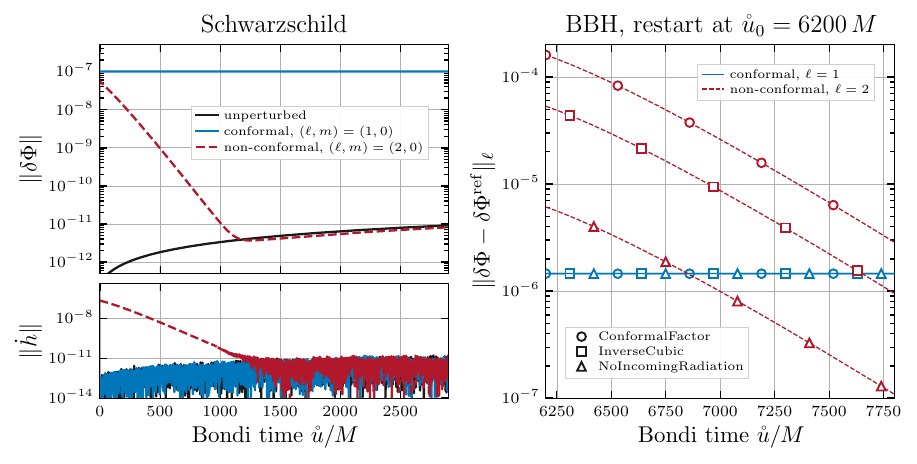}
    \end{center}
    \caption{The conformal part of the initial angular map is free; the
    non-conformal part is not. Blue denotes the conformal part and red the
    non-conformal part in every panel.
    \emph{Left}: three CCE evolutions of Schwarzschild with identical worldtube
    data. Their initial angular maps are unperturbed, perturbed by a conformal
    $(\ell,m)=(1,0)$ mode, or perturbed by a non-conformal $(\ell,m)=(2,0)$
    mode. Both perturbations have magnitude $\lVert\delta\Phi\rVert=10^{-7}$,
    Eq.~\eqref{eq:deltaPhi_def}. The upper panel shows $\lVert\delta\Phi\rVert$
    and the lower panel the news, on the same time axis. Only the non-conformal
    perturbation radiates as it relaxes toward the unperturbed run.
    \emph{Right}: the same split in the equal-mass binary, for the three
    pre-existing schemes restarted at $\mathring u_{0}=6200\,M$; plotted is the
    difference of the initial angular map from that of a reference run started at
    $t_{0}=0$, in the per-$\ell$ norm
    $\lVert\cdot\rVert_{\ell}$ of Eq.~\eqref{eq:angular_norm_ell}. The markers
    identify the scheme. All three hold
    $\ell=1$ fixed at a common offset and relax $\ell=2$ at a common rate.}
    \label{fig:AngularMap}
\end{figure*}
}
\newcommand{\figRadialCoeffs}{%
\begin{figure}[t]
    \begin{center}
    \includegraphics[width=\columnwidth]{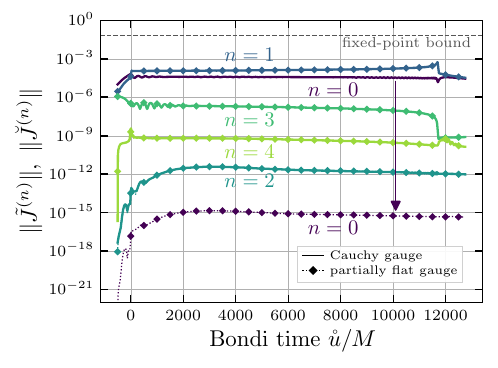}
    \end{center}
    \caption{Angular norms $\lVert\tilde J^{(n)}\rVert$,
    Eq.~\eqref{eq:angular_norm}, of the radial expansion coefficients of $J$.
    The coefficients are obtained from a reference CCE evolution using
    Eq.~\eqref{eq:coefffromder}, for an equal-mass, non-spinning BBH at
    $R=500\,M$. Solid lines denote the Cauchy gauge and diamonds the partially
    flat gauge; the
    arrow joins the $n=0$ coefficient in the two gauges. The dashed grey line
    marks the amplitude bound $7\times10^{-2}$ of
    Appendix~\ref{app:fixed_point}, under which the fixed point of
    Eq.~\eqref{eq:fixedpoint} exists and is unique. The bound applies to the
    pointwise maxima of $\lvert\tilde J^{(0)}\rvert$ and
    $\lvert\tilde J^{(1)}\rvert$, which in this evolution remain below
    $3\times10^{-4}$.}
    \label{fig:RadialCoeffs}
\end{figure}
}
\newcommand{\figTeukolskyWave}{%
\begin{figure}[t]
    \begin{center}
    \includegraphics[width=\columnwidth]{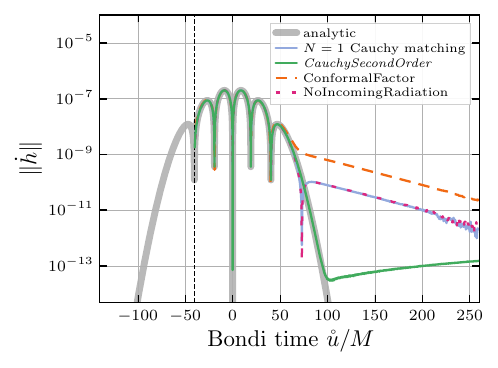}
    \end{center}
    \caption{Norm of the $(2,0)$ news radiated by the initial data for a linear
    Teukolsky wave of amplitude $\mathcal{A}=10^{-2}$ and duration $T=20\,M$. The
    colored curves are CCE evolutions started $40\,M$ before
    the peak of the pulse, in a radiative phase; the pale grey reference is the
    closed-form news of the wave, Eqs.~\eqref{eq:tw_news}
    and~\eqref{eq:tw_profile}. The black dashed line marks the start time
    $\mathring u_{0}=-40\,M$.}
    \label{fig:TeukolskyWave}
\end{figure}
}
\newcommand{\figBBHStrain}{%
\begin{figure*}[t]
    \begin{center}
    \includegraphics[width=\textwidth]{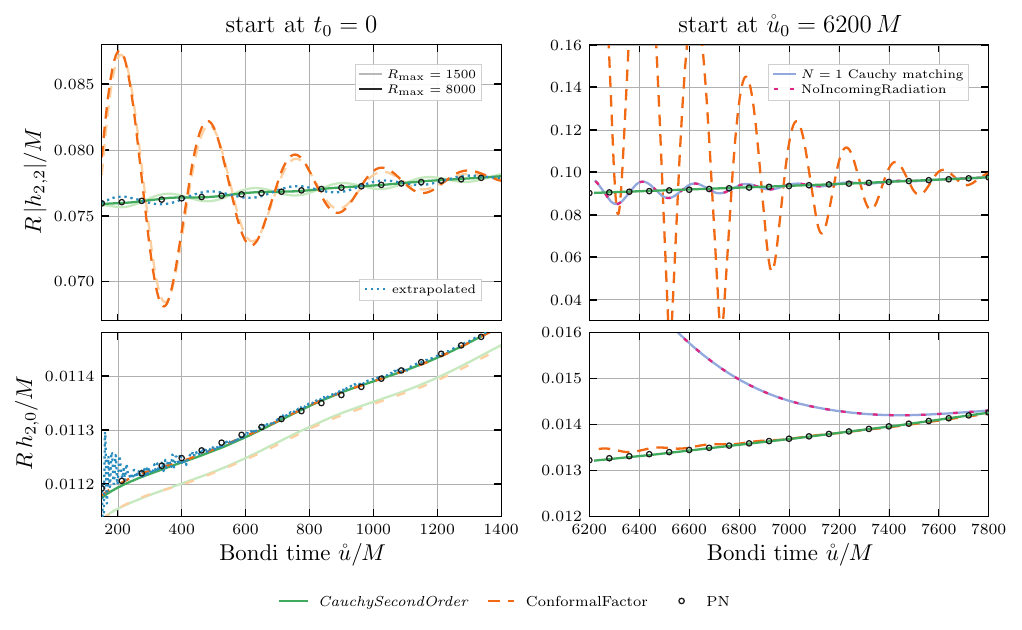}
    \end{center}
    \caption{Strain at $\Igroup$ for the equal-mass, non-spinning BBH with
    $R=300\,M$. The rows show $R|h_{2,2}|/M$ and $R\,h_{2,0}/M$; the latter is purely real for this system. Hollow circles show the post-Newtonian prediction.
    \emph{Left}: the production configuration, $t_{0}=0$, with
    \emph{CauchySecondOrder} and \emph{ConformalFactor} with
    $R_{\mathrm{max}}=1500$ and $8000\,M$ (light to dark), and the extrapolated
    waveform~\cite{Boyle:2009vi, Iozzo:2020jcu}
    of the $R_{\mathrm{max}}=1500\,M$ simulation shown by a dotted curve.
    \emph{Right}: the restart at $\mathring{u}_{0}=6200\,M$ with
    $R_{\mathrm{max}}=3000\,M$, with \emph{ConformalFactor},
    \emph{NoIncomingRadiation}, $N=1$ Cauchy matching, and
    \emph{CauchySecondOrder}.}
    \label{fig:BBHStrain}
\end{figure*}
}
\newcommand{\figTails}{%
\begin{figure}[t]
    \begin{center}
    \includegraphics[width=\columnwidth]{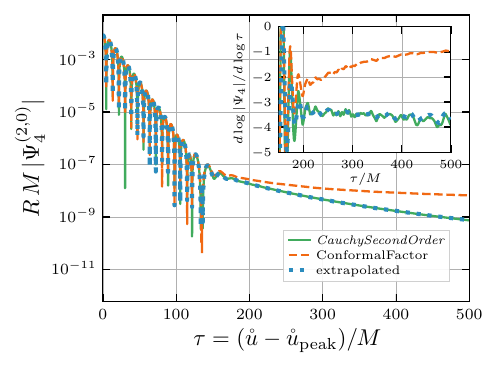}
    \end{center}
    \caption{Late-time tail of the head-on collision \texttt{SXS:BBH:3995} ($q=1$,
    initial separation $d_{0}=200\,M$, worldtube at $R=600\,M$), in the
    $(\ell,m)=(2,0)$ mode; each curve is aligned to its own peak.
    \emph{Main panel}: $RM|\Psi_{4}^{(2,0)}|$ against
    $\tau=\mathring u-\mathring u_{\mathrm{peak}}$, with the extrapolated
    reference shown as a dotted curve.
    \emph{Inset}: the local power-law index $d\log|\Psi_{4}|/d\log\tau$ over the
    tail.}
    \label{fig:Tails}
\end{figure}
}
\newcommand{\figIDAccuracy}{%
\begin{figure*}[t]
    \begin{center}
    \includegraphics[width=\textwidth]{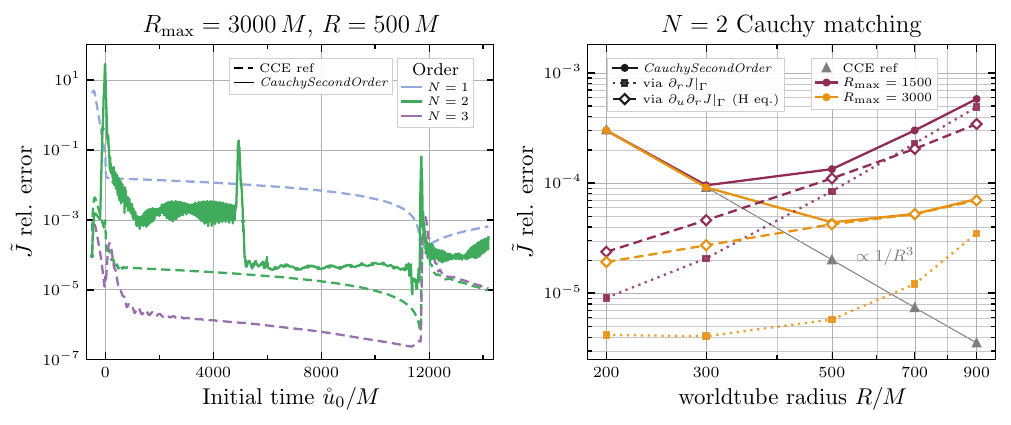}
    \end{center}
    \caption{Relative error $\epsilon(\tilde J)$ between the matching ansatz and
    the solution of a reference CCE evolution initialized with
    \emph{CauchySecondOrder} at $t_0=0$, evaluated on the same slice,
    Eq.~\eqref{eq:eps_def}.
    \emph{Left}: against initial time at $R_{\mathrm{max}}=3000\,M$ and
    $R=500\,M$, with color indicating matching order $N$. Dashed curves (\emph{CCE ref})
    take every worldtube derivative from the reference run; the solid curve takes every worldtube derivative from the worldtube itself and should be viewed as
    the production \emph{CauchySecondOrder}.
    \emph{Right}: the same errors against worldtube radius, averaged over
    $\mathring u_{0}\in[6500,8000]\,M$, with color indicating
    $R_{\mathrm{max}}$. Grey triangles show the \emph{CCE ref} truncation floor
    and circles the production data. Hollow diamonds and filled squares show
    the $\partial_{\tilde u}\partial_{r}J|_{\Gamma}$ and
    $\partial_{r}J|_{\Gamma}$ contributions, measured by replacing
    each input with its reference value. Together with the truncation error,
    these are the three terms of Eq.~\eqref{eq:eps_total}.}
    \label{fig:IDAccuracy}
\end{figure*}
}

\newcommand{\CornellPhysics}{\affiliation{Department of Physics, Cornell University, Ithaca, NY 14853, USA}}
\newcommand{\Cornell}{\affiliation{Cornell Center for Astrophysics and Planetary Science, Cornell University, Ithaca, NY 14853, USA}}
\newcommand{\CornellLepp}{\affiliation{Laboratory for Elementary Particle Physics, Cornell University, Ithaca, NY 14853, USA}}
\newcommand{\Caltech}{\affiliation{Theoretical Astrophysics 350-17, California Institute of Technology, Pasadena, CA 91125, USA}}
\newcommand{\UMiss}{\affiliation{Department of Physics and Astronomy, University of Mississippi, University, MS 38677, USA}}

\begin{document}

\title{Accurate initial data for Cauchy-characteristic evolution}

\author{Guido Da Re \orcidlink{0009-0007-2006-094X}} \email{gdare@caltech.edu} \Caltech
\author{Keefe Mitman \orcidlink{0000-0003-0276-3856}} \email{kem343@cornell.edu} \Cornell
\author{Leo C. Stein \orcidlink{0000-0001-7559-9597}} \UMiss
\author{Nils Deppe \orcidlink{0000-0003-4557-4115}} \CornellLepp \CornellPhysics \Cornell
\author{Himanshu Chaudhary \orcidlink{0000-0002-4101-0534}} \Caltech
\author{Dongze Sun \orcidlink{0000-0003-0167-4392}} \Caltech
\author{Michael Boyle \orcidlink{0000-0002-5075-5116}} \Cornell
\author{Kyle C.~Nelli \orcidlink{0000-0003-2426-8768}} \Caltech
\author{Mark A. Scheel \orcidlink{0000-0001-6656-9134}} \Caltech
\author{Saul A. Teukolsky~\orcidlink{0000-0001-9765-4526}} \Caltech\Cornell

\date{\today}

\begin{abstract}
Accurate gravitational-wave (GW) templates from numerical-relativity (NR)
simulations are essential for GW science. These templates are typically computed
at finite radii and extrapolated to future null infinity $\Igroup$, the
asymptotic boundary of spacetime. However, this procedure is known to not accurately capture
momentum fluxes and memory effects, limiting the accuracy of waveform models and data
analyses that rely on them. Cauchy-characteristic evolution (CCE) instead integrates Einstein's equations along outgoing null hypersurfaces from
a Cauchy simulation's finite-radius worldtube to $\Igroup$. Its accuracy,
though, has been limited by initial data that produce long-lived spurious
transients, which contaminate several cycles of the waveform emitted by the binary coalescence.
We identify incorrect initial angular coordinates as the source of these
transients. We construct initial data that matches the Cauchy worldtube data through
second order in the radial derivatives and solve for the angular coordinates consistent with the resulting radial profile. These improvements remove the
unphysical transient: a $10\%$
modulation of a binary black hole's $(2,2)$ mode amplitude lasting over
$1000\,M$ disappears, and the late-time tail of a head-on black hole collision is
recovered. The construction is independent of the Cauchy formulation
and allows archived simulations to be reprocessed with CCE, paving the way
for improved GW catalogs.
\end{abstract}

\maketitle

\section{Introduction}
\label{sec:introduction}
Gravitational-wave (GW) observations probe the populations and dynamics of
compact binaries. The catalog of compact-binary coalescences observed by the
LIGO--Virgo--KAGRA Collaboration~\cite{LIGOScientific:2014pky, VIRGO:2014yos,
KAGRA:2020tym} supports inference of stellar-mass black-hole mass and spin
distributions, tests of general relativity (GR) in the strong-field regime,
and standard-siren measurements of the expansion
history~\cite{LIGOScientific:2018mvr, LIGOScientific:2021djp, LIGOScientific:2026wfs, LIGOScientific:2026ctl, LIGOScientific:2026oim, LIGOScientific:2026uyd}. Individual events, such as
GW250114, can provide the most stringent tests of GR to
date~\cite{LIGOScientific:2025rid, LIGOScientific:2025wao}.
Each analysis compares observed data with a predicted waveform, so its accuracy
is limited by the accuracy of waveform models.

As detector sensitivities improve, the increasing signal-to-noise ratio (SNR)
requires more accurate waveforms~\cite{Lindblom:2008cm, Toubiana:2024car}. This demand will grow
with next-generation observatories such as Cosmic
Explorer~\cite{Reitze:2019iox, Evans:2021gyd} and the Einstein
Telescope~\cite{Punturo:2010zz, Maggiore:2019uih}. Space-based missions such as
LISA~\cite{LISA:2017pwj, Colpi:2024xhw} will observe long-duration, high-SNR
signals against a background of overlapping sources, making waveform accuracy
essential for separating individual systems. Waveform models are calibrated to
numerical-relativity (NR) simulations and inherit their limitations. Recent
studies~\cite{Purrer:2019jcp, Ferguson:2020xnm, Pompili:2024yec, Scheel:2025jct, Knapp:2025ecw}
indicate that current NR waveform errors can
bias parameter recovery with these instruments.

Part of this error comes from waveform extraction, the procedure used to obtain
a waveform from an NR simulation. Gravitational radiation is defined
unambiguously only at null infinity, where the distinct radial falloffs
separate the radiative and Coulombic degrees of freedom~\cite{Bondi:1962px,
Sachs:1962wk}. This separation occurs in gauge theories generally and, in a
generalized sense, in GR.\footnote{In GR the separation is unique only up to
the global Bondi--Metzner--Sachs (BMS) symmetries of future null infinity, under
which the radiation transforms covariantly.} In the bulk, local gauge freedom
makes the split ambiguous. Future null infinity $\Igroup$ also admits an
unambiguous idealization of an inertial detector and is therefore the natural
place to define the measured GW strain.

Standard NR simulations do not reach $\Igroup$. The most widely used methods
for solving Einstein's equations in the strong-field
regime~\cite{Pretorius:2005gq, Campanelli:2005dd, Baker:2005vv} evolve Cauchy
data on spacelike hypersurfaces within a finite domain, with approximate
outgoing-radiation boundary conditions. Waveforms are usually extracted by
extrapolation~\cite{Boyle:2009vi, Iozzo:2020jcu}: the strain or Weyl scalar
$\Psi_{4}$ is evaluated on finite-radius spheres and fit with polynomials in
$1/r$, or corrected with black-hole perturbation
theory~\cite{Nakano:2015pta}. This inexpensive method underlies most existing waveform
catalogs~\cite{Boyle:2019kee, Scheel:2025jct, Healy:2017psd, Healy:2019jyf, Healy:2020vre, Healy:2022wdn, Ficarra:2026cej, Jani:2016wkt, Ferguson:2023vta, Hamilton:2023qkv, Huerta:2019oxn}, but approximates both the gauge
and the dynamics in the wave zone. It does not accurately capture momentum
fluxes and nonlinear memory~\cite{Favata:2010zu, Mitman:2024uss, DaRe:2025glj}.

Cauchy-characteristic evolution (CCE) evolves the full nonlinear Einstein
equations along outgoing null hypersurfaces from a timelike worldtube supplied
by a Cauchy simulation.\footnote{An alternative is hyperboloidal
evolution, which uses spacelike slices that approach future null infinity
(see~\cite{Rinne:2009Hyperboloidal,Peterson:2024Hyperboloidal} and references
therein).} It yields the strain, news, and Weyl scalars in an
asymptotically inertial frame at $\Igroup$~\cite{Bishop:1996gt, Winicour:2005ge,
Babiuc:2011qi, Handmer:2016mls, Barkett:2019uae, Moxon:2020gha, Moxon:2021gbm},
recovering the physics missed by extrapolation~\cite{Pollney:2010hs,
Mitman:2020pbt, DaRe:2025glj}. It has been applied to both
generalized-harmonic~\cite{Mitman:2020pbt, Moxon:2021gbm} and
moving-puncture~\cite{Reisswig:2009us, Rashti:2024yoc} simulations. It also enables the BMS frame-fixing
procedures~\cite{Mitman:2021xkq, Mitman:2024uss} needed for waveform comparison,
post-Newtonian hybridization, and black-hole spectroscopy~\cite{Berti:2009kk,
Mitman:2022kwt, Berti:2025hly}. The \texttt{SpECTRE}~\cite{spectreCode}
implementation~\cite{Moxon:2020gha, Moxon:2021gbm} produces highly accurate
waveforms at a cost compatible with large NR campaigns.

CCE accuracy has nevertheless been limited by its initial data. Unlike Cauchy
initial data, which obey the Hamiltonian and momentum constraints, the
characteristic initial data are freely specifiable on a single null slice;
the constraints instead take the form of flux-balance laws in retarded time.
Physically, these data must encode both outgoing radiation and ingoing
radiation back-scattered from the system's causal past. The Cauchy simulation
provides neither the exterior metric nor the history before its initial time,
so existing prescriptions approximate this information. They produce a
long-lived, low-frequency transient that contaminates several inspiral cycles
and degrades post-ringdown features such as tails~\cite{DeAmicis:2024eoy}.
Such waveform errors can bias parameter estimation~\cite{Purrer:2019jcp}.
We call this transient \emph{characteristic junk},
by analogy with the junk from Cauchy initial data. It makes part of an expensive
Cauchy simulation unusable for accurate waveform extraction.

We identify the initial angular coordinates as the source of characteristic
junk and construct initial data free of it. Our scheme, \emph{CauchySecondOrder},
matches the radial profile to the Cauchy data through second order, enforces
asymptotic regularity, and determines the angular coordinates from that profile.
A companion Letter~\cite{companion} presents the main result; here we give the
derivation, implementation, and error budget. The construction is
independent of the Cauchy formulation and allows archived simulations with
stored worldtube data, including those in existing
catalogs~\cite{Boyle:2019kee,Scheel:2025jct,Healy:2017psd,Healy:2019jyf,
Healy:2020vre,Healy:2022wdn,Ficarra:2026cej,Jani:2016wkt,Ferguson:2023vta,
Hamilton:2023qkv,Huerta:2019oxn}, to be reprocessed at
$\Igroup$ without repeating their Cauchy evolutions. This includes both
generalized-harmonic and moving-puncture simulations. Surrogate, effective-one-body, and
phenomenological models can then be trained on waveforms evolved with
Einstein's equations all the way to $\Igroup$, improving their accuracy for
next-generation detectors.

We test the scheme on an analytic Teukolsky wave, an equal-mass, non-spinning
binary, and a head-on collision whose post-ringdown tail provides a stringent
test. In all three cases, the characteristic junk is removed and the accuracy
of the worldtube data becomes the limiting factor.
Section~\ref{sec:cce_formulation} fixes notation and reviews CCE. Readers
familiar with Refs.~\cite{Moxon:2020gha, Moxon:2021gbm} may start with
Sec.~\ref{sec:cce_id_problem}, which states the initial-data problem and reviews
existing schemes. Section~\ref{sec:characteristic_junk} distinguishes Cauchy
and characteristic junk; Sec.~\ref{sec:angular_problem} identifies the origin
of the latter and expresses the angular gauge condition as a Beltrami problem.
Section~\ref{sec:cauchy_matching_scheme} constructs the initial data and states
the conditions for a unique small-amplitude solution at fixed matching order.
Section~\ref{sec:results} presents the tests, Sec.~\ref{sec:id_accuracy}
quantifies the error budget, and Sec.~\ref{sec:disc-concl} summarizes our findings and envisioned future work.

\section{CCE formulation}
\label{sec:cce_formulation}
Cauchy-characteristic evolution propagates gravitational radiation from the
boundary of a finite computational domain to future null infinity~$\Igroup$
using the fully nonlinear Einstein equations. A standard Cauchy evolution
supplies the spacetime metric $g_{\mu\nu}$ and its derivatives throughout that
domain. Once the simulation is complete, the metric data are stored on a
coordinate sphere of fixed radius at a sequence of times. These spheres sweep
out a timelike worldtube, denoted $\Gamma$.

The worldtube typically lies in the radiation zone, where the field is weak.
The Cauchy coordinates, however, are chosen for numerical stability in the
strong-field region near the source and do not guarantee asymptotic regularity
at $\Igroup$. Moreover, nonlinearities beyond the Cauchy domain can make the
finite-radius data differ from those at $\Igroup$. CCE addresses both issues
by using the Cauchy data as inner boundary conditions for Einstein's equations
on outgoing null hypersurfaces extending to future null infinity. This evolution
also constructs the asymptotically inertial frame in which gravitational
radiation is computed.

\begin{figure}[t]
		\centering
		\sbox{\prlSchematicBox}{%
			%% Scri: routed through \Igroup so this file and the body of the paper cannot
%% disagree about the symbol.  \providecommand, not \newcommand, so the
%% paper's own definition wins when this file is \input into it, and the
%% standalone build still has one.
\providecommand{\Igroup}{\mathscr{I}^{+}}

% Color palette (same as junkpic.tex)
\definecolor{darkblue}{RGB}{0, 48, 100}
\definecolor{myblue}{RGB}{0, 114, 189}
\definecolor{mygreen}{RGB}{0, 150, 80}
\definecolor{myorange}{RGB}{230, 140, 0}
\definecolor{myred}{RGB}{220, 30, 30}
\definecolor{mypurple}{RGB}{130, 60, 180}

\begin{tikzpicture}[
    line cap=round,
    outgw/.style={line width=2.0pt, -{Stealth[length=3.2mm, width=2.6mm]},
      decorate, decoration={snake, amplitude=1.1mm, segment length=5mm,
        post length=3mm}},
    ingw/.style={line width=1.6pt, -{Stealth[length=3.2mm, width=2.6mm]},
      decorate, decoration={snake, amplitude=1.1mm, segment length=5mm,
        post length=3mm}},
    cauchyfoliation/.style={mypurple!40, line width=0.9pt},
    ccefoliation/.style={myorange!40, line width=0.9pt},
  ]

    % ------------------------------------------------------------------
  % Fainter foliations forward in time (drawn first, behind everything)
  % ------------------------------------------------------------------
  
  % First matched pair of slices
  \draw[cauchyfoliation] (-4.3,0.6) -- (0,0.6);
  \draw[ccefoliation]    (0,0.6) -- (4.3,4.9);
  
  % Second matched pair of slices
  \draw[cauchyfoliation] (-4.3,1.2) -- (0,1.2);
  \draw[ccefoliation]    (0,1.2) -- (4.0,5.2);

  % ------------------------------------------------------------------
  % Cauchy domain (left of the worldtube)
  % ------------------------------------------------------------------
  % Inspiraling binary worldlines from the Peters--Mathews formulas
  % (separation ~ (t_merg - t)^(1/4), phase ~ (t_merg - t)^(5/8)),
  % merging into a remnant
  \begin{scope}[shift={(-2.2,0)}]
    \newcommand{\Tmerg}{5.6}
    \newcommand{\Tscale}{30}
    \newcommand{\Rscale}{0.16}
    % The trailing `r' inside sin() tells TikZ we're using radians
    \draw[black, very thick, variable=\t, domain=0:\Tmerg, samples=500]
      plot ({\Rscale * (\Tscale*(\Tmerg-\t))^(1/4)
             * sin((\Tscale*(\Tmerg-\t))^(5/8) r)}, {\t});
    \draw[black, very thick, variable=\t, domain=0:\Tmerg, samples=500]
      plot ({-\Rscale * (\Tscale*(\Tmerg-\t))^(1/4)
             * sin((\Tscale*(\Tmerg-\t))^(5/8) r)}, {\t});
    % Remnant worldline after merger
    \draw[black, very thick] (0,\Tmerg) -- (0,7.0);
  \end{scope}
  % Backup: previous hand-drawn braid
  %\begin{scope}[shift={(-2.5,0)}]
  %  \draw[black, very thick]
  %    (-0.5,0) to[out=90,in=-90] (0.5,1) to[out=90,in=-90] (-0.4,1.8)
  %    to[out=90,in=-90] (0.3,2.5) to[out=90,in=-90] (-0.2,3.1)
  %    to[out=90,in=-90] (0.1,3.6) to[out=90,in=-90] (0,4.1) -- (0,7);
  %  \draw[black, very thick]
  %    (0.5,0) to[out=90,in=-90] (-0.5,1) to[out=90,in=-90] (0.4,1.8)
  %    to[out=90,in=-90] (-0.3,2.5) to[out=90,in=-90] (0.2,3.1)
  %    to[out=90,in=-90] (-0.1,3.6) to[out=90,in=-90] (0,4.1);
  %\end{scope}

  % Cauchy initial data surface
  \draw[mypurple, line width=1.4pt, double] (-4.3,0) --
    node[pos=0.45, below=3pt, font=\large\bfseries] {Cauchy initial data}
    (0,0);

  % Radiation from the binary crossing the worldtube (45 degrees)
  \draw[myblue, outgw] (-1.55, 0.80) -- (-0.25, 2.10);
  \draw[myblue, outgw] (-1.55, 2.75) -- (-0.25, 4.05);

  % ------------------------------------------------------------------
  % Characteristic domain (right of the worldtube)
  % ------------------------------------------------------------------
  % Worldtube
  \draw[darkblue, line width=2.5pt] (0,0) --
    node[pos=0.92, left=4pt, black, font=\Large] {$\Gamma$} (0,7);

  % Future null infinity
  \draw[black, line width=2.5pt] (1.4,7.8) --
    node[pos=0.05, above right=2pt, black, font=\Large] {$\Igroup$}
    (5.2,4.0);

  % CCE initial data slice
  \draw[myorange, line width=1.4pt, double] (0,0) -- (4.6,4.6);
  \node[myorange, font=\large\bfseries, rotate=45] at (2.29,1.41)
    {CCE initial data};

  % Radial coordinate along the slice
  %\draw[black, line width=1.2pt, -{Stealth[length=2.8mm, width=2.2mm]}]
  %  (0.35,1.55) -- (1.20,2.40) node[right=2pt, font=\large] {$r$};

  % Outgoing radiation (45 degrees, parallel to the null slices)
  \draw[myblue, outgw] ($(0.5,0.55)$) -- ($(1.9,1.85)$);
    
  \node[myblue, font=\small\bfseries,
  rotate=45,
  align=center] at (1.0,4.0) {Outgoing\\ radiation};

  % Back-scattered ingoing radiation crossing the slice (45 degrees)
  \draw[mygreen, ingw] (4.05,3.15) -- (3.05,4.15);
  \draw[mygreen, ingw] (4.60,3.70) -- (3.60,4.70);
  \node[mygreen, font=\small\bfseries,
  rotate=-45,
  align=center] at (2.1,5.8)
    {Back-scattered\\ ingoing radiation};

  % Matching point on the worldtube
  \fill[myred] (0,0) circle (0.15);
  \node[myred, anchor=north, font=\large\bfseries] at (1.2,-0.12) {Matching};
  \node[black, anchor=north, font=\large] at (1.2,-0.62)
    {$\partial_r^n J\rvert_\Gamma$};

\end{tikzpicture}%
		}%
		\resizebox{\columnwidth}{!}{\usebox{\prlSchematicBox}}%
		\caption{The characteristic initial-data problem. A Cauchy evolution
			(left) supplies data on a worldtube $\Gamma$; a characteristic
			evolution on null slices propagates radiation to future null infinity
			$\Igroup$. Orange is the initial slice, on which $J$ is specified; blue
			and green are outgoing and back-scattered ingoing radiation; red is the
			matching point, where $J|_{\Gamma}$ and the radial derivatives
			$\partial_r^n J\rvert_\Gamma$ are supplied.}
		\label{fig:schematic}
	\end{figure}
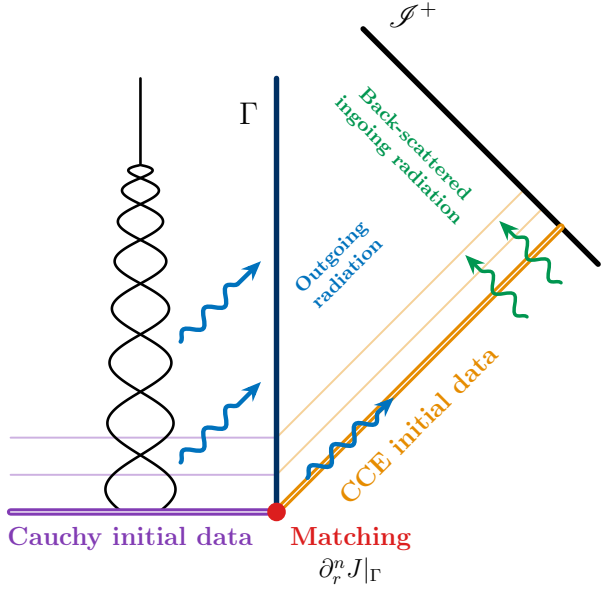

Figure~\ref{fig:schematic} shows the Cauchy and characteristic domains,
separated by the worldtube $\Gamma$. The worldtube boundary data alone do not
determine the characteristic evolution: initial data must also be supplied on
the first null hypersurface. These data encode both the dominant outgoing
radiation and the ingoing radiation back-scattered from the system's entire
causal past.

\begin{figure*}[t]
    \begin{center}
    \resizebox{\textwidth}{!}{%% Scri: routed through \Igroup so this file and the body of the paper cannot
%% disagree about the symbol.  \providecommand, not \newcommand, so the
%% paper's own definition wins when this file is \input into it, and the
%% standalone build still has one.
\providecommand{\Igroup}{\mathscr{I}^{+}}
\begin{tikzpicture}[
  %% Rounded teal content boxes
  mainbox/.style={
    rounded corners=10pt,
    fill=darkteal,
    text=white,
    minimum width=3.9cm,
    minimum height=2.7cm,
    text width=3.5cm,
    align=center,
    font=\normalsize\sffamily
  },
  %% Thick arrows between boxes
  arrowstyle/.style={
    -Stealth,
    ultra thick,
    draw=gray!55!black
  },
  %% Small dashed annotation boxes; the argument sets both frame and text color
  annotbox/.style={
    draw=#1,
    text=#1,
    dashed,
    fill=white,
    rounded corners=2pt,
    text width=3.05cm,
    font=\scriptsize,
    align=center,
    inner sep=3.5pt,
    line width=1.0pt
  },
  annotbox/.default=red,
  %% Labels on arrows
  eqlabel/.style={
    rounded corners=2pt,
    font=\footnotesize,
    midway,
    fill=white,
    draw=gray!55,
    line width=0.4pt,
    inner sep=3.2pt
  }
]

%% ─── Layout variables ─────────────────────────────────────────────────────────
%% Panel inner sep (same for all four panels)
\newcommand{\panelinnersep}{0.52cm}
%% Column x-centers (cm)
\pgfmathsetmacro{\xleft}{2.1}
\pgfmathsetmacro{\xmid}{7.7}
\pgfmathsetmacro{\xright}{14.1}
%% Row y-centers (cm)
\pgfmathsetmacro{\ytop}{-0.4}
\pgfmathsetmacro{\ybot}{-5.0}
%% Distance above a row center for the panel-title phantom node (cm)
\pgfmathsetmacro{\yphantom}{2.0}
%% Precomputed phantom y-coordinates
\pgfmathsetmacro{\ytopphantom}{\ytop + \yphantom}   %% = 1.75
\pgfmathsetmacro{\ybotphantom}{\ybot + \yphantom}   %% = -3.25
%% Panel title yshift (same for all four title labels)
\newcommand{\titleyshift}{-5pt}
%% Annotation box offsets from host-box corner: inward (x) and upward (y), in cm
\pgfmathsetmacro{\annotinset}{-0.30}
\pgfmathsetmacro{\annotlift}{0.35}
%% Arrow label gap from arrow line
\newcommand{\arrowlabeloffset}{6pt}
\newcommand{\maxtextwidth}{6cm}

%% ─── Main content boxes ───────────────────────────────────────────────────────

%% Left column
\node[mainbox] (adm) at (\xleft, \ytop) {
  Original ADM\\coordinates\\[5pt]
  $\{t',\, r',\, \phi'^{A'}\}$
};
\node[mainbox] (bs) at (\xleft, \ybot) {
  Bondi-Sachs\\coordinates\\[5pt]
  $\{\mathring{u},\, \mathring{r},\, \mathring{\phi}^{\mathring{A}}\}$
};

%% Middle column
\node[mainbox] (blc) at (\xmid, \ytop) {
  Bondi-like Cauchy\\coordinates\\[5pt]
  $\{u,\, r,\, \phi^{A}\}$
};
\node[mainbox] (blpf) at (\xmid, \ybot) {
  Bondi-like partially flat\\coordinates\\[5pt]
  $\{\hat u,\, \hat{r},\, \hat{\phi}^{\hat{A}}\}$
};

%% Right column
\node[mainbox] (nbc) at (\xright, \ytop) {
  Numerical Bondi-like\\Cauchy coordinates\\[5pt]
  $\{\tilde{u},\, \tilde{y},\, \tilde{\phi}^{\tilde{A}}\}$
};
\node[mainbox] (nbpf) at (\xright, \ybot) {
  Numerical Bondi-like\\partially flat coordinates\\[5pt]
  $\{\breve{u},\, \breve{y},\, \breve{\phi}^{\breve{A}}\}$
};

%% ─── Phantom anchor nodes to reserve vertical space for panel titles ──────────
%% These zero-size nodes push the panel backgrounds upward so title text fits.
\node[minimum width=0pt, minimum height=0pt, inner sep=0pt]
  (green-top)  at (\xleft,  \ytopphantom) {};
\node[minimum width=0pt, minimum height=0pt, inner sep=0pt]
  (cyan-top)   at (\xleft,  \ybotphantom) {};
\node[minimum width=0pt, minimum height=0pt, inner sep=0pt]
  (yellow-top) at (\xmid,   \ytopphantom) {};
\node[minimum width=0pt, minimum height=0pt, inner sep=0pt]
  (peach-top)  at (\xright, \ytopphantom) {};

%% ─── Colored background panels (drawn below everything) ──────────────────────
\begin{scope}[on background layer]
  %% Green — Cauchy coordinates
  \node[fill=cauchygreen, rounded corners=8pt,
        fit=(adm)(green-top), inner sep=\panelinnersep] (greenpanel) {};
  %% Cyan — Asymptotic inertial coordinates
  \node[fill=asyncyan, rounded corners=8pt,
        fit=(bs)(cyan-top), inner sep=\panelinnersep] (cyanpanel) {};
  %% Yellow — Bondi-like coordinates
  \node[fill=bondiyellow, rounded corners=8pt,
        fit=(blc)(blpf)(yellow-top), inner sep=\panelinnersep] (yellowpanel) {};
  %% Peach — Numerical Bondi-like coordinates
  \node[fill=numpeach, rounded corners=8pt, minimum width=\maxtextwidth+\panelinnersep,
        fit=(nbc)(nbpf)(peach-top), inner sep=\panelinnersep] (peachpanel) {};
\end{scope}

%% ─── Panel title labels (drawn on top of backgrounds) ────────────────────────
\node[align=center, font=\small\bfseries, anchor=north, yshift=\titleyshift, text width=\maxtextwidth]
  at (greenpanel.north) {Cauchy Coordinates};

\node[align=center, font=\small\bfseries, anchor=north, yshift=\titleyshift, text width=\maxtextwidth]
  at (cyanpanel.north) {Asymptotic inertial\\coordinates};

\node[align=center, anchor=north, yshift=\titleyshift, text width=\maxtextwidth] at (yellowpanel.north) {%
  {\normalsize\bfseries Bondi-like coordinates}\\[-1pt]%
  {\scriptsize [Metric in the form of Eq.~\eqref{eq:Bondilikemetric}]}%
};

\node[align=center, anchor=north, yshift=\titleyshift, text width=\maxtextwidth] at (peachpanel.north) {%
  {\normalsize\bfseries Numerical Bondi-like coordinates}\\[-1pt]%
  {\scriptsize [Compactified radial coordinate]}%
};

%% ─── Arrows ──────────────────────────────────────────────────────────────────
%% ADM  →  Bondi-like Cauchy  (horizontal right)
\draw[arrowstyle] (adm.east) -- (blc.west);
%  node[eqlabel, above=\arrowlabeloffset] {Eq.~\eqref{eq:adm_to_blc}};

%% Bondi-like Cauchy  →  Numerical Bondi-like Cauchy  (horizontal right)
\draw[arrowstyle] (blc.east) -- (nbc.west)
  node[eqlabel, above=\arrowlabeloffset] {Eq.~\eqref{eq:rtoy}};

%% Bondi-like Cauchy  →  Bondi-like partially flat  (vertical down)
\draw[arrowstyle] (blc.south) -- (blpf.north)
  node[eqlabel, right=\arrowlabeloffset] {Eqs.~\eqref{eq:generalangtrans}, \eqref{eq:angEOM}};

%% Bondi-like partially flat  →  Bondi-Sachs  (horizontal left)
\draw[arrowstyle] (blpf.west) -- (bs.east)
  node[eqlabel, above=\arrowlabeloffset] {Sec.~\ref{subsec:inertialBondiSachs}};

%% Bondi-like partially flat  →  Numerical Bondi-like partially flat  (horizontal right)
\draw[arrowstyle] (blpf.east) -- (nbpf.west)
  node[eqlabel, above=\arrowlabeloffset] {Eq.~\eqref{eq:rtoy}};

%% ─── Annotation boxes ────────────────────────────────────────────────────────
%% Placed so their top-left sits slightly inside the bottom-left of the host box,
%% causing them to overlap the host box's bottom edge (matching the PDF layout).

\node[annotbox, anchor=north west]
  at ($(adm.south west) + (\annotinset, \annotlift)$) {
  Coordinates from the Cauchy simulation
};

\node[annotbox, anchor=north west]
  at ($(bs.south west) + (\annotinset, \annotlift)$) {
  Strain is extracted at $\Igroup$ in these coordinates
};

\node[annotbox, anchor=north east]
  at ($(nbpf.south east) + (-\annotinset, \annotlift)$) {
  CCE is evolved in these coordinates
};

\node[annotbox=annotgreen, anchor=north east]
  at ($(nbc.south east) + (-\annotinset, \annotlift)$) {
  Cauchy matching is performed in these coordinates
};

\end{tikzpicture}}%
    \end{center}
    \caption{Coordinate systems used by the \texttt{SpECTRE} CCE module. The
    Cauchy simulation supplies metric data on the worldtube $\Gamma$; a
    Bondi-like system $\{x^{\alpha}\}$ is built along the outgoing null rays
    leaving it, and compactified to numerical coordinates
    $\{\tilde x^{\tilde\alpha}\}$. The evolution is carried out in the partially
    flat system $\{\hat x^{\hat\alpha}\}$, reached by the angular map
    $\hat\phi^{\hat A}(u,\phi^{A})$, with numerical counterpart
    $\{\breve x^{\breve\alpha}\}$; an inertial Bondi--Sachs system $\{\mathring x^{\mathring\alpha}\}$ is
    constructed near $\Igroup$. Each arrow is labeled with the corresponding map.}
    \label{fig:coordinateplot}
\end{figure*}
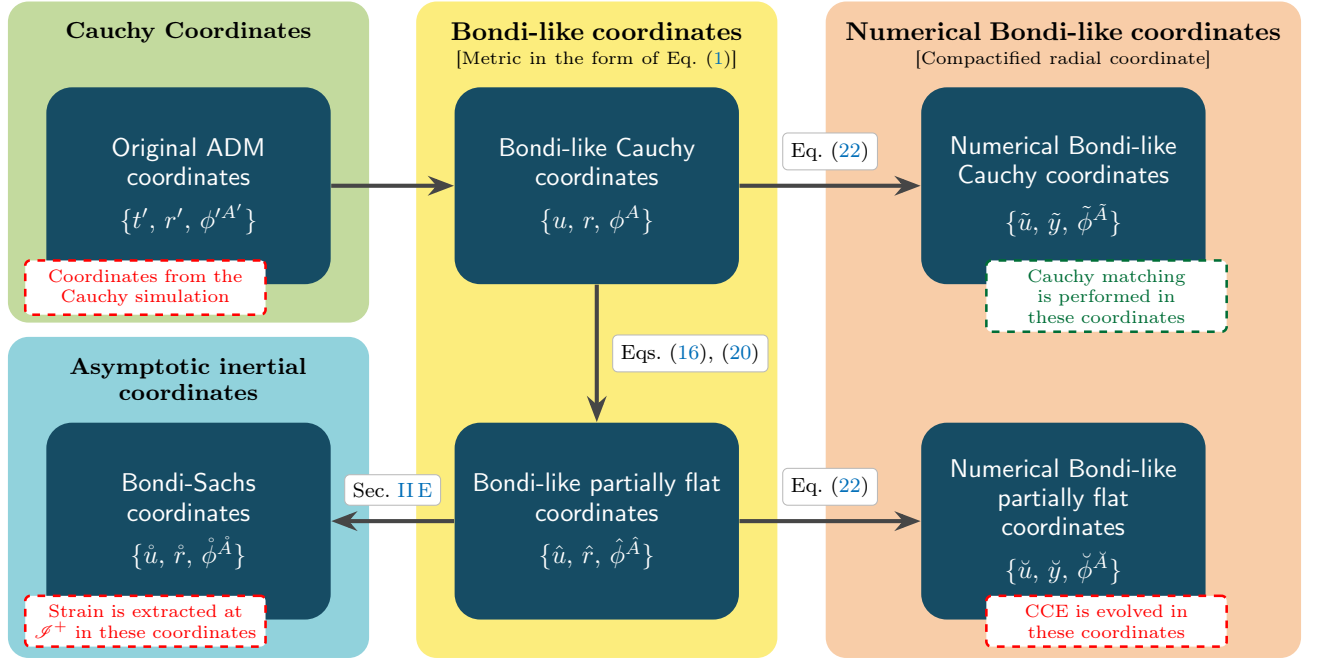

This section fixes notation and reviews the main ingredients of CCE.
Figure~\ref{fig:coordinateplot} provides an overview of the
coordinate systems and transformations introduced below. Readers
familiar with Refs.~\cite{Moxon:2020gha, Moxon:2021gbm} may skip to
Sec.~\ref{sec:cce_id_problem}.

\subsection{Bondi--Sachs metric}
\label{subsec:bondi_sachs_metric}
CCE uses a null foliation in a Bondi-like gauge. The coordinates
$(u,r,\phi^A)$ are constructed from outgoing radial null geodesics emanating
from the worldtube $\Gamma$. Geodesics originating at the same Cauchy time $t$
on $\Gamma$ generate a null hypersurface. We label it by the emission time,
setting $u=t$ on $\Gamma$ and keeping $u$ constant along each ray.
The radial coordinate $r$ is the areal (luminosity) radius along these geodesics,
defined by $r = (\det g_{AB}/\det q_{AB})^{1/4}$, where $q_{AB}$ is the metric
on the unit $S^2$. The angular coordinates $\phi^{A}$ remain constant along
each ray and label the points on each topological 2-sphere of constant $u$ and
$r$. The metric then takes the general Bondi-like form
\begin{align}
    ds^2 & = -\left(\frac{V}{r}e^{2\beta} - r^2 h_{AB} U^A U^B\right)du^2
    - 2 e^{2\beta} du\,dr \nonumber                                       \\
         & \quad - 2 r^2 h_{AB} U^B du\,d\phi^A
    + r^2 h_{AB} d\phi^A d\phi^B.
    \label{eq:Bondilikemetric}
\end{align}
In these coordinates the worldtube $\Gamma$ is not a surface of constant $r$;
it is defined by $r = R(u,\phi^{A})$. The Cauchy metric data are typically
given in Arnowitt-Deser-Misner (ADM) form. Expressions for converting them to a Bondi-like slicing at
the worldtube are given in Sec.~II\,B of Ref.~\cite{Moxon:2020gha} and Sec.~4 of
Ref.~\cite{Bishop:2016lgv}.

These coordinates do not yet define a Bondi--Sachs gauge: they do not approach
an asymptotically inertial, or Bondi--Metzner--Sachs (BMS), frame at
$\Igroup$~\cite{Bondi:1962px, Sachs:1962wk, Madler:2016xju}, so they cannot be
used directly for waveform extraction. The construction does not enforce the
falloff conditions required for the metric to approach the Minkowski metric
in inertial coordinates as $r\to\infty$,
\begin{subequations}\label{eq:asymptotics}
        \begin{align}
        \beta(x^{\alpha})  & = \mathcal{O}(r^{-1}),
        \label{eq:asymptotics:a}                                                 \\[4pt]
        \frac{V(x^{\alpha})}{r}      & = 1 + \mathcal{O}(r^{-1}),
        \label{eq:asymptotics:b}                                                 \\[4pt]
        U^{A}(x^{\alpha})  & = \mathcal{O}(r^{-2}),
        \label{eq:asymptotics:c}                                                 \\[4pt]
        h_{AB}(x^{\alpha}) & = q_{AB} + \mathcal{O}(r^{-1}),
        \label{eq:asymptotics:d}
    \end{align}
\end{subequations}
Instead, the Bondi-like coordinates inherit the time and angular coordinate
choices of the Cauchy simulation, which generically give the metric coefficients
a non-inertial falloff. CCE recovers the inertial behavior in two steps. It
first transforms the angular coordinates into the so-called \emph{partially flat
gauge}, so named because it enforces $\hat U^{(0)\hat A}=0$ and $\hat J^{(0)}=0$, while leaving
the asymptotic time coordinate unfixed. Here the superscript $(0)$ denotes the radially constant
term in the asymptotic expansion. This transformation is applied to the worldtube data, and
the gauge conditions are preserved by the evolution equations throughout the characteristic domain.
The remaining Bondi--Sachs conditions
are then imposed by transforming the time coordinate in a neighborhood of
$\Igroup$ so that it matches the inertial time measured by asymptotic observers,
with accompanying subleading corrections to the angular and radial coordinates.
\subsection{Spin-weighted formalism and \texorpdfstring{conventions}{conventions}}
\label{subsec:spin_weighted_formalism}

We use the spin-weighted formalism to describe the gauge transformations and
Einstein's equations in Bondi-like coordinates.

A complex null dyad $q^A$ satisfying the normalization conditions $q_A q^A = 0$ and $q_A \bar q^A = 2$ allows one to
write the unit $S^{2}$ metric as
\begin{align}
    q_{AB} = \tfrac12(q_A \bar q_B + \bar q_A q_B).
    \label{eq:dyad_metric}
\end{align}
Adopting standard spherical coordinates $\phi^{A} = (\theta,\phi)$, we take
\begin{align}
    q^{A} = \left( -1, -\frac{i}{\sin\theta}\right).
    \label{eq:dyad_choice}
\end{align}
Any angular tensor may be encoded as a spin-weighted scalar via
\begin{align}
    v = v_{A_1 \dots A_n B_1 \dots B_m}q^{A_1}\dots q^{A_n}\bar{q}^{B_1}\dots \bar{q}^{B_m},
    \label{eq:spin_weighted_def}
\end{align}
where the spin weight is $s = n - m$.

Angular derivatives of spin-weighted scalars are expressed using the spin-raising and
spin-lowering operators $\eth$ and $\bar\eth$,
\begin{subequations}\label{eq:spin_derivatives}
    \begin{align}
        \eth v
         & = q^{A_1}\dots q^{A_n}\bar{q}^{B_1}\dots \bar{q}^{B_m} \, q^{C} D_{C} v_{A_1 \dots A_n B_1 \dots B_m},
        \label{eq:spin_derivatives:a}                                                                                   \\[6pt]
        \bar{\eth} v
         & = q^{A_1}\dots q^{A_n}\bar{q}^{B_1}\dots \bar{q}^{B_m} \, \bar{q}^{C} D_{C} v_{A_1 \dots A_n B_1 \dots B_m},
        \label{eq:spin_derivatives:b}
    \end{align}
\end{subequations}
where $D_A$ is the covariant derivative on $S^2$.
Acting on a field of spin weight $s$, $\eth$ returns one of spin weight
$s+1$ and $\bar\eth$ one of spin weight $s-1$.

Any spin-weighted scalar $v$ may be expanded in spin-weighted spherical
harmonics (SWSHs), $v = \sum_{\ell,m} v_{\ell,m}\, {}_{s}Y_{\ell,m}$, and throughout this
paper $\lVert\cdot\rVert$ denotes the $L^{2}$ norm of such a field over the sphere,
taken at fixed retarded time and radius,
\begin{align}
    \lVert v\rVert \;\equiv\; \Big(\sum_{\ell,m}\lvert v_{\ell,m}\rvert^{2}\Big)^{1/2}.
    \label{eq:angular_norm}
\end{align}
Single bars $\lvert\cdot\rvert$ are reserved for the pointwise modulus of a
field.
We also use two subscripted variants. The first
restricts the sum to fixed $\ell$,
\begin{align}
    \lVert v\rVert_{\ell} \;\equiv\; \Big(\sum_{m=-\ell}^{\ell}\lvert v_{\ell,m}\rvert^{2}\Big)^{1/2}.
    \label{eq:angular_norm_ell}
\end{align}
The second,
$\lVert\cdot\rVert_{\Sigma}$, integrates over an entire null slice $\Sigma$, including both the radial and angular
directions.

We also use the supremum over the sphere,
\begin{align}
    \lVert v\rVert_{\infty} \;\equiv\; \sup_{S^{2}}\,\lvert v\rvert .
    \label{eq:sup_norm}
\end{align}
\subsection{Characteristic field equations}
\label{subsec:characteristic_field_equations}
The spin-weighted ``Bondi'' scalars
encode the Bondi-like metric components and serve as the numerical CCE
variables:
\begin{subequations}\label{eq:spinweights}
    \begin{align}
        U       & \equiv U^{A} q_{A}, \label{eq:spinweights:a} \\[4pt]
        Q       & \equiv r^{2} e^{-2\beta} q^{A} h_{AB} \partial_{r} U^{B},
                \label{eq:spinweights:b} \\[4pt]
        r^{2} W & \equiv V - r, \label{eq:spinweights:c} \\[4pt]
        J       & \equiv \tfrac12 q^{A} q^{B} h_{AB}, \label{eq:spinweights:d} \\[4pt]
        K       & \equiv \tfrac12 q^{A} \bar q^{B} h_{AB}
        = \sqrt{1 + J\bar J}, \label{eq:spinweights:e}
    \end{align}
\end{subequations}
together with the auxiliary variable $H = \partial_u J$ and the
spin-weight-zero function $\beta$ of Eq.~\eqref{eq:Bondilikemetric}.
The scalar $\beta$ acts as a lapse for the null foliation, relating the
retarded time that labels the hypersurfaces to the affine parameter along
their generators. The corresponding shift $U$ measures the angular drift of
the outgoing rays. The auxiliary variable $Q$ carries its radial derivative,
making the Einstein equation for $U$ first order. The scalar $W$ encodes the
Bondi mass aspect through $r^{2}W \to -2M_{B}$ as $r\to\infty$, with $M_{B}$
the Bondi mass.

Of the two spin-weighted scalars built from the angular metric, $K$ is not
independent: the unit-determinant condition $\det h_{AB} = \det q_{AB}$ fixes
it as $K=\sqrt{1+J\bar J}$. The leading asymptotic behavior of $H$ determines the Bondi news. The remaining
scalar, $J$, is the field specified by the initial data. It measures the
departure of the angular metric from the unit-sphere metric. Its $1/r$
coefficient at $\Igroup$ in the Bondi--Sachs gauge is the gravitational-wave
strain.

In the Bondi-like coordinates of Sec.~\ref{subsec:bondi_sachs_metric}, and using
the spin-weighted scalars defined above, Einstein's equations reduce to a
hierarchy of first-order radial ordinary differential equations (ODEs) on each
null hypersurface $u=\mathrm{const}$, the hypersurface equations:
\begin{subequations}\label{eq:hypersurface}
    \begin{align}
        \partial_{r} \beta                        & = S_{\beta}(J), \label{eq:hypersurface_beta}                 \\
        \partial_{r}\!\left[r^{2} Q\right]        & = S_{Q}(J,\beta), \label{eq:hypersurface_Q}                  \\
        \partial_{r} U                            & = S_{U}(J,\beta,Q), \label{eq:hypersurface_U}                \\
        \partial_{r}\!\left[r^{2} W\right]        & = S_{W}(J,\beta,Q,U), \label{eq:hypersurface_W}              \\
        \partial_{r}\!\left[r H\right] + L_{H} H + L_{\bar{H}} \bar{H}
                                                  & = S_{H}(J,\beta,Q,U,W), \label{eq:hypersurface_H}
    \end{align}
\end{subequations}
followed by the evolution equation
\begin{align}
    \partial_{u} J = H .
    \label{eq:Jevol}
\end{align}
Equations~\eqref{eq:hypersurface} follow from the metric
form~\eqref{eq:Bondilikemetric} alone, and therefore hold in any Bondi-like
gauge; only the asymptotic conditions~\eqref{eq:asymptotics}
distinguish the Bondi--Sachs gauge. Both Bondi-like systems introduced below are
consequently governed by the same hierarchy.
The remaining components of the Einstein equations are constraints encoding
conservation laws across null slices; their $r\to\infty$ limits, combined with the Bianchi
identities, correspond to the BMS balance laws~\cite{Madler:2016xju,
Flanagan:2015pxa, Strominger:2017zoo}.

The CCE evolution begins with prescribed data on an initial null hypersurface
$\Sigma_{0}$ at $u = u_{0}$. Starting from $J$ on that surface, one integrates
the hypersurface equations in sequence to obtain $H$, using the worldtube data
as boundary conditions for each equation. Equation~\eqref{eq:Jevol} then
advances $J$ to the next hypersurface. Repeating these steps determines the
metric throughout the domain from $\Gamma$ to $\Igroup$.
\subsection{The partially flat gauge}
\label{subsec:partially_flat_gauge}
The partially flat gauge is the first step toward the asymptotically inertial
frame described in Sec.~\ref{subsec:bondi_sachs_metric}.
At fixed retarded time, the worldtube $\Gamma$ is a coordinate sphere of the
Cauchy simulation, with boundary data expressed in the Cauchy gauge. The
construction of Sec.~\ref{subsec:bondi_sachs_metric} carries its angular
coordinates $\phi^{A}$ unchanged along the outgoing null rays to $\Igroup$.
These coordinates are chosen for strong-field stability rather than for their
behavior at large $r$. Consequently, the null-hypersurface cross-sections are
not asymptotically round: $J$ retains an $\mathcal{O}(1)$ contribution instead
of falling off as $\mathcal{O}(1/r)$.

In this gauge, the constant and $r^{-2}$ contributions to $J$
can produce a net $1/r$ contribution to the right-hand sides of
Eqs.~\eqref{eq:hypersurface}~\cite{Moxon:2020gha}, as they contain derivatives and nonlinear
combinations of the fields weighted by powers of $r$. Radial integration
then generates $r^{-n}\log r$ terms, degrading spectral convergence~\cite{Moxon:2020gha, Moxon:2021gbm}.
Earlier CCE implementations~\cite{Bishop:1996gt,
Babiuc:2011qi, Reisswig:2009us, Handmer:2014qha, Handmer:2016mls} evolved in
such coordinates and applied both the angular and time transformations only
near $\Igroup$, at the extraction step. Such transformations at the extraction step do not remove logarithms
arising during the evolution, which can limit waveform accuracy.

The partially flat gauge of Ref.~\cite{Moxon:2020gha}, used in
\texttt{SpECTRE}'s CCE module, instead applies the angular transformation at
the worldtube, directly to the evolved variables. Together with the initial-data regularity
conditions below, this
prevents logarithmic terms from arising in the bulk evolution
and leaves the transformation to inertial Bondi--Sachs coordinates to be performed
near $\Igroup$. The resulting coordinates $\{\hat x^{\hat \alpha}\}$ are
designed to preserve
\begin{align}
    \hat U^{(0)\hat A}=0
    \label{eq:pfgaugeUasymptotics}
\end{align}
throughout the evolution. In the partially flat gauge, $\hat U^{\hat A}=\mathcal{O}(\hat r^{-1})$ in general.
The Bondi--Sachs falloff in Eq.~\eqref{eq:asymptotics:c} is obtained after
the final transformation to inertial Bondi--Sachs coordinates described in
Sec.~\ref{subsec:inertialBondiSachs}~\cite{Moxon:2020gha}.

Reference~\cite{Moxon:2020gha} proves that in this gauge
the falloff of $\hat J$ is preserved by the evolution: if $\hat J$ on the
initial hypersurface contains neither a constant term $\hat J^{(0)}$ nor a
second-order term $\hat J^{(2)}$ in its $1/\hat r$ expansion, i.e.
\begin{align}
    \hat J(u_0,\hat r,\hat \phi^{\hat A}) = \sum_{\substack{n = 1 \\ n \ne 2}}^{\infty} \frac{\hat J^{(n)}(u_0,\hat \phi^{\hat A})}{\hat r^n} ,
    \label{eq:pfconstraintJ}
\end{align}
then
\begin{align}
    \partial_{\hat u}\hat J^{(0)}=0,
    \qquad \partial_{\hat u}\hat J^{(2)}=0.
    \label{eq:pfgaugeJpreservation}
\end{align}
Both the initially vanishing coefficients
therefore remain zero throughout the evolution.
Together with $\hat U^{(0)\hat A}=0$, these conditions
ensure that the net $1/\hat r$ contributions to the radial integrands vanish,
preventing logarithmic terms. They also yield simple expressions for the
leading asymptotic Weyl scalars.

The transformation for mapping to partially
flat coordinates has the general form
\begin{equation}
    \begin{cases}
        \hat{u} = u,                            \\
        \hat{r} = \hat{\omega}(u,\phi^{A})\, r, \\
        \hat{\phi}^{\hat{A}} = \hat{\phi}^{\hat{A}}(u,\phi^{A}),
    \end{cases}
    \label{eq:generalangtrans}
\end{equation}
where the conformal rescaling of the radial coordinate is required to preserve
the Bondi-like form~\eqref{eq:Bondilikemetric} of the metric, and $\hat\omega$
is obtained from the Jacobian determinant of the angular transformation. We
encode the angular Jacobian by the two spin-weighted scalars
\begin{subequations}\label{eq:angjacobians}
    \begin{align}
        \hat{a} & = \hat{q}^{\hat{A}}\partial_{\hat{A}}\phi^{A}\, q_{A}, \label{eq:angjacobians:a}       \\
        \hat{b} & = \bar{\hat q}^{\hat{A}}\partial_{\hat{A}}\phi^{A}\, q_{A}, \label{eq:angjacobians:b}
    \end{align}
\end{subequations}
in terms of which
\begin{equation}
    \hat{\omega} = \tfrac{1}{2}\sqrt{\hat{b}\bar{\hat b} - \hat{a}\bar{\hat a}}.
    \label{eq:omega_def}
\end{equation}
Under this transformation the Bondi scalars transform nonlinearly; in particular
\begin{subequations}\label{eq:pftrans}
    \begin{align}
        \hat{\beta} & = \beta - \tfrac{1}{2}\ln{\hat{\omega}},\label{eq:pftrans:a}                                                 \\
        \hat{J}     & = \frac{\bar{\hat b}^2 J + \hat{a}^2\bar{J} + 2\hat{a}\bar{\hat b}K}{4\hat{\omega}^2}.\label{eq:pftrans:b}
    \end{align}
\end{subequations}
The angular map is conformal precisely when $\hat a=0$, in which
case it is a M\"obius transformation of $S^2$. Although the M\"obius group is
isomorphic to the Lorentz group $\mathrm{SO}^{+}(1,3)$, the coordinate transformation in
Eq.~\eqref{eq:generalangtrans} leaves time unchanged. Completing the
corresponding Lorentz transformation generally requires the time rescaling
discussed in Sec.~\ref{subsec:inertialBondiSachs}.

To enforce the gauge condition~\eqref{eq:pfgaugeUasymptotics}, the angular
transformation in Eq.~\eqref{eq:generalangtrans} must satisfy the evolution
equation~\cite{Moxon:2020gha, Moxon:2021gbm}
\begin{align}
    \partial_{u}\hat{\phi}^{\hat{A}} = - \partial_{A}\hat{\phi}^{\hat{A}}\, U^{(0)A},
    \label{eq:angEOM}
\end{align}
where $U^{(0)A}$ is the leading, $r$-independent, term in
the radial expansion of $U^A$. A brief derivation is given in Appendix~\ref{sec:pf_gauge_eom}. This ODE
is solved numerically alongside the field equations, and therefore requires an
initial angular map $\hat{\phi}^{\hat{A}}(u = u_0, \phi^{A})$ to be specified as
part of the initial data.

 Equation~\eqref{eq:pfgaugeUasymptotics} is maintained dynamically
by Eq.~\eqref{eq:angEOM}, whereas the regularity conditions in
Eq.~\eqref{eq:pfconstraintJ} must be imposed on
the initial slice. The conditions $\hat J^{(0)}=0$ and
$\hat J^{(2)}=0$ play distinct roles in the initial-data construction.
In this work, we show that $\hat J^{(0)}=0$ constrains the non-conformal
part of the initial angular map, as developed in
Sec.~\ref{sec:angular_problem}.

\subsection{Inertial Bondi--Sachs coordinates at \texorpdfstring{$\Igroup$}{scri+}}
\label{subsec:inertialBondiSachs}
The second step completes the inertial gauge. A time
reparametrization, together with subleading angular and radial coordinate
corrections, enforces the remaining Bondi--Sachs conditions near
$\Igroup$. Denoting the resulting inertial Bondi--Sachs coordinates by
$\{\mathring{x}^{\mathring \alpha}\}$, the transformation Jacobian gives
\begin{align}
    \partial_{\hat u}\mathring{u}\big|_{\Igroup} = e^{2\hat \beta^{(0)}},
    \label{eq:inertialtime}
\end{align}
where $\hat\beta^{(0)}$ is the leading-order term in the $1/\hat r$ expansion
of $\hat\beta$. Equation~\eqref{eq:inertialtime} evolves in retarded time and
requires an initial condition,
$\mathring{u}|_{\Igroup}(u_0, \hat \phi^{\hat A}) = \alpha(\hat \phi^{\hat A})$.
This function on the sphere is freely specifiable.
Note that if the angular map $\hat\phi^{\hat A}(\phi^{A})$ is a conformal
transformation with $\hat\omega\ne1$, it changes $\hat\beta$ through
Eq.~\eqref{eq:pftrans:a} and therefore rescales the Bondi--Sachs time through
Eq.~\eqref{eq:inertialtime}. This completes the correspondence between the
conformal part of the angular map and a Lorentz transformation of the
Bondi--Sachs coordinates. The initial value in Eq.~\eqref{eq:inertialtime}
corresponds instead to a supertranslation, which together with the conformal
part of the initial angular map selects the BMS frame associated with the
characteristic initial slice.

\subsection{Numerical coordinates}
\label{subsec:numerical_coordinates}
In a Bondi-like coordinate system the radial coordinate $r$ ranges from the
worldtube radius $R(u,\phi^{A})$ to $+\infty$. Because an infinite domain cannot
be represented numerically, we apply the radial compactification
\begin{equation}
    \begin{cases}
        \tilde{u} = u,                              \\
        \tilde{y} = 1 - \dfrac{2R(u, \phi^{A})}{r}, \\
        \tilde{\phi}^{\tilde{A}} = \delta_{ A}^{\tilde{A}} \phi^{ A},
    \end{cases}
    \label{eq:rtoy}
\end{equation}
so that $r \in [R, +\infty) \to \tilde y \in [-1, 1]$. The worldtube is at
the fixed grid location $\tilde y = -1$, and $\Igroup$ is at $\tilde y = +1$.
The associated Jacobian is
\begin{equation}
    \begin{cases}
        \partial_{u} = \partial_{\tilde{u}} - (1 - \tilde{y})\dfrac{\partial_{\tilde{u}}R}{R}\partial_{\tilde{y}}, \\
        \partial_{r} = \dfrac{(1-\tilde{y})^2}{2R}\partial_{\tilde{y}},                                            \\
        \eth = \tilde{\eth} - (1 - \tilde{y})\dfrac{\tilde{\eth}R}{R}\partial_{\tilde{y}},
    \end{cases}
    \label{eq:jacobians}
\end{equation}
together with $\partial_{\tilde u}R = \partial_{u}R$ and
$\tilde\eth R = \eth R$. All Bondi scalars are invariant under this coordinate transformation except
for $H$:
\begin{align}
    \tilde{H} = H + \frac{r}{R}\partial_{u}R \, \partial_{r}J.
    \label{eq:Hbreve_def}
\end{align}
This is because $\tilde H$ is the retarded-time derivative of $J$
taken at fixed grid location, which follows the worldtube as $R$ changes,
whereas $H$ is taken at fixed Bondi radius.

The two Bondi-like
coordinate systems defined so far are the \emph{Cauchy} coordinates, $\{x^{\alpha}\}$, built
on the worldtube from the Cauchy angular coordinates, and the
\emph{partially flat} coordinates, $\{\hat x^{\hat\alpha}\}$, used for the
evolution. Their compactified counterparts are $\{\tilde
x^{\tilde\alpha}\}$ and $\{\breve x^{\breve \alpha}\}$, respectively.
Thus
$\tilde J$ denotes the Cauchy-gauge $J$ on the numerical grid, and $\breve J$
denotes its partially flat counterpart.
Note that since $\hat r=\hat\omega r$ and
$\hat R=\hat\omega R$, the two compactified radial coordinates in fact coincide,
$\breve y=\tilde y$.

Expansion coefficients follow the same convention: $\tilde J^{(n)}$ and
$\breve J^{(n)}$ are the coefficients of $(1-\tilde y)^{n}$ and
$(1-\breve y)^{n}$, while $J^{(n)}$ and $\hat J^{(n)}$ are those of $1/r^{n}$ and
$1/\hat r^{n}$, so that $\tilde J^{(n)} = J^{(n)}/(2R)^{n}$ and
$\breve J^{(n)} = \hat J^{(n)}/(2\hat R)^{n}$. Compactification leaves the
zeroth coefficients unchanged.

The \texttt{SpECTRE} CCE implementation expresses all Bondi scalars and
Einstein's equations in partially flat numerical coordinates.
Under Eq.~\eqref{eq:rtoy}, each equation in the
hierarchy~\eqref{eq:hypersurface} becomes an ODE in $\breve y$ on the fixed
grid $[-1,1]$. The radial weights $r^{2}$ and $r$ become powers of
$(1-\breve y)$, while the hierarchy retains its structure. Expressions for
the resulting source functions and the operators $L_{\breve H}$,
$L_{\bar{\breve H}}$ are given in Sec.~IV [Eqs.~(52)--(66)] of
Ref.~\cite{Moxon:2020gha}; the hierarchy in the present notation is
Eq.~(9) of Ref.~\cite{Moxon:2021gbm}.

\section{The characteristic initial-data problem}
\label{sec:cce_id_problem}
This section describes the data required on the first null hypersurface,
their formal freedom and physical constraints, and the three pre-existing
schemes used to construct them.

As in the Cauchy problem, characteristic initial data represent the system at
a finite time: we cannot evolve indefinitely from past null infinity. We must
specify $J$ on an entire null hypersurface $\Sigma_{0}$ at an initial time $u = u_0$. This specification is
necessarily approximate. Determining $J$ on $\Sigma_{0}$ requires knowledge
of the spacetime from $\Gamma$ to $\Igroup$, whereas the Cauchy simulation
supplies data only on the worldtube.

The initial-data problem in the characteristic formulation is fundamentally
different from its Cauchy counterpart. In the Cauchy problem, Einstein's
equations yield elliptic constraints on the initial spacelike hypersurface. In a
Bondi-like gauge there is no analogous radial constraint on a single null
hypersurface; the constraints appear instead as retarded-time flux-balance laws
that must hold on each timelike worldtube of fixed $r$. The initial data is
therefore, in principle, a freely specifiable field $J$ on the initial null
slice.

Physical consistency constrains this freedom. As discussed in
Sec.~\ref{sec:cce_formulation}, $J$ on $\Sigma_{0}$ encodes outgoing radiation
and ingoing radiation back-scattered from the system's entire causal past.
Neither is known throughout the slice: the Cauchy simulation provides the
metric neither outside $\Gamma$ nor before its own initial slice. We can
nevertheless seek consistency between the Cauchy and characteristic initial
slices by requiring them to describe the same ingoing and outgoing radiation.
The radial derivatives $\partial_r^{n}J|_{\Gamma}$ determine the radial
profile of $J$ on $\Sigma_{0}$ order by order. Matching the differential
structure of $J$ at the worldtube therefore provides a local proxy for the
radiation content of the entire null slice.

Given the partially flat formulation, once the initial null slice $\Sigma_0$ is
chosen the following initial data must be provided:
\begin{enumerate}
    \item the full specification of $\breve{J}_{\Sigma_0}$ on $\Sigma_0$ in
      numerical partially flat coordinates;
    \item the angular map from Cauchy to partially flat gauge,
      $\hat{\phi}^{\hat {A}}(u_0, \phi^{A})$, at the initial time $u_0$;
    \item the initial inertial time
      $\mathring{u}|_{\Igroup} (u_0, \hat \phi^{\hat A}) = \alpha(\hat \phi^{\hat A})$.
\end{enumerate}
The conformal part of $\hat{\phi}^{\hat {A}}(u_0, \phi^{A})$ and the function
$\alpha(\hat \phi^{\hat A})$ constitute the BMS freedom of the initial data,
which is fixed in post-processing~\cite{Mitman:2021xkq, Mitman:2024uss}.
In \texttt{SpECTRE}, we initialize the Bondi time with
$\alpha=0$. When loading the waveforms into \texttt{scri}, we shift
the time origin to adopt the convention $\mathring u\approx t-R$.
For evolutions started at $t_0=0$, the initial slice therefore
corresponds to $\mathring u_0=-R$ in the waveform data and plots,
before any further alignment specified in the captions.
The construction therefore reduces to specifying $\breve J_{\Sigma_0}$ and
the non-conformal part of $\hat{\phi}^{\hat {A}}(u_0, \phi^{A})$.

Given the expected smoothness of $J_{\Sigma_0}$ at $\Igroup$, in any Bondi-like
gauge this field can be approximated by an asymptotic expansion in $1/r$ about
future null infinity,
\begin{align}
    J_{\Sigma_0}(u_{0}, r,\phi^{A}) = \sum_{n = 0}^{\infty}\frac{J_{\Sigma_0}^{(n)}(u_{0},\phi^{A})}{r^n}.
    \label{eq:JSigma0_asymptotic}
\end{align}
Because \texttt{SpECTRE} evolves in the partially flat gauge, the constructed
$J$ must additionally satisfy Eq.~\eqref{eq:pfconstraintJ}, so the initial data
must take the form
\begin{align}
    \breve J_{\Sigma_0}(\breve{u}_0, \breve y, \breve \phi^{\breve A}) = \sum_{\substack{n = 1 \\ n \ne 2}}^{N+2} \breve J^{(n)}_{\Sigma_0}(\breve{u}_0, \breve \phi^{\breve A})(1 - \breve y)^n,
    \label{eq:generalIDAnsatzpf}
\end{align}
where $\breve y$ is the compactified coordinate of Eq.~\eqref{eq:rtoy}, and
the coefficients follow the convention of
Sec.~\ref{subsec:numerical_coordinates}. The sum is truncated at $N+2$, with
$N$ the highest available worldtube radial-derivative order (usually $N=1$).

In principle, the coefficients can be obtained from post-Minkowskian
calculations~\cite{Bini:2017wfr, Bern:2019nnu}, assumptions about the ingoing
radiation, or differentiable matching to the worldtube data. At minimum, the
ansatz must be continuous with the boundary data for stability; accuracy
requires higher-order differentiability~\cite{Moxon:2020gha, Moxon:2021gbm}.
The existing schemes, summarized in the following sections, use either an assumption about ingoing
radiation or matching to the worldtube data.
\subsection{\texorpdfstring{Schemes with a freely prescribed non-conformal map}%
{Schemes with a freely prescribed non-conformal map}}
\label{subsec:pfmatchingscheme}
This class of schemes prescribes the non-conformal part of the initial
map $\hat\phi^{\hat A}(u_{0},\phi^{A})$ as a free choice. The worldtube
Bondi scalars are then transformed into these
new coordinates and used to constrain the coefficients of
Eq.~\eqref{eq:generalIDAnsatzpf}. All such schemes adopt first-order matching,
$N=1$, giving a cubic ansatz
\begin{align}
    \breve J_{\Sigma_0}(\breve{u}_0, \breve y, \breve \phi^{\breve A}) = \breve J^{(1)}(\breve \phi^{\breve A})(1 - \breve y) + \breve J^{(3)}(\breve \phi^{\breve A})(1 - \breve y)^3,
    \label{eq:cubic_ansatz}
\end{align}
with $\breve J^{(1)}$ and $\breve J^{(3)}$ obtained by requiring first-order
differentiability at the worldtube ($r = R \to \breve y = -1$):
\begin{align}
    \begin{cases}
        2\breve J^{(1)} + 8 \breve J^{(3)} = \breve J|_{\Gamma}(u_0, \breve \phi^{\breve A}), \\
        \breve J^{(1)} + 12\breve J^{(3)} = - \partial_{\breve y} \breve J|_{\Gamma}(u_0, \breve \phi^{\breve A}),
    \end{cases}
    \label{eq:cubic_matching_constraints}
\end{align}
where $\partial_{\breve y}\breve J|_{\Gamma}$ follows from
Eq.~\eqref{eq:jacobians},
\begin{align}
    \partial_{\breve y} \breve J|_{\Gamma}(u_0, \breve \phi^{\breve A}) = \frac{\hat R(u_0,\hat \phi^{\hat A})}{2}\partial_{\hat r} \hat J|_{\Gamma}(u_0,  \hat \phi^{\hat A}).
    \label{eq:dybreveJ_from_drhatJ}
\end{align}
The \emph{InverseCubic} scheme chooses the identity map, so that the Cauchy
Bondi-like coordinates initially coincide with the evolution coordinates.
The current default, \emph{ConformalFactor}, instead chooses
$\hat\phi^{\hat A}(u_0,\phi^{A})$ to set
$\hat\beta(u_0,\hat\phi^{\hat A})$ to zero at the worldtube. This sets the
null-foliation lapse to unity there, bringing the frame closer to the
Bondi--Sachs gauge. From Eq.~\eqref{eq:pftrans:a}, imposing
\begin{align}
    1 = e^{2\hat{\beta}} = \hat{\omega}^{-1}e^{2\beta}
    \label{eq:CF_minimization}
\end{align}
yields $\hat{\omega} = e^{2\beta}$, which is solved iteratively for the
corresponding initial angular map; that solve is described in
Appendix~\ref{app:angular_solve}.
Despite its name, this scheme generally produces an angular map
that is not conformal with respect to the unit-sphere metric.
\subsection{No-incoming-radiation scheme}
\label{subsec:NIR}
The \emph{NoIncomingRadiation} scheme imposes $\Psi_0=0$ on
$\Sigma_0$ in the null tetrad associated with the Cauchy Bondi-like
coordinates. This is an approximate no-incoming-radiation condition: at finite
radius, $\Psi_0$ does not unambiguously separate incoming radiation from other
curvature contributions. The condition can be written as~\cite{Moxon:2021gbm, Newman:1961qr}
\begin{equation}
    \begin{aligned}
        \partial_{\tilde y}^{2} \tilde J
         & = \frac{1}{16 \tilde K^2}
        \Big(
        \bar{\tilde J}^2 (\partial_{\tilde{y}} \tilde{J})^2
        + \tilde{J}^2 (\partial_{\tilde{y}} \bar{\tilde J})^2               \\
         & \qquad
        - 2(2 + \tilde{J}\bar{\tilde J})\, \partial_{\tilde{y}} \tilde{J}\, \partial_{\tilde{y}} \bar{\tilde J}
        \Big)               \\
         & \quad\times\left(
        -4\tilde{J} - (1 - \tilde{y})\partial_{\tilde{y}} \tilde{J}
        \right),
        \label{eq:psi0eq}
    \end{aligned}
\end{equation}
in the Cauchy Bondi-like gauge. This second-order radial ODE for $\tilde J$
is integrated outward from the boundary values $\tilde J|_{\Gamma}$ and
$\partial_{\tilde y}\tilde J|_{\Gamma}$, ensuring first-order differentiability
at the worldtube. It also fixes the second radial derivative
$\partial_{\tilde y}^{2}\tilde J$ in terms of the field and its first
derivative. Section~\ref{sec:cauchy_matching_scheme} shows why a more accurate
value of this second derivative is crucial.

The resulting $\tilde J$ may contain a zeroth-order term in its $1-\tilde y$ expansion
(though not a second-order one~\cite{Moxon:2020gha}), and so is not in a
partially flat gauge. Expanding the compactified form of Eq.~\eqref{eq:pftrans:b} in $1-\tilde y$ gives to
leading order the condition
\begin{align}
    0 = \breve J^{(0)} = \frac{\bar{\hat b}^2 \tilde J^{(0)} + \hat a^2\bar{\tilde J}^{(0)} + 2\hat a\bar{\hat b} \tilde K^{(0)}}{4\hat\omega^{2}},
    \label{eq:NIRangtranscondition}
\end{align}
The angular map is obtained by solving this condition iteratively, as
described in Appendix~\ref{app:angular_solve}. Thus
\emph{NoIncomingRadiation} is the only pre-existing scheme that derives its
angular map from the worldtube data. In this work we show that solving a form of
Eq.~\eqref{eq:NIRangtranscondition} is crucial to constructing correct
characteristic initial data.
\section{Cauchy and characteristic junk radiation}
\label{sec:characteristic_junk}
\figJunkOverview
Two unphysical transients contaminate CCE waveforms. The \emph{Cauchy junk}
originates in the relaxation of the Cauchy initial data and the imperfect
treatment of the finite outer boundary; it reaches $\Igroup$ through the
worldtube data. The \emph{characteristic junk} originates on the initial null
hypersurface. This section distinguishes the two and shows that the long-lived
transient produced by every existing scheme is characteristic junk, caused by
the initial-data construction, not the Cauchy junk.

Throughout we use an equal-mass ($q=1$), non-spinning, quasi-circular binary black hole (BBH)
simulation, extracted on worldtubes at several radii $R$ and run with several
placements $R_{\mathrm{max}}$ of the Cauchy outer boundary. To compare waveforms
consistently, all results are mapped to the post-Newtonian (PN) BMS
frame.\footnote{The PN waveform is generated with the \textsc{PostNewtonian.jl}
package, \url{https://github.com/moble/PostNewtonian.jl}.}

All configurations use \texttt{SpEC}~\cite{Boyle:2019kee} Cauchy evolutions at \texttt{Lev3}
resolution, differing only in $R_{\mathrm{max}}$ and the worldtube extraction
radii. Worldtube data are stored through $\ell_{\max}=16$. Every
characteristic evolution in this paper uses \texttt{SpECTRE}~\cite{spectreCode}
with $\ell_{\max}=20$ and $15$ radial collocation points. At this resolution,
Ref.~\cite{Moxon:2021gbm} finds that the characteristic solution has converged;
we have also confirmed that increasing either resolution leaves the results
below unchanged. We therefore do not present a characteristic convergence
study: the error budget, discussed in Sec.~\ref{sec:id_accuracy}, is dominated by the
worldtube data and Cauchy outer boundary.

The Cauchy junk has two components. First, the relaxation of the Cauchy initial
data releases a high-frequency burst that sweeps outward past the worldtube.
Second, the burst reflects off the finite outer boundary and returns to the
worldtube at times set by $R$ and $R_{\mathrm{max}}$. Back-scattering at the
Cauchy subdomain interfaces fills the intervening interval with lower-amplitude
noise. The characteristic junk is instead generated on $\Sigma_{0}$ by an
inconsistency between the initial data and the worldtube data to which they
are matched. It is low frequency and decays slowly.

Figure~\ref{fig:JunkOverview} compares the three pre-existing schemes through
the waveform amplitudes at $\Igroup$ of the dominant quadrupolar $(2,2)$ mode
and the dominant memory $(2,0)$ mode, which is purely real for this system. For
evolutions started on the Cauchy initial slice, the first null slice leaves
the worldtube at Cauchy time $t_{0}=0$, corresponding to Bondi time
$\mathring u \approx -R$. The Cauchy junk crosses $\Gamma$ at
$\mathring u \approx 0$, followed shortly by the physical radiation. The
$(2,2)$ amplitude then exhibits long-lived, damped oscillations about its
expected value. \emph{InverseCubic} and \emph{ConformalFactor} behave
similarly, while \emph{NoIncomingRadiation} produces the largest oscillations.

For the $(2,0)$ mode, \emph{ConformalFactor} produces a negligible transient.
The other two schemes introduce spurious radiation that decays on a timescale
comparable to the $(2,2)$ transient. Its success in suppressing memory-mode
junk made \emph{ConformalFactor} the default scheme until now.

While not explicitly shown, the curves contain both transients. Outside the plotted window, the main Cauchy junk pulse leaves a low-frequency residue whose frequency and decay closely
resemble the characteristic junk. They can be distinguished only through their
response to the Cauchy outer boundary: the characteristic junk is essentially independent of
$R_{\mathrm{max}}$, while the Cauchy residue shrinks as $R_{\mathrm{max}}$ is
moved outward.

For the pre-existing schemes, the much larger characteristic junk hides this
residue and prevents that comparison. The residue becomes visible once the
scheme introduced here eliminates the characteristic junk, as shown in Sec.~\ref{subsec:bbh_results}.
\figBondiConstraints
The Bondi--Sachs constraints at $\Igroup$ reveal the duration and structure of
the Cauchy junk more clearly than the strain; see
Refs.~\cite{Madler:2016xju, Moxon:2020gha} for their detailed formulation.
Figure~\ref{fig:BondiConstraints} shows the constraint-violation norm associated
with $\Psi_{2}$; the other constraints behave similarly. Solid curves denote
evolutions started at $t_{0}=0$. The violations at $\Igroup$ are inherited from
the Cauchy worldtube data and transported outward along the null
hypersurfaces, so they trace the causal structure of the Cauchy junk.

The first spike, at $\mathring u = 0$, marks the initial junk crossing of
$\Gamma$. High-frequency noise then persists for $2(R_{\mathrm{max}}-R)$,
as back-scattering from the Cauchy subdomain interfaces fills the region
between the worldtube and outer boundary. This interval ends when the first
outer-boundary reflection returns to the worldtube. The feature at
$2R_{\mathrm{max}}$ marks the third crossing: the reflected pulse has passed
through the center and crosses the worldtube outward again. Later reflections
are progressively damped and harder to see. The worldtube data thus remain
measurably contaminated well after the strain appears clean.

To test whether the characteristic transient originates in the Cauchy junk,
we restart CCE after $2R_{\mathrm{max}}$, once the dominant Cauchy burst and
all significant reflections have crossed the worldtube. A transient inherited
from the Cauchy junk should then be absent. Figures~\ref{fig:JunkOverview}
and~\ref{fig:BondiConstraints} include such a restart at
$\mathring{u}_{0}=6200\,M$, with $\mathring u_{0} > 2R_{\mathrm{max}}$.

The transient reappears in full for every scheme and is now visible in
isolation. The trends are similar to those at $t_{0}=0$, except that
\emph{NoIncomingRadiation} now performs best in the $(2,2)$ amplitude. The
constraint violations show the distinction especially clearly: characteristic
junk is visible in the restart (dashed curves), whereas at $t_{0}=0$ it is
masked by the much larger Cauchy contribution.

The long-lived transient therefore originates in the pre-existing schemes'
characteristic initial data. The following sections identify the source of
this error.

\section{The initial angular coordinate problem and the origin of junk}
\label{sec:angular_problem}
The initial data include both $J$ and the angular map
$\hat\phi^{\hat A}(u_{0},\phi^{A})$ of Eq.~\eqref{eq:angEOM}. Every existing
scheme except \emph{NoIncomingRadiation} chooses this map \emph{a priori}.
Only its conformal part is free, however: the partially flat evolution drives
the non-conformal part toward the configuration fixed by the worldtube data.
An incorrect choice for this part of the initial map produces characteristic
junk.
\subsection{A toy model for the junk mechanism}
\label{subsec:schwarzschild_test}
\figAngularMap
The mechanism is clearest in a nonradiating spacetime. In the Bondi--Sachs
gauge, the Schwarzschild worldtube data are known analytically,
\begin{align}
    J|_{\Gamma} = \beta|_{\Gamma} = U|_{\Gamma} = Q|_{\Gamma} = 0,
    \qquad
    W|_{\Gamma} = -\frac{2M}{R^2},
    \label{eq:Schwarzschildwt}
\end{align}
for any placement of $\Gamma$. The exact solution has $J = 0$ everywhere on
every null slice. Starting from these worldtube coordinates, a generic initial
angular map transforms $J$ according to Eq.~\eqref{eq:pftrans:b},
\begin{align}
    \hat J_{\Sigma_0} = \frac{\hat a\,\bar{\hat b}}{2\hat\omega^{2}}.
    \label{eq:Jhat_schwarzschild_wt}
\end{align}
Any physically equivalent initial data within this coordinate family must
have this form. The partially flat condition $\hat J^{(0)} = 0$ then requires
$\hat{a} = 0$, so only purely conformal maps are allowed.

Now follow the existing schemes: apply an angular map at the worldtube, then
match the radial profile of $\hat{J}$ to leading order in $1/\hat r$, imposing
$\hat J^{(0)} = 0$ directly on the series expansion. This gives

\begin{align}
    \hat J_{\Sigma_0} = \frac{\hat a\,\bar{\hat b}}{2\hat\omega^{2}}\,\frac{\hat R}{\hat r}.
    \label{eq:Jhat_schwarzschild_matched}
\end{align}

This profile describes a gauge-inequivalent spacetime with radiation absent
from the static solution. Higher-order matching
with the same prescribed angular map does not resolve the problem,
because it still gives $\hat J = \mathcal{O}(1/\hat r)$ at every order.

The example shows why the non-conformal part of the initial angular map is
constrained by the partially flat gauge. The condition $\hat J^{(0)}=0$ has
a unique solution up to conformal maps; for Schwarzschild, that solution is
purely conformal. A non-conformal part incompatible with the worldtube data
and the gauge condition introduces junk radiation. To demonstrate the ensuing
evolution, we first rewrite the angular map.

Consider the angular map from the partially flat gauge
to the Cauchy gauge $\phi^{A}(\hat \theta, \hat \phi) = \left(\theta(\hat \theta,
\hat \phi), \phi(\hat \theta, \hat \phi)\right)$.
Since a map between points of $S^{2}$ is not itself a tangent vector, it
cannot be expanded in the tangent basis $\{\vec{e}_{\hat A}\}$ of $S^{2}$.
We instead define
\begin{align}
    \delta \vec\phi = \delta \phi^{\hat A} \vec e_{\hat A},
    \qquad
    \delta\phi^{\hat A} = \phi^{A}(\hat \theta, \hat \phi)\delta^{\hat A}_{A} - \mathrm{id}^{\hat A},
    \label{eq:delta_phi_vec}
\end{align}
where $\mathrm{id}^{\hat A} = (\hat\theta, \hat\phi)$ denotes the identity
map. For a small perturbation about the identity, this displacement lies
approximately in the tangent space of $S^2$. We can therefore treat
$\delta\phi^{\hat A}$ approximately as a vector field, form a spin-weight $1$ complex
scalar, and expand it in spin-weight $1$ spherical harmonics,
\begin{align}
    \delta\Phi(\hat{\theta}, \hat{\phi})
     & \equiv \delta\phi^{\hat A}\, q_{\hat A} = \sum_{\ell,m}  \delta\Phi_{\ell,m}\, {}_1 Y_{\ell,m}(\hat{\theta}, \hat{\phi}).
    \label{eq:deltaPhi_def}
\end{align}
This decomposition separates the conformal and non-conformal parts of the
map. A spin-weight-$1$ field $\eta$ representing a conformal Killing vector
on $S^2$ satisfies $\eth \eta = 0$. Since $\eth \, {}_1 Y_{1,m} = 0$, the
$\ell = 1$ modes of $\delta\Phi$ completely encode the conformal part, and
the $\ell \ge 2$ modes encode the non-conformal part.
We use this decomposition to prescribe Schwarzschild initial coordinates
with either a conformal perturbation $\delta\Phi_{\ell=1}$ or a
non-conformal perturbation $\delta\Phi_{\ell\ge2}$. The left column of
Fig.~\ref{fig:AngularMap} compares three CCE runs with identical Schwarzschild
worldtube data: one has an unperturbed initial map, one has a conformal
$(\ell,m)=(1,0)$ perturbation (a boost along $z$), and one has a
non-conformal $(\ell,m)=(2,0)$ perturbation of the same magnitude.

The conformal perturbation excites only $\ell=1$ modes and yields stable
coordinate evolution with no Bondi news $\dot h$, as required for
Schwarzschild. The non-conformal $(\ell,m)=(2,0)$ perturbation produces
$\hat
J(u_0)\ne0$. The map then relaxes and emits junk with the same transient
signature seen in the binary.
For a BBH, the initial angular map has a nontrivial non-conformal part.
The right panel of Fig.~\ref{fig:AngularMap} shows the three pre-existing
schemes restarted at $\mathring u_{0}=6200\,M$. Each map is compared with
that of a reference run started at $t_{0}=0$, whose junk has long decayed.
The differences are measured at fixed $\ell$ using the
norm in Eq.~\eqref{eq:angular_norm_ell}.

The conformal $\ell=1$ curves are flat and coincide for all three schemes.
Their common offset represents the freely specifiable conformal transformation
relative to the reference run. Each scheme starts its angular solve from the
identity map, so none includes the rigid rotation accumulated by the reference
run by $\mathring u_{0}$.

The $\ell=2$ curves exhibit the same relaxation as the BBH strain and the
Schwarzschild Bondi news. These non-conformal modes approach the value fixed by
the Cauchy worldtube data and the partially flat condition. Higher multipoles
behave similarly, with relaxation rates increasing with $\ell$; only
$\ell=2$ is shown. Thus the conformal part of the map is freely specifiable,
whereas an incorrect non-conformal part relaxes and produces characteristic
junk.

\subsection{Initial angular map as a solution to the Beltrami equation}
\label{sec:angular_coord_transformation}
As we have just seen, the non-conformal part of the
angular coordinates is fully constrained by the equation in compactified coordinates
\begin{align}
    0 = \breve J^{(0)} = \frac{\bar{\hat b}^2 \tilde J^{(0)} + \hat a^2\bar{\tilde J}^{(0)} + 2\hat a\bar{\hat b} \tilde K^{(0)}}{4\hat\omega^{2}}.
    \label{eq:pfgaugeconstraint0}
\end{align}
It is useful to define the scalar
\begin{align}
    \mu = \frac{\hat a}{\bar{\hat b}},
    \label{eq:mu_def}
\end{align}
which controls the non-conformal nature of the coordinates: when $\mu=0$ the map is conformal. Then
Eq.~\eqref{eq:pfgaugeconstraint0} is equivalent to the complex quadratic equation
\begin{align}
    \mu^{2}\bar{\tilde J}^{(0)} + 2\mu \tilde K^{(0)} + \tilde J^{(0)}= 0.
    \label{eq:muquad_main}
\end{align}
The identity $\tilde K^{2}- \tilde J
\bar{\tilde J}=1$ simplifies the discriminant, giving the roots
$\mu_{\pm}=(-\tilde K^{(0)}\pm1)/\bar{\tilde J}^{(0)}$, or equivalently
$\mu_{+} = -\frac{\tilde J^{(0)}}{1 + \tilde K^{(0)}}$ and
$\mu_{-} = -\frac{1 + \tilde K^{(0)}}{\bar{\tilde J}^{(0)}}$. Their magnitudes satisfy
\begin{align}
    |\mu_{+}| &= \left| \frac{1 - \tilde K^{(0)}}{\bar{\tilde J}^{(0)}} \right| < 1 , \\
    |\mu_{-}| &= \left| \frac{1 + \tilde K^{(0)}}{\bar{\tilde J}^{(0)}} \right| > 1 .
    \label{eq:coeffsign}
\end{align}
From Eq.~\eqref{eq:omega_def},
$\hat\omega^{2} = \tfrac{1}{4}|\hat b|^{2}(1-|\mu|^{2})$, so $\mu_{-}$ is
unphysical: by Eq.~\eqref{eq:coeffsign} it would give $\hat\omega_{-}^{2}<0$. We
therefore set
\begin{align}
    \mu \equiv \mu_{+} = -\frac{\tilde J^{(0)}}{1 + \tilde K^{(0)}}.
    \label{eq:beltramicoefficient}
\end{align}
The condition $\breve J^{(0)} = 0$ states that the asymptotic angular metric is
proportional to the round unit-sphere metric in the partially flat
coordinates, i.e., that those coordinates are \emph{isothermal} for it~\cite{Chern:1955}.
Finding the coordinates that solve the partially flat gauge constraint is
mathematically equivalent to solving the Beltrami equation on $S^{2}$.

Let $z$ and $\hat z$ be local complex stereographic coordinates on the Cauchy
and partially flat spheres, respectively. Up to a common factor, the Jacobian
scalars in Eq.~\eqref{eq:angjacobians} are the Wirtinger derivatives of the
map $z(\hat z)$: $\hat a\propto\partial_{\bar{\hat z}}z$ and
$\bar{\hat b}\propto\partial_{\hat z}z$. Their ratio $\mu$, defined in
Eq.~\eqref{eq:mu_def}, is therefore a Beltrami coefficient~\cite{LehtoVirtanen:1973}.

We pose the problem for the inverse map $\hat z(z)$. Its coefficient $\nu$ satisfies
$|\nu|=|\mu|$ pointwise and, unlike $\mu$, is a known function of its
argument, because $\tilde J^{(0)}$ is supplied on the Cauchy sphere. The gauge
condition then becomes the Beltrami equation~\cite{Ahlfors:1966}
\begin{align}
    \partial_{\bar z}\, \hat z = \nu \, \partial_{z}\, \hat z ,
    \label{eq:beltrami}
\end{align}
a first-order complex partial differential equation for the angular map, with
Beltrami coefficient $\nu$. Since $\mu$ measures the pointwise departure
from conformality, solving this equation determines the non-conformal part of the
initial coordinates.

The measurable Riemann mapping theorem~\cite{AhlforsBers:1960} guarantees a
solution whenever the coefficient $\nu$ satisfies
$\lVert\nu\rVert_{\infty}<1$, unique up to post-composition with conformal
maps of the sphere. To verify this bound, write
$\tilde K^{(0)}=\sqrt{1+|\tilde J^{(0)}|^{2}}$. Equation~\eqref{eq:beltramicoefficient}
then gives
\begin{align}
    |\nu| \;=\; |\mu| \;=\; \sqrt{\frac{\tilde K^{(0)}-1}{\tilde K^{(0)}+1}} ,
    \label{eq:mu_magnitude}
\end{align}
This magnitude is below unity at every point and approaches unity only as
$|\tilde J^{(0)}|\to\infty$. The uniform bound therefore holds for any bounded
worldtube data.\footnote{This makes Eq.~\eqref{eq:beltrami} equivalent to a
first-order elliptic system of partial differential equations.}

Because the conformal maps act on $\hat z$, the remaining freedom is exactly
a conformal change of the partially flat coordinates. Existence and uniqueness
of the quasi-conformal solution establish the decomposition of the angular
map into conformal and non-conformal parts, and the well-posedness of the
partially flat gauge for any strain magnitude at the worldtube.

The \emph{NoIncomingRadiation} scheme was already solving the Beltrami
equation, through its Newton-based iteration.
Appendix~\ref{app:angular_solve} gives the solver we use, which recovers the map
from the closed-form coefficient~\eqref{eq:beltramicoefficient} rather than
linearizing the gauge condition.

The remaining task is to determine the Beltrami
coefficient accurately. It depends only on $\tilde J^{(0)}$, the value
at $\Igroup$ of the initial Cauchy-gauge $\tilde J$ ansatz.

Schemes that freely prescribe the
non-conformal part introduce junk because their map does not solve the
Beltrami problem. \emph{NoIncomingRadiation} solves the problem, but with an
inaccurate coefficient: it obtains the radial profile and its
$\tilde J^{(0)}$ by integrating $\Psi_{0}=0$ outward from the worldtube,
Eq.~\eqref{eq:psi0eq}. The resulting $\tilde J^{(0)}$ inherits this
approximation and omits radial information supplied by higher worldtube
derivatives. Higher-order matching incorporates this information.

The next section constructs $\tilde{J}_{\Sigma_0}$ by higher-order matching
in Cauchy coordinates, which is crucial to obtaining an accurate
$\tilde J^{(0)}$ and eliminating CCE junk.

\section{The correct construction of characteristic initial data}
\label{sec:cauchy_matching_scheme}
We construct the Cauchy-gauge radial profile of $J$ from the worldtube data
so that the Beltrami problem for the angular coordinates can be solved to high
precision. We first give the general construction of $\tilde J$ by matching
to the worldtube derivatives while preserving the remaining partially flat
gauge condition $\breve J^{(2)} = 0$. We then specialize to second order and
present the scheme implemented in \texttt{SpECTRE}~\cite{spectreCode} as
\emph{CauchySecondOrder}.
\subsection{Asymptotic expansion of \texorpdfstring{$J$}{J} in the Cauchy gauge}
\label{sec:asymptotic_expansion}
\figRadialCoeffs
Section~\ref{sec:angular_coord_transformation} established that the
Cauchy-gauge radial profile of $J$ determines the Beltrami coefficient through
$\tilde J^{(0)}$. The Cauchy gauge itself imposes no conditions on the asymptotic
series: in particular, $\tilde J^{(0)}$ and $\tilde J^{(2)}$ must be determined
as part of the construction. The two partially flat conditions play different
roles. While $\breve J^{(0)}=0$ fixes the angular coordinates entirely,
$\breve J^{(2)}=0$ constrains the radial expansion coefficients in the Cauchy
gauge.

Given the worldtube derivatives $\partial_{\tilde y}^{\,n}\tilde J|_\Gamma$
up to order $N$, we write the Cauchy-matched radial profile as
\begin{align}
    \tilde J_{[N]}(\tilde y;x_{[N]})
        &=\sum_{\substack{n=0\\n\ne2}}^{N+1}
          \tilde J_{[N]}^{(n)}(x_{[N]})\,(1-\tilde y)^n+x_{[N]}(1-\tilde y)^2 .
    \label{eq:generalansatz}
\end{align}
Here $x_{[N]}=\tilde J_{[N]}^{(2)}$ is initially a free parameter. The subscript $[N]$
labels the matching order, while the superscript $(n)$ labels the radial
power. The highest power is
$N+1$, rather than the $N+2$ of Eq.~\eqref{eq:generalIDAnsatzpf}, because the
Cauchy-gauge ansatz retains the $n=0$ coefficient. Requiring the profile to
match the supplied worldtube derivatives gives
\begin{align}
    \left.\partial_{\tilde y}^{n}\tilde J_{[N]}(\tilde y;x_{[N]})\right|_\Gamma
        =\left.\partial_{\tilde y}^{n}\tilde J\right|_\Gamma,
        \qquad 0\le n\le N .
    \label{eq:Cauchymatching}
\end{align}

We refer to this construction as $N$th-order Cauchy matching. These $N+1$
conditions leave one of the $N+2$ coefficients to be determined by the
remaining partially flat condition, $\breve J^{(2)}=0$.

To express that condition in the Cauchy gauge, define the inverse Jacobian
factors
\begin{align}
    a = -\frac{\hat a}{\hat{\omega}^2}, \qquad
    b = \frac{\bar{\hat b}}{\hat{\omega}^2}.
    \label{eq:inverse_jacobians}
\end{align}
Hats on $a$, $b$, and $\omega$ denote the Jacobian data of
$\phi^A(\hat\phi^{\hat A})$, defined in
Eqs.~\eqref{eq:angjacobians}--\eqref{eq:omega_def}. The unaccented symbols
refer to the inverse map $\hat\phi^{\hat A}(\phi^A)$, with
$\omega=1/\hat\omega$. Since the angular transformation is independent of
radius, compactification does not change these quantities:
$\breve a\equiv\hat a$ and $\tilde a\equiv a$, and likewise for $b$ and
$\omega$. We therefore use only the hatted and unaccented forms.

Applying Eq.~\eqref{eq:pftrans:b} to the inverse map at second order gives
$\tilde J^{(2)}$ in terms of the partially flat coefficients. The condition
$\breve J^{(2)}=0$ removes the terms containing that coefficient and its
complex conjugate. Meanwhile, $\breve J^{(0)}=0$ gives
$\breve K^{(0)}=1$, $\breve K^{(1)}=0$, and
$\breve K^{(2)}=\tfrac12|\breve J^{(1)}|^2$ in the expansion of
$\breve K=\sqrt{1+\breve J\bar{\breve J}}$. Only one term remains:
\begin{align}
    \tilde{J}^{(2)} = \frac{a\bar{b}}{2\omega^2}\,\breve K^{(2)}
        = \frac{a\bar{b}}{4\omega^2}\,\big|\breve J^{(1)}\big|^2 .
    \label{eq:j2inverse}
\end{align}
The closed-form Beltrami coefficient~\eqref{eq:beltramicoefficient}, together
with Eq.~\eqref{eq:inverse_jacobians}, simplifies the prefactor to
\begin{align}
    \frac{a\bar b}{4\omega^2} = -\frac{\hat a\hat b}{4\hat\omega^2}
        = -\frac{\mu}{1 - |\mu|^{2}} = \frac{\tilde J^{(0)}}{2} ,
    \label{eq:muratio}
\end{align}
so that
\begin{align}
    \tilde{J}^{(2)} = \tfrac{1}{2}\,\tilde J^{(0)}\,\big|\breve J^{(1)}\big|^2 ,
    \label{eq:j2exact}
\end{align}
The angular transformation is linear in the components $(K,J,\bar J)$
and preserves the quadratic form $K^2-|J|^2$. Since it is independent of
radius and $\breve y=\tilde y$, the first radial coefficients satisfy
\begin{equation}
    \big|\breve J^{(1)}\big|^2-\big(\breve K^{(1)}\big)^2
    =\big|\tilde J^{(1)}\big|^2-\big(\tilde K^{(1)}\big)^2.
    \label{eq:first_radial_invariant}
\end{equation}
Using $\breve K^{(1)}=0$,
we can therefore express
this relation entirely in Cauchy-gauge quantities:
\begin{align}
    \tilde{J}^{(2)} = \frac{1}{2}\tilde J^{(0)}\left[\big|\tilde{J}^{(1)}\big|^2
        - \left(\tilde{K}^{(1)}\right)^{2}\right] .
    \label{eq:j2}
\end{align}
where,
\begin{align}
    \tilde K^{(1)}
        = \frac{\Re\big(\tilde J^{(0)}\bar{\tilde J}^{(1)}\big)}{\tilde K^{(0)}}
    \label{eq:K1}
\end{align}
Equation~\eqref{eq:j2} is the remaining nonlinear constraint. Together with
the $N+1$ worldtube conditions~\eqref{eq:Cauchymatching}, it closes the system
of $N+2$ equations for the $N+2$ coefficients of
Eq.~\eqref{eq:generalansatz}.

For each prescribed $x_{[N]}$, the worldtube matching conditions in
Eq.~\eqref{eq:Cauchymatching} form a linear system for the remaining
coefficients. Its solution is
\begin{align}
    \tilde J_{[N]}^{(n)}(x_{[N]}) = \mathcal{T}^{(n)}_{[N]}
        + \alpha_{[N]}^{(n)}\big(x_{[N]} - \mathcal{T}^{(2)}_{[N]}\big) ,
    \label{eq:linearsolution}
\end{align}
where
\begin{align}
    \mathcal{T}^{(n)}_{[N]}
    = \frac{(-1)^{n}}{n!}\sum_{k=n}^{N}\frac{2^{\,k-n}}{(k-n)!}\,
        \partial_{\tilde y}^{\,k}\tilde J\big|_{\Gamma}
    \label{eq:tk}
\end{align}
and
\begin{align}
    \alpha^{(n)}_{[N]}
    = (-1)^{n}\,2^{\,3-n}\;\frac{(N-1)!}{n!\,(N+1-n)!} .
    \label{eq:alphak}
\end{align}
Substituting these coefficients into the constraint~\eqref{eq:j2} leaves a
nonlinear equation for $x_{[N]}$,
\begin{align}
    x_{[N]} = \Phi_{[N]}(x_{[N]}) ,
    \label{eq:fixedpoint}
\end{align}
with $\Phi_{[N]}$ a function stemming from Eqs.~\eqref{eq:j2} and~\eqref{eq:linearsolution}.
Appendix~\ref{app:fixed_point} proves that a solution exists, is unique within
the prescribed small-amplitude region, and is reached by iterating
$\Phi_{[N]}$ from $x_{[N]}=0$ whenever the $x_{[N]}=0$ values of
$\tilde J_{[N]}^{(0)}$ and $\tilde J_{[N]}^{(1)}$ are bounded in magnitude by $7\times10^{-2}$.
This sufficient bound ensures that the fixed-point map preserves
the prescribed region and that the iteration is contractive, as shown in
Appendix~\ref{app:fixed_point}.
Physical worldtube data in the radiation zone comfortably satisfy this bound.
Together with the Beltrami solution established above, this determines the
initial data within the chosen ansatz and amplitude bound at fixed order $N$,
up to the residual Lorentz freedom.

Figure~\ref{fig:RadialCoeffs} shows the size of the radial coefficients in
a reference CCE evolution started at $t_0=0$, our best available approximation
to the exact solution in the characteristic domain. We transform its partially
flat $\breve J$ back into the Cauchy gauge using Eq.~\eqref{eq:pftrans:b} and the inverse
Jacobians in Eq.~\eqref{eq:inverse_jacobians}. We then interpolate onto the Cauchy
grid using the inverse angular map $\phi^A(\hat u,\hat\phi^{\hat A})$, which
we evolve alongside $\hat\phi^{\hat A}(\hat u,\phi^A)$, as in the Cauchy-characteristic matching (CCM)
code~\cite{Ma:2023qjn, Ma:2024qcv}. Radial derivatives with respect to the
compactified coordinate are unaffected by the angular map, so
$\partial_{\breve y}^{n}\tilde J=\partial_{\tilde y}^{n}\tilde J$. The
coefficients therefore follow from the numerical radial derivatives at
$\Igroup$:
\begin{align}
    \tilde{J}^{(n)} = \frac{(-1)^{n}}{n!}\, \partial_{\tilde y}^{n}\tilde J\big|_{\Igroup}.
    \label{eq:coefffromder}
\end{align}
For typical binaries like the equal-mass, non-spinning one shown, the
quadratic coefficient $\tilde J^{(2)}$ is small compared with the other
coefficients:
$\lVert\tilde J^{(2)}\rVert$ is at most $\sim10^{-12}$, whereas
$\lVert \tilde J^{(0)}\rVert\sim4\times10^{-5}$ and
$\lVert\tilde J^{(1)}\rVert\sim10^{-4}$; every other coefficient up to $n=4$
lies well above it. High-order matching, which fixes these larger
coefficients, therefore matters most for the initial data. The next
subsection specializes the construction to $N=2$ and explains why going from
first to second order changes the initial data qualitatively.

\subsection{\texorpdfstring{$N=2$ Cauchy matching:
  \emph{CauchySecondOrder}}{N=2 Cauchy matching: CauchySecondOrder}}
\label{subsec:cauchy_matching_scheme}
The matching order controls which physics the initial data contains, as well as
its accuracy. At $N=1$, the ansatz uses only $J|_\Gamma$ and
$\partial_rJ|_\Gamma$. We show in the next section that this gives results
equivalent to solving the approximate no-incoming-radiation condition $\Psi_0=0$,
Eq.~\eqref{eq:psi0eq}, with the same two worldtube quantities as boundary
conditions. The reason is that the right-hand side of Eq.~\eqref{eq:psi0eq}
starts at cubic order in the field amplitude.
Here we count both $\tilde J$ and $\partial_{\tilde y}\tilde J$ as
first order in that amplitude, consistent with the comparable sizes of
$\tilde J^{(0)}$ and $\tilde J^{(1)}$ in Fig.~\ref{fig:RadialCoeffs}.
At the amplitudes of interest, imposing $\Psi_0=0$
therefore amounts to $\partial_{\tilde y}^2\tilde J\approx0$. The $N=1$
truncation makes the same approximation: its second derivative,
$\partial_{\tilde y}^2\tilde J_{[1]}(\tilde y;x_{[1]})=2x_{[1]}$, also starts at cubic order in the field amplitude by Eq.~\eqref{eq:j2}.

Supplying $\partial_r^2J\rvert_{\Gamma}$ independently retains radial
information omitted by this approximation, including contributions from
back-scattered radiation. We therefore advocate for using $N=2$,
implemented in the open-source \texttt{SpECTRE} code~\cite{spectreCode} as
\emph{CauchySecondOrder}.

At this order, the general solution~\eqref{eq:linearsolution} gives the
following ansatz and coefficients in the compactified coordinate:
\begin{subequations}\label{eq:cso_ansatz}
    \begin{align}
        \tilde J_{[2]}(\tilde y;x_{[2]}) &= \tilde J_{[2]}^{(0)}(x_{[2]}) + \tilde J_{[2]}^{(1)}(x_{[2]})(1-\tilde y)
            + x_{[2]}\,(1-\tilde y)^{2} \nonumber\\
        &\quad + \tilde J_{[2]}^{(3)}(x_{[2]})(1 - \tilde{y})^3 , \\[4pt]
        \tilde J_{[2]}^{(0)}(x_{[2]}) &=  \tilde J|_{\Gamma} + 2\partial_{\tilde{y}}\tilde{J}|_{\Gamma}
            + \tfrac43 \partial_{\tilde y}^{2}\tilde J|_{\Gamma} + \tfrac43\,x_{[2]} , \\
        \tilde J_{[2]}^{(1)}(x_{[2]}) &= - \left(\partial_{\tilde{y}}\tilde{J}|_{\Gamma}
            + \partial_{\tilde y}^{2}\tilde J|_{\Gamma}\right) - 2x_{[2]} , \\
        \tilde J_{[2]}^{(3)}(x_{[2]}) &= \tfrac{1}{12}\,\partial_{\tilde y}^{2}\tilde J|_{\Gamma}
            - \tfrac16\,x_{[2]} ,
    \end{align}
\end{subequations}
Every value of $x_{[2]}$ preserves the matching to the three worldtube quantities.
The constraint fixes $x_{[2]}$ through Eq.~\eqref{eq:fixedpoint}, with $\Phi_{[2]}$ the
right-hand side of Eq.~\eqref{eq:j2} evaluated on $\tilde J_{[2]}^{(0)}(x_{[2]})$ and
$\tilde J_{[2]}^{(1)}(x_{[2]})$. We solve it by iterating $x_{[2],k+1}=\Phi_{[2]}(x_{[2],k})$ from
$x_{[2],0}=0$ until the largest change in $x_{[2]}$ over the angular collocation points
falls below a prescribed tolerance. In practice, one or two iterations suffice,
consistent with the small contraction factor of the map for radiation-zone
data established in Appendix~\ref{app:fixed_point}. The converged profile
$\tilde J_{[2]}(\tilde y;x_{[2]})$ gives the initial field $\tilde J_{\Sigma_0}$.

To use Eq.~\eqref{eq:cso_ansatz}, we need the second radial derivative at the
worldtube. Of the required radial derivatives, the current \texttt{SpECTRE}
implementation receives only $\partial_rJ|_\Gamma$, or equivalently
$\partial_{\tilde y}\tilde J|_\Gamma=(R/2)\partial_rJ|_\Gamma$, from the
Cauchy simulation. This requires only the ADM variables and their first spacetime
derivatives, which are available in commonly used Cauchy formulations such as
generalized harmonic~\cite{Friedrich:1985afv, Pretorius:2005gq, Lindblom:2005qh} or BSSN~\cite{Shibata:1995we, Baumgarte:1998te}.
The \texttt{SpECTRE} worldtube preprocessor~\cite{spectreCode}
constructs the required Bondi--Sachs input from either the spacetime metric
of generalized-harmonic evolutions or the ADM variables of moving-puncture
evolutions~\cite{Campanelli:2005dd, Baker:2005vv}. Moving-puncture codes
typically output the extrinsic curvature rather than time derivatives of the
metric. The preprocessor reconstructs the required time derivatives of the
spatial metric, lapse, and shift from the extrinsic curvature and the
$1+\log$ and Gamma-driver gauge conditions.

There are two ways to obtain $\partial_{\tilde y}^2\tilde J|_\Gamma$. Computing
second derivatives of the ADM metric would require modifying the Cauchy code
and rerunning every simulation. We instead invert the characteristic
$H$-hypersurface equation~\eqref{eq:hypersurface_H} at the worldtube. The
procedure in Appendix~\ref{Appendix:Dy2J} uses only quantities already stored by existing simulations, together with the time
derivative $\partial_{\tilde u}\partial_rJ|_\Gamma$. Archived simulations
with stored worldtube data, whether from generalized-harmonic or
moving-puncture evolutions, can therefore be reprocessed with the new scheme
without rerunning an expensive
Cauchy evolution. This single time derivative is the accuracy bottleneck of
the method, as we discuss in Sec.~\ref{sec:id_accuracy}.

In practice, we compute $\partial_{\tilde u}\partial_rJ|_\Gamma$ by
interpolating the stored time series of $\partial_rJ|_\Gamma$ and
differentiating the interpolant. The \texttt{SpECTRE} CCE module uses a
high-order barycentric interpolant for worldtube data by default. Empirically,
this choice can introduce numerical artifacts when the evolution starts at
$t_0=0$ with large worldtube data at very small extraction radii. These
artifacts can distort the SWSH mode content of
$\partial_{\tilde u}\partial_rJ|_\Gamma$ enough to prevent the angular solver
from finding a solution. Second-order interpolation resolves these problems
while giving numerically indistinguishable results for well-behaved
simulations.

With the required derivatives available, the complete construction consists
of three steps:
\begin{enumerate}
    \item Match the initial $\tilde J$ on $\Sigma_0$ to the worldtube value
      and its first two radial derivatives in the Cauchy gauge, using
      Eq.~\eqref{eq:cso_ansatz}, and solve the fixed-point
      problem~\eqref{eq:fixedpoint} for $x_{[2]}=\tilde J_{[2]}^{(2)}$.
    \item Use the resulting $\tilde J_{[2]}^{(0)}$ to solve the Beltrami
      problem~\eqref{eq:pfgaugeconstraint0}, with
      coefficient~\eqref{eq:beltramicoefficient}, for the initial angular map
      $\hat\phi^{\hat A}(u_0,\phi^A)$, using the iterative method of
      Appendix~\ref{app:angular_solve}.
    \item Transform the whole profile into the partially flat gauge with
      this map, via Eq.~\eqref{eq:pftrans:b}.
\end{enumerate}
The transformed profile preserves the worldtube matching and satisfies both
partially flat conditions: $\breve J^{(0)}=0$ through the angular map and
$\breve J^{(2)}=0$ through the root $x_{[2]}$ (Appendix~\ref{app:fixed_point}). The
radiative content of the worldtube data thus fixes the non-conformal part of
the initial coordinates. The conformal (Lorentz) part remains free, as
Sec.~\ref{sec:angular_coord_transformation} showed.

\section{Numerical results}
\label{sec:results}
We test the scheme on three systems: a Teukolsky wave with an analytic
solution; the quasi-circular BBH used throughout, both from $t_{0}=0$ and from a later time; and a head-on collision, whose post-ringdown tail
provides a stringent test. In all three, second-order Cauchy matching removes
the characteristic junk.
\subsection{Teukolsky wave}
\label{subsec:teukolsky_wave}
\figTeukolskyWave
A Teukolsky wave is a linear perturbation of Minkowski spacetime, for which
semi-analytic worldtube data can be computed directly from the linearized
metric. The corresponding linearized news at $\Igroup$
is~\cite{Teukolsky:1982nz, Barkett:2019uae, Moxon:2020gha}
\begin{align}
    \dot h = \frac{3}{4}\sin^{2}\!\theta \, \partial_{\mathring{u}}^{5} f(\mathring{u}).
    \label{eq:tw_news}
\end{align}
Following Refs.~\cite{Barkett:2019uae, Moxon:2020gha}, we take a strictly
outgoing $(2,0)$ perturbation with a Gaussian profile
\begin{align}
    f(\mathring u) = \mathcal{A}\, e^{-\mathring u^{2}/T^{2}},
    \label{eq:tw_profile}
\end{align}
of amplitude $\mathcal{A} = 10^{-2}$ and duration $T = 20\,M$, small enough that
the linearized solution remains close to the fully nonlinear CCE
solution.\footnote{There is no Cauchy simulation in this test; the ``Cauchy
gauge'' worldtube data come from the analytic perturbative expressions. We
nevertheless refer to them as Cauchy gauge for consistency.}

Figure~\ref{fig:TeukolskyWave} compares the analytic solution (grey),
Eq.~\eqref{eq:tw_news}, with CCE evolutions (colored curves) started $40\,M$
before the peak, when the characteristic initial data already contain
nontrivial radiation.

All schemes closely follow the analytic curve during the pulse, but their
behavior differs once the wave decays. The pre-existing schemes leave an
exponentially decaying junk signal in the Bondi news, qualitatively similar
to the Schwarzschild and BBH cases. Among them,
\emph{NoIncomingRadiation} performs best. First-order Cauchy matching closely reproduces its behavior, for the
reason given in Sec.~\ref{subsec:cauchy_matching_scheme}. Second-order
matching essentially eliminates the junk: the news closely follows the
analytic solution and plateaus at a numerical floor once the signal becomes
too small to resolve.

Incorporating higher radial derivatives therefore improves the
initial radial profile beyond the $\Psi_0=0$ approximation.
\subsection{Equal-mass, non-spinning binary}
\label{subsec:bbh_results}
\figBBHStrain
 The left column of
Fig.~\ref{fig:BBHStrain} compares the strain from \emph{CauchySecondOrder} and
\emph{ConformalFactor} at two Cauchy outer-boundary locations. The evolutions
start at $t_{0}=0$, as in production runs. With the previous scheme, the
$(2,2)$ amplitude oscillates by up to $\sim10\%$ about its expected value for
more than $1000\,M$. This oscillation is essentially independent of
$R_{\mathrm{max}}$, identifying it as characteristic junk by the criterion of
Sec.~\ref{sec:characteristic_junk}. With the new scheme, the waveform follows
the post-Newtonian prediction from $\mathring{u} \approx 0$ onward.

A small modulation remains from Cauchy junk due to the finite outer boundary,
as anticipated in Sec.~\ref{sec:characteristic_junk}. It decreases
systematically as that boundary moves outward and also appears in the
extrapolated waveform of the $R_{\mathrm{max}}=1500\,M$ simulation, which involves no characteristic
evolution. The same contamination affects the $(2,0)$ mode, which shows a
discrepancy that
shrinks as $R_{\mathrm{max}}$ moves outward. The new scheme suppresses the
characteristic transient below this Cauchy-side contamination. The error budget
at $\Igroup$ is thus dominated by the Cauchy outer boundary, rather than the
characteristic initial data. CCM provides a natural solution to this separate
problem~\cite{Ma:2023qjn}.

 The right column repeats the comparison for
evolutions started at $\mathring{u}_{0}=6200\,M$. By this time, every Cauchy-side
transient has passed, isolating the characteristic junk, as shown in
Sec.~\ref{sec:characteristic_junk}. Once again, \emph{NoIncomingRadiation}
and first-order Cauchy matching overlap and both leave a transient, while
second-order matching removes it.

We show only the two most relevant modes. The higher modes contain far less
characteristic junk, and \emph{ConformalFactor} and \emph{CauchySecondOrder}
agree for them, so they do not distinguish the two schemes.
\subsection{Head-on collision}
\label{subsec:tails}
\figTails
Characteristic junk is most significant in short simulations, where it
occupies a large fraction of the post-merger signal. Figure~\ref{fig:Tails} shows
the late-time tail of the head-on collision studied in
Ref.~\cite{DeAmicis:2024eoy}.

We define
$\tau = \mathring u - \mathring u_{\mathrm{peak}}$, measuring time from the waveform peak.
In the late-time tail regime, $|\Psi_4^{(2,0)}|\sim\tau^k$, with $k<0$.
The tail's small amplitude makes it susceptible to contamination by an
additive junk floor. We track its evolution through the local power-law
index $d\log|\Psi_{4}^{(2,0)}|/d\log\tau$, obtained from a sliding
least-squares fit of $\log|\Psi_{4}^{(2,0)}|$ against $\log\tau$ over a
window of $201$ samples. With \emph{CauchySecondOrder}, the index remains around
$-3.5$, in agreement with predictions in the
literature~\cite{Ma:2024qcv, DeAmicis:2024eoy}. With \emph{ConformalFactor},
the signal saturates at its junk floor, and the index settles at a higher,
incorrect value.
\section{Accuracy of the initial data}
\label{sec:id_accuracy}
\figIDAccuracy
We quantify the initial-data error by comparing the Cauchy-gauge ansatz
$\tilde J^{\mathrm{ansatz}}$, constructed at each candidate initial time using
the procedure in Sec.~\ref{sec:asymptotic_expansion}, with the solution of a reference CCE run started
at $t_{0}=0$. The error at $t_{0}=0$ itself cannot be measured directly because
no reference initial slice is available there.
Both panels of Fig.~\ref{fig:IDAccuracy} show the relative distance between
$\tilde{J}$ constructed at a given $\mathring{u}_0$ and the reference solution
on the same slice,
\begin{align}
    \epsilon(\tilde J) = \frac{\big\lVert \tilde J^{\mathrm{ansatz}}
                        - \tilde J^{\mathrm{ref}}\big\rVert_{\Sigma}}
                    {\big\lVert \tilde J^{\mathrm{ref}}\big\rVert_{\Sigma}},
    \label{eq:eps_def}
\end{align}
where $\lVert\cdot\rVert_{\Sigma}$ is the unweighted Euclidean norm over the
radial collocation points and angular modes of the whole initial slice.
Throughout this section, $\epsilon(X)$ denotes the relative error of $X$ in
the norm appropriate to that quantity. To obtain the reference field, we
transform the partially flat $\breve{J}$ of the reference CCE run back into
the Cauchy gauge, as in Sec.~\ref{sec:asymptotic_expansion}.

The left panel of Fig.~\ref{fig:IDAccuracy} shows $\epsilon(\tilde J)$ as a
function of the initial time for $R_{\mathrm{max}}=3000\,M$ and $R=500\,M$.
The dashed curves use reference-run values for every worldtube quantity in
Eq.~\eqref{eq:Cauchymatching}. They therefore isolate the truncation error of
the $1/r$ expansion at first, second, and third order. Each additional order
reduces the residual by roughly two orders of magnitude, quantifying the
improvement from additional radial derivatives discussed in
Sec.~\ref{sec:asymptotic_expansion}.
This provides numerical evidence for convergence of the
radial reconstruction toward the reference solution as the matching order $N$
increases over the range tested.

The solid curve uses the production second-order initial data, constructed
solely from the Cauchy worldtube data. It lies above the second-order
truncation curve throughout the settled window, so truncation no longer
limits the accuracy at $R=500\,M$. Its variation with $\mathring u_{0}$
tracks the Bondi--Sachs constraint violations in
Fig.~\ref{fig:BondiConstraints} feature for feature. The limiting error therefore comes from the
worldtube data and the constraint violations inherited from the Cauchy
simulation.

Three main sources contribute to $\epsilon(\tilde J)$ for
\emph{CauchySecondOrder}: truncation of the radial ansatz, error in the
worldtube $\partial_{r}J|_{\Gamma}$, and error in its retarded-time derivative
$\partial_{\tilde u}\partial_{r}J|_{\Gamma}$, which is needed to compute
$\partial_{\tilde{y}}^2\tilde{J}|_{\Gamma}$. To isolate how these errors scale
with $R$, we express the expansion of $J$ using the physical radius $r$
rather than the compactified coordinate $\tilde y$, so that its coefficients do not depend on
$R$. The two sets of coefficients are related by
$\tilde J^{(n)} = J^{(n)}/(2R)^{n}$.

In this expansion,
$\lVert \tilde J\rVert_{\Sigma}\simeq \lVert J^{(1)}\rVert/R$.
Although Sec.~\ref{sec:asymptotic_expansion} calls $J^{(0)}$ the leading
coefficient, this refers to its \emph{order} in $1/r$, not its magnitude. In
the Cauchy gauge, $\lVert J^{(0)}\rVert$ decreases with $R$ faster than
$\lVert J^{(1)}\rVert/R$, so the $n=1$ term sets the norm at the worldtube
radii used here.

For the $N=2$ ansatz~\eqref{eq:cso_ansatz}, the truncation error relative to
$\lVert\tilde J\rVert_{\Sigma}$ is then
\begin{align}
    \epsilon_{\mathrm{trunc}}(\tilde J) \;\sim\;
        \frac{\lVert J^{(4)}\rVert}{\lVert J^{(1)}\rVert}\; R^{-3}.
    \label{eq:eps_trunc}
\end{align}

For the error induced by the first radial derivative, the scaling
$\lVert\partial_{r}J|_{\Gamma}\rVert\sim R^{-2}\lVert
J^{(1)}\rVert$ gives
\begin{align}
    \epsilon_{\partial_{r}J}(\tilde J) \;\sim\; \epsilon\big(\partial_{r}J|_{\Gamma}\big),
    \label{eq:eps_direct}
\end{align}
where $\epsilon_{\partial_{r}J}(\tilde J)$ is the relative error in the initial
data due to $\partial_{r}J|_{\Gamma}$, and
$\epsilon\big(\partial_{r}J|_{\Gamma}\big)$ is the relative error in
$\partial_{r}J|_{\Gamma}$ itself.
The $H$-hypersurface equation supplies the required second derivative,
$\partial_{\tilde y}^{2}\tilde J|_{\Gamma}$ (Appendix~\ref{Appendix:Dy2J}).
Absolute variations in this equation satisfy
\begin{align}
    \delta\big(\partial_{\tilde y}^{2}\tilde J|_{\Gamma}\big)
      \approx \frac{2}{c_{1}}\;\delta\big(\partial_{\tilde y}\tilde H|_{\Gamma}\big),
    \label{eq:conditioning}
\end{align}
where $c_{1}$ multiplies $\partial_{\tilde y}^{2}\tilde J$ in the
$H$-hypersurface equation. To leading order, $c_1$ is
\begin{align}
    c_{1} \;&\approx\; \frac{2}{R}\Big(1 + R\,W + 2\,\partial_{u}R\Big)
    \nonumber\\
    &\;\simeq\; \frac{2}{R}\left(1 - \frac{2M_{B}}{R}\right)\nonumber\\
    &\approx \frac{2}{R} ,
    \label{eq:c1}
\end{align}
using $r^{2}W\to-2M_{B}$, where $M_{B}$ is the Bondi mass. This is an
approximation because $\partial_{\tilde y}^{2}\tilde J$ also enters through
$\partial_{\tilde y}\tilde W$ and the angular-Jacobian conversions that
reintroduce $\partial_{\tilde y}^{2}\tilde\beta$ and
$\partial_{\tilde y}^{2}\tilde U$ (Appendix~\ref{Appendix:Dy2J}). These
contributions are small here: the exact coefficient obtained by numerically
probing the residual agrees closely with Eq.~\eqref{eq:c1}.

Equation~\eqref{eq:Hbreve_def} gives
$\partial_{\tilde y}\tilde H|_{\Gamma}=\tfrac12(\partial_{u}R\,\partial_{r}J +
R\,\partial_{\tilde u}\partial_{r}J)|_{\Gamma}$. The second term dominates, so
an error in $\partial_{\tilde u}\partial_{r}J|_{\Gamma}$ is amplified by
$R^2/2$ in the ansatz. Combining this amplification with
$\lVert\partial_{\tilde u}\partial_{r}J|_{\Gamma}\rVert\sim
R^{-2}\lVert\partial_{u}J^{(1)}\rVert\sim\omega_{\mathrm{GW}}
R^{-2}\lVert J^{(1)}\rVert$ gives
\begin{align}
    \epsilon_{\partial_{u}\partial_{r}J}(\tilde J) \;\sim\;
        \omega_{\mathrm{GW}} R\;\epsilon\big(\partial_{\tilde u}\partial_{r}J|_{\Gamma}\big).
    \label{eq:eps_indirect}
\end{align}
where $\omega_{\mathrm{GW}}$ denotes the strain frequency. Combining the
three contributions gives
\begin{align}
    \epsilon(\tilde J)\;\sim\;
      \frac{\lVert J^{(4)}\rVert}{\lVert J^{(1)}\rVert}R^{-3}
      \;+\; \epsilon\big(\partial_{r}J|_{\Gamma}\big)
      \;+\; \omega_{\mathrm{GW}} R\,\epsilon\big(\partial_{\tilde u}\partial_{r}J|_{\Gamma}\big).
    \label{eq:eps_total}
\end{align}
We omit numerical prefactors because they depend on the norm convention of
Eq.~\eqref{eq:eps_def}, whereas the scalings do not.
We measure each term in Eq.~\eqref{eq:eps_total} separately by reconstructing
the initial data with individual worldtube inputs replaced by their reference
values. The right panel of Fig.~\ref{fig:IDAccuracy} shows the results,
averaged over $\mathring u_{0}\in[6500,8000]\,M$. As the left panel shows,
this window begins after the last of the Cauchy junk and its main reflections have
passed.

The grey triangles use the reference CCE values of $J$ and its radial
derivatives at the worldtube. They follow the $1/R^{3}$ truncation-error
scaling of Eq.~\eqref{eq:eps_trunc}. The solid curves show the production
initial data from \emph{CauchySecondOrder}. At smaller radii, they coincide
with the truncation floor; beyond $R\simeq400\,M$, they rise above it.

The dashed and dotted curves separate the other two contributions in
Eq.~\eqref{eq:eps_total}. The dashed curves show the error propagated from
$\partial_{\tilde u}\partial_{r}J|_{\Gamma}$ through the $H$-hypersurface
equation: the $\omega_{\mathrm{GW}}R\,
\epsilon(\partial_{\tilde u}\partial_{r}J|_{\Gamma})$ term. The dotted curves
show the error from $\partial_{r}J|_{\Gamma}$ entering the ansatz directly:
the $\epsilon(\partial_{r}J|_{\Gamma})$ term. Each is measured by replacing
one worldtube input with its reference value while retaining the other from
the worldtube data.

Both $\epsilon(\partial_{r}J|_{\Gamma})$ and
$\epsilon(\partial_{\tilde u}\partial_{r}J|_{\Gamma})$ depend on the extraction
radius $R$ and outer-boundary location $R_{\mathrm{max}}$. Both decrease as
the outer boundary moves away from the worldtube and are small at the largest
outer boundary.

In these comparisons, the norm of $\partial_{r}J|_{\Gamma}$ decreases
approximately as $R^{-2}$, whereas its absolute discrepancy with the reference
CCE solution levels off and then increases at larger extraction radii. As a result, the error $\epsilon(\partial_{r}J|_{\Gamma})$ grows with $R$.

The retarded-time derivative has the smaller relative error:
at $R_{\mathrm{max}}=3000\,M$, it is $\sim 10^{-6}$ across the full range of
worldtube radii. Its propagated error on the initial slice, however, grows with $R$
and exceeds the input error by nearly an order of magnitude at $R=900\,M$.

This error becomes dominant because $\partial_{\tilde u}\partial_{r}J|_{\Gamma}$
enters the ansatz through the $H$-equation solve, which amplifies it by the
factor $\omega_{\mathrm{GW}}R$ in Eq.~\eqref{eq:eps_indirect}. This factor is
of order $40$ at the largest radius and is independent of
$R_{\mathrm{max}}$.

For the smallest outer boundary, both contributions exceed the truncation
error beyond the middle of the range of worldtube radii shown. For the largest outer
boundary, the amplified $H$-equation error dominates throughout and is the
leading limitation on accuracy. Extrapolating its scaling with
$R_{\mathrm{max}}$ indicates that an outer boundary of
$R_{\mathrm{max}}\sim10^{4}\,M$ would be needed to reduce it to the truncation
floor.

With the new initial-data scheme, accuracy is therefore limited primarily by
the worldtube data and the method used to obtain
$\partial_{\tilde y}^{2}\tilde J|_{\Gamma}$. Both can be improved. CCM removes
the outer-boundary error entirely. Computing
$\partial_{\tilde y}^{2}\tilde J|_{\Gamma}$ directly from the Cauchy equations
would avoid the $H$-hypersurface inversion and its error amplification. This
direct calculation is required for CCM in any case: the initial data are
specified during the Cauchy evolution, before a worldtube time series is
available to interpolate and differentiate $\partial_{r}J|_{\Gamma}$.
\section{Discussion and conclusions}
\label{sec:disc-concl}
Characteristic evolution is the only robust method for solving Einstein's
equations all the way to future null infinity, $\Igroup$, but its initial data have limited its
accuracy. Existing prescriptions fail to match the Cauchy solution and produce a
transient that contaminates several inspiral cycles. We have identified the
source of this mismatch and constructed initial data free of it. Our main
findings are as follows.
\begin{itemize}
\item \textit{Characteristic junk originates in the initial angular
coordinates.} Every pre-existing scheme produces long-lived, low-frequency
junk [Sec.~\ref{sec:characteristic_junk}]. It arises from an inaccurate initial angular
map from the Cauchy gauge to the partially flat gauge used in characteristic evolution, rather than from
imperfect Cauchy initial data
[Sec.~\ref{sec:angular_problem}].
\item \textit{The initial angular map is determined by a Beltrami equation.}
The condition that the radially constant part of $J$ in the evolved coordinate system is zero yields a Beltrami
problem on the sphere $S^{2}$ in Eq.~\eqref{eq:beltrami}. For bounded worldtube data, the
measurable Riemann mapping theorem~\cite{AhlforsBers:1960} gives a
quasi-conformal solution unique up to conformal maps of the sphere, which
represent the Lorentz freedom of the Bondi--Metzner--Sachs (BMS) group. Existing schemes fail to
determine this map accurately, and the mismatch produces characteristic junk
[Sec.~\ref{subsec:pfmatchingscheme}].
\item \textit{The nonlinear worldtube matching admits a unique solution
within a small-amplitude region.}
Requiring the inverse-square radial term of $J$ to vanish in the
evolution coordinates constrains the coefficients of its Cauchy-gauge radial
expansion. Together with matching the ansatz's value and radial derivatives
to the worldtube data, this closes the nonlinear system at each fixed
matching order.
Under the amplitude bound of
Appendix~\ref{app:fixed_point}, the iteration converges to the unique fixed
point in the prescribed region. Together with the Beltrami solution, this
determines the initial data within the chosen ansatz, up to residual conformal
freedom [Sec.~\ref{sec:asymptotic_expansion}].
\item \textit{CauchySecondOrder infers the angular coordinates by matching
$J$ and its first two radial derivatives at the worldtube.}
The scheme solves the full nonlinear matching system in the
Cauchy gauge and uses the radially constant part of $J$ in the Cauchy coordinate system to determine the angular map
[Sec.~\ref{sec:cauchy_matching_scheme}].
First-order matching reproduces the no-incoming-radiation
approximation at leading order, whereas matching the second radial derivative
of $J$ independently retains radial information omitted by that approximation.
We recover this derivative by inverting the
hypersurface equation for the time derivative of $J$ at the worldtube [Appendix~\ref{Appendix:Dy2J}], using
only quantities already stored by Cauchy simulations.
\item \textit{CauchySecondOrder removes characteristic junk in all tested configurations}
[Sec.~\ref{sec:results}]. The Teukolsky wave's news follows the linearized
solution down to a numerical floor. The binary loses the ${\sim}10\%$
modulation of its $(2,2)$ amplitude that lasted over $1000\,M$. The head-on
collision's late-time tail is resolved, with a local power-law index around
$-3.5$, in agreement with predictions~\cite{Ma:2024qcv, DeAmicis:2024eoy}.
The previous scheme instead saturates at its junk floor.
\item \textit{Worldtube errors limit waveform accuracy once characteristic
junk is removed.} The small residual modulation in Fig.~\ref{fig:BBHStrain}
converges away as the Cauchy outer boundary moves outward and also appears in the
extrapolated waveform, identifying it as outer-boundary contamination. Within
the scheme's error budget, Eq.~\eqref{eq:eps_total}, the truncation error falls
as $R^{-3}$, where $R$ is the worldtube radius, and dominates only at small extraction radii
[Sec.~\ref{sec:id_accuracy}]. Beyond $R\simeq400\,M$, worldtube errors dominate;
the reconstruction of the second radial derivative amplifies one of these contributions by
a factor proportional to the worldtube radius. Third-order matching
would lower the truncation floor, but errors in the same first- and
second-derivative worldtube inputs would still limit the accuracy in these
configurations.
\end{itemize}
The \emph{CauchySecondOrder} implementation in the open-source
\texttt{SpECTRE} code~\cite{spectreCode} allows archived
simulations with stored worldtube data from existing
catalogs~\cite{Boyle:2019kee,Scheel:2025jct,Healy:2017psd,Healy:2019jyf,
Healy:2020vre,Healy:2022wdn,Ficarra:2026cej,Jani:2016wkt,Ferguson:2023vta,
Hamilton:2023qkv,Huerta:2019oxn} to be reprocessed at $\Igroup$
without rerunning the Cauchy evolutions. This reprocessing is possible
for both generalized-harmonic and moving-puncture simulations. Waveform models can then be trained on
solutions of Einstein's equations extending to $\Igroup$. Short simulations
benefit most because junk occupies a large fraction of their signal. Removing
it enables precise measurements of head-on tails [Sec.~\ref{subsec:tails}]
and opens the way to more accurate waveforms for unbound black-hole scattering, allowing closer
comparisons with analytical predictions~\cite{Damour:2014afa, Hopper:2022rwo, Damour:2022ybd, Rettegno:2023ghr, Long:2025tvk, Bern:2021yeh, Dlapa:2022lmu, Driesse:2024xad}.

The construction also extends to Cauchy-characteristic
matching (CCM)~\cite{Ma:2023qjn}, which removes the outer boundary and the dominant
residual contamination it introduces into the waveforms.

In forthcoming work, we will investigate why standard Regge--Wheeler--Zerilli (RWZ) strain extraction and
 extrapolation to future null infinity do not recover nonlinear displacement memory.
 This investigation is motivated by the recovery of memory from data and radial derivatives
  on a single worldtube without characteristic evolution, using the matching and asymptotic coordinate transformations
  developed here. We will compare this reconstruction
   with full CCE waveforms and analyze how RWZ extraction combines the underlying metric
    information, to understand which contributions to the asymptotic strain it retains.

\begin{acknowledgments}
  This work was supported in part by the Sherman
  Fairchild Foundation; by NSF Grants PHY-2407742, PHY-2308615,
  PHY-2607486, OAC-2513338, and NASA award 80NSSC26K0340 at Cornell;
  by NSF Grants
  PHY-2309211, PHY-2309231, PHY-2607927, and OAC-2513339, and
  NASA award 80NSSC26K0340 at Caltech;
  and by NSF Grant PHY-2608279 and NASA award 80NSSC26K0340 at the University of
  Mississippi.
  K.M. is supported by NASA through the NASA Hubble Fellowship grant
  \#HST-HF2-51562.001-A awarded by the Space Telescope Science Institute, which is
  operated by the Association of Universities for Research in Astronomy,
  Incorporated, under NASA contract NAS5-26555. The simulations were carried
    out on the mbot cluster at Cornell and CaltechHPC at Caltech.
\end{acknowledgments}
\appendix

\section{Derivation of the partially flat gauge evolution equation}
\label{sec:pf_gauge_eom}
To obtain the partially flat gauge, we seek an angular coordinate transformation
of the form Eq.~\eqref{eq:generalangtrans} that makes $\hat U^{(0)\hat A}$ vanish.
This is the constant term in the $1/\hat r$ expansion of $\hat U^{\hat A}$ at
$\Igroup$. Starting from
\begin{align}
    e^{-2\hat\beta}\hat U^{\hat A} & = -\hat g^{\hat r\hat A}
    \nonumber\\
    & = -g^{\alpha\beta}\partial_{\alpha}\hat r\,\partial_{\beta}\hat\phi^{\hat A} \nonumber      \\
                                   & = -g^{ru}\partial_{r}\hat r\,\partial_{u}\hat\phi^{\hat A}
    - g^{rA}\partial_{r}\hat r\,\partial_{A}\hat\phi^{\hat A} \nonumber \\
    & \quad - g^{AB}\partial_{A}\hat r\,\partial_{B}\hat\phi^{\hat A},
    \label{eq:appendixA_inverse_metric}
\end{align}
and substituting the Bondi inverse metric components together with the Jacobian
of Eq.~\eqref{eq:generalangtrans}, one obtains
\begin{align}
    \hat U^{\hat A}
     & = e^{2(\hat\beta-\beta)}\hat\omega\big(\partial_{u}\hat\phi^{\hat A} + U^{A}\partial_{A}\hat\phi^{\hat A}\big) \nonumber \\
     & \quad-\ \frac{e^{2\hat\beta}}{r}\,h^{AB}\partial_{A}\hat\omega\,\partial_{B}\hat\phi^{\hat A}.
    \label{eq:appendixA_Uhat}
\end{align}
Since $\hat\omega$ and $\partial_{A}\hat\phi^{\hat A}$ are independent of $r$
and $e^{2(\hat\beta-\beta)}$ is $\mathcal{O}(1)$ at $\Igroup$,
taking $r\to\infty$ yields
\begin{align}
    \hat U^{(0)\hat A}
    = e^{2(\hat\beta^{(0)}-\beta^{(0)})}\hat\omega\big(\partial_{u}\hat\phi^{\hat A} + U^{(0)A}\partial_{A}\hat\phi^{\hat A}\big),
    \label{eq:appendixA_U0}
\end{align}
so that the partially flat condition $\hat U^{(0)\hat A}=0$ is equivalent to
\begin{align}
    \partial_{u}\hat\phi^{\hat A} = -\,U^{(0)A}\partial_{A}\hat\phi^{\hat A},
    \label{eq:angEOM_appendix}
\end{align}
which coincides with Eq.~\eqref{eq:angEOM}.

Equation~\eqref{eq:angEOM_appendix} makes the angular coordinates constant along
the null generators of $\Igroup$, so they are stationary at null infinity.
To show this, introduce the coordinates $(u,l,\phi^{A})$ with $l=1/r$ and the
conformally rescaled inverse metric
$\tilde g^{\alpha\beta}=l^{-2}g^{\alpha\beta}$, which is regular at null
infinity. In this appendix only, the tilde denotes conformal rescaling; in the
body of the paper it denotes the compactified Cauchy coordinate system. The null generators
at $\Igroup$ are
\begin{align}
    \tilde n^{\alpha}\big|_{\Igroup} = \tilde g^{\alpha\beta}\partial_{\beta}l\big|_{\Igroup}
    = e^{-2\beta^{(0)}}\big(1,0,U^{(0)A}\big),
    \label{eq:appendixA_null_generator}
\end{align}
and requiring the angular coordinates to be constant along them,
$\tilde n^{\alpha}|_{\Igroup}\partial_{\alpha}\hat\phi^{\hat A}=0$, recovers
Eq.~\eqref{eq:angEOM_appendix}.

In the partially flat gauge the generators take the form
$\tilde n^{\hat\alpha}|_{\Igroup} = (e^{-2\hat\beta^{(0)}},0,0)$. Inertial
Bondi--Sachs coordinates follow by rescaling the parameter along these
generators so that
\begin{align}
    \tilde n^{\hat\alpha}\big|_{\Igroup}\partial_{\hat\alpha}\mathring u^{(0)} = 1
    \quad\Longleftrightarrow\quad
    \partial_{\hat u}\mathring u^{(0)} = e^{2\hat\beta^{(0)}},
    \label{eq:appendixA_inertial_time}
\end{align}
which is Eq.~\eqref{eq:inertialtime}. With this rescaling,
$\tilde n^{\mathring\alpha}|_{\Igroup} = (1,0,0)$, and
$\partial_{\mathring u}$ %
is the null generator at $\Igroup$.

\section{Computation of \texorpdfstring{$\partial_{\tilde y}^{2}\tilde J$}{dy2J}
at the worldtube}
\label{Appendix:Dy2J}
The \emph{CauchySecondOrder} scheme requires
$\partial_{\tilde y}^{2}\tilde J|_{\Gamma}$, which is not supplied by the
worldtube data. It can be obtained from second derivatives of the ADM metric,
which would require modifying the Cauchy code, or by inverting the Einstein
equations at the worldtube. This appendix describes the latter approach and
its implementation in \texttt{SpECTRE}.

We use Eq.~\eqref{eq:hypersurface_H}, the last equation in the hypersurface
hierarchy, compactified as in Sec.~\ref{subsec:numerical_coordinates} and
written in the Cauchy gauge. This equation for $\tilde H$ is the only one
that explicitly contains $\partial_{\tilde y}^{2}\tilde J$. Evaluating it at
the worldtube $\tilde y = -1$ and collecting the second radial derivatives of
$J$ gives the complex algebraic identity
\begin{equation}
    c_{1}\,\partial_{\tilde y}^{2}\tilde J
    + c_{2}\,\partial_{\tilde y}^{2}\bar{\tilde J}
    + c_{3} = 0,
    \label{eq:appendixB_algebraic}
\end{equation}
in which $c_{1}$, $c_{2}$ and $c_{3}$ depend only on quantities known at
$\Gamma$. The equation can be solved algebraically,
\begin{align}
    \partial_{\tilde y}^{2}\tilde J\big|_{\Gamma}
    = \frac{\bar c_{1}\,c_{3} - c_{2}\,\bar c_{3}}
           {c_{2}\bar c_{2} - c_{1}\bar c_{1}} ,
    \label{eq:appendixB_solution}
\end{align}
which is well posed wherever $|c_{1}| \neq |c_{2}|$.

Symbolic expressions for $c_{1,2,3}$ are cumbersome because the concise form of the hypersurface
equations in Ref.~\cite{Moxon:2020gha} uses the physical angular operator $\eth$ at constant $r$, whereas
worldtube calculations use the numerical operator $\tilde\eth$ at constant
$\tilde y$. Converting between the two introduces radial derivatives
through the Jacobian in Eq.~\eqref{eq:jacobians}. For a Bondi variable $f$ evaluated at $r=R(u,\phi^{A})$, the
chain rule gives
$(\eth f)|_{\Gamma} = \tilde\eth(\tilde f|_{\Gamma}) - \partial_{r}f|_{\Gamma}\,\eth R$,
i.e.,
\begin{align}
    (\eth f)|_{\Gamma}
    = \tilde{\eth}\tilde f|_{\Gamma}
    - 2\frac{\eth R}{R}\,\partial_{\tilde{y}}\tilde f|_{\Gamma},
    \label{eq:appendixB_eth_jacobian}
\end{align}
with corresponding rules for second-order angular derivatives
\begin{align}
    (\eth\eth f)|_{\Gamma}
     & = \tilde{\eth}\tilde{\eth}\tilde f|_{\Gamma}
    - 4\frac{\tilde{\eth}R}{R}\,\tilde{\eth}\partial_{\tilde{y}}\tilde f|_{\Gamma}
    - 2\frac{\tilde{\eth}\tilde{\eth}R}{R}\,\partial_{\tilde{y}}\tilde f|_{\Gamma} \nonumber            \\
     & \quad + 4\!\left(\frac{\tilde{\eth}R}{R}\right)^{\!2}\!\partial_{\tilde{y}}^2\tilde f|_{\Gamma},
    \label{eq:appendixB_etheth}
\end{align}
\begin{align}
    (\eth\bar{\eth} f)|_{\Gamma}
     & = \tilde{\eth}\bar{\tilde\eth}\tilde f|_{\Gamma}
    - 2\frac{\tilde{\eth}R}{R}\,\bar{\tilde\eth}\partial_{\tilde{y}}\tilde f|_{\Gamma}
    - 2\frac{\bar{\tilde\eth}R}{R}\,\tilde{\eth}\partial_{\tilde{y}}\tilde f|_{\Gamma} \nonumber \\
     & \quad - 2\frac{\tilde{\eth}\bar{\tilde\eth}R}{R}\,\partial_{\tilde{y}}\tilde f|_{\Gamma}
    + 4\frac{\tilde{\eth}R\,\bar{\tilde\eth}R}{R^2}\,\partial_{\tilde{y}}^2\tilde f|_{\Gamma},
    \label{eq:appendixB_ethethbar}
\end{align}
which can introduce $\partial_{\tilde y}^{2}\tilde f$ into other hypersurface
equations. Obtaining $c_{1,2,3}$ therefore requires combining equations from
the hierarchy. For example, the $H$-equation contains
$\partial_{\tilde y}\tilde W$, which is supplied by the $W$-equation. The
$W$-equation in turn contains $\eth\partial_{\tilde y}\tilde U$, which
introduces $\partial_{\tilde y}^{2}\tilde U$. Both the $W$- and $Q$-equations
also contain second angular derivatives of $\beta$, which introduce
$\partial_{\tilde y}^{2}\tilde\beta$. Each resulting radial derivative must be
replaced using its hypersurface equation or a derivative of that equation.
For example,
\begin{align}
    \partial_{\tilde{y}}^2\tilde{U}
     & = \frac{e^{2\tilde{\beta}}\partial_{\tilde{y}}\tilde{\beta}}{R}
    (\tilde{K}\tilde{Q} - \tilde{J}\bar{\tilde Q}) \nonumber          \\
     & \quad+ \frac{e^{2\tilde{\beta}}}{2R}
    \!\left(
    \partial_{\tilde{y}}\tilde{K}\,\tilde{Q}
    + \tilde{K}\partial_{\tilde{y}}\tilde{Q}
    - \partial_{\tilde{y}}\tilde{J}\,\bar{\tilde Q}
    - \tilde{J}\partial_{\tilde{y}}\bar{\tilde Q}
    \right),
    \label{eq:appendixB_dy2U}
\end{align}
and
\begin{subequations}\label{eq:appendixB_dy2beta}
\begin{align}
        \partial_{\tilde{y}}^2\tilde{\beta}\big|_{\Gamma}
        &= -\tfrac{1}{2}\partial_{\tilde{y}}\tilde{\beta}
        + \tfrac{1}{2}\Big[
          \mathrm{Re}\big(\partial_{\tilde y}^{2}\tilde J\,
                          \partial_{\tilde y}\bar{\tilde J}\big)
          - \partial_{\tilde y}\tilde K\,\partial_{\tilde y}^{2}\tilde K
        \Big] , \\[4pt]
        \partial_{\tilde y}^{2}\tilde K
        &= \frac{2\,\mathrm{Re}\big(\bar{\tilde J}\partial_{\tilde y}^{2}\tilde J\big)
                 + 2\,\partial_{\tilde y}\tilde J\,\partial_{\tilde y}\bar{\tilde J}
                 - 2\big(\partial_{\tilde y}\tilde K\big)^{2}}{2\tilde K} ,
    \end{align}
\end{subequations}
where the latter expressions introduce $\partial_{\tilde
y}^{2}\tilde J$ and its conjugate, but no other second derivative. Collecting
the coefficients of $\partial_{\tilde{y}}^2\tilde{J}|_{\Gamma}$ and its
conjugate symbolically therefore requires all of these substitutions.

The implementation instead reuses the Jacobian conversions performed during
the CCE evolution. Define
\begin{align}
    F\big(\partial_{\tilde y}^{2}\tilde J\big)
    = \Big[\mathrm{rhs}(H) - \mathrm{lhs}(H)\Big]_{\tilde y = -1},
    \label{eq:appendixB_residual}
\end{align}
the right-minus-left side of Eq.~\eqref{eq:hypersurface_H} evaluated at the
worldtube with a trial value substituted for $\partial_{\tilde y}^{2}\tilde J$
(and its complex conjugate for $\partial_{\tilde y}^{2}\bar{\tilde J}$). By
construction $F(X) = c_{1}X + c_{2}\bar X + c_{3}$, so three residual evaluations determine the coefficients explicitly:
\begin{subequations}\label{eq:appendixB_probe}
\begin{align}
        c_{3} &= F(0), \\
        c_{1}
            &= \frac{F(1)-iF(i)-(1-i)F(0)}{2}, \\
        c_{2}
            &= \frac{F(1)+iF(i)-(1+i)F(0)}{2},
    \end{align}
\end{subequations}
after which Eq.~\eqref{eq:appendixB_solution} gives the required derivative
without a symbolic expression for $c_{1,2,3}$.

The evaluation of $F$ uses the CCE evolution routines throughout. Angular
derivatives are computed with the SWSH differentiation routines and converted
from numerical to physical coordinates using the code's Jacobian application,
which implements
Eqs.~\eqref{eq:appendixB_eth_jacobian}--\eqref{eq:appendixB_ethethbar}.
Hypersurface right-hand sides use the same integrand specializations as the
evolution. This applies the substitutions above through code already tested
against the evolution, retaining every dependence on
$\partial_{\tilde y}^{2}\tilde J$. Substituting the resulting derivative back
into $F$ reproduces the $H$-hypersurface equation to machine precision.

All quantities needed to evaluate $F$ are available at the worldtube except
$\partial_{\tilde u}\partial_{r}J|_{\Gamma}$. This derivative cannot be obtained
from first-order ADM derivatives alone, so we compute it by differentiating
the stored $\partial_{r}J|_{\Gamma}$ in retarded time. As
Sec.~\ref{sec:id_accuracy} shows, this step limits the accuracy of the scheme.

\section{Solving for the partially flat angular map}
\label{app:angular_solve}

Section~\ref{sec:angular_coord_transformation} reduced the partially flat gauge
condition to an algebraic equation at each angular point and solved it in
closed form. The resulting Beltrami coefficient,
$\mu=\hat a/\bar{\hat b}$ in Eq.~\eqref{eq:beltramicoefficient}, specifies the
target angular Jacobian at every point of the sphere:
\begin{align}
    \hat a_{\star} = -\,\frac{\bar{\hat b}_{\star}\,\tilde J^{(0)}}{1 + \tilde K^{(0)}}
    \label{eq:astar}
\end{align}
This appendix describes how to recover the associated map. The unknown is the forward map $\phi^{A}(\hat\phi^{\hat A})$, with Beltrami
coefficient $\mu$. Its inverse solves Eq.~\eqref{eq:beltrami}, and the two
coefficients satisfy $|\nu|=|\mu|$. Because $\tilde J^{(0)}$ is supplied on the
Cauchy sphere, $\mu$ enters through a pullback that depends on the unknown
map. The iteration below resolves this dependence.

In \texttt{SpECTRE} the angular map is represented in Cartesian form
\begin{subequations}\label{eq:angtocart}
\begin{align}
        \hat x^1 & = \sin\hat\theta\cos\hat\phi, \label{eq:angtocart:a} \\
        \hat x^2 & = \sin\hat\theta\sin\hat\phi, \label{eq:angtocart:b} \\
        \hat x^3 & = \cos\hat\theta, \label{eq:angtocart:c}
    \end{align}
\end{subequations}
from which $\hat a$ and $\hat b$ follow by differentiation~\cite{Moxon:2021gbm},
with $\eth$ acting on the collocation sphere on which the current map is defined. A
displacement of $\hat x^{i}$ changes the Jacobians according to the chain rule,
\begin{align}
    \eth\,\delta\hat x^{i}
      = \tfrac12\!\left(\delta\hat a\,\overline{\eth \hat x^{i}}
                      + \overline{\delta\hat b}\,\eth \hat x^{i}\right),
    \label{eq:chainrule}
\end{align}
and we generate the displacement from a single spin-weight-$1$ potential
$\eta$, ensuring that it is the variation of a map:
\begin{align}
    \delta \hat x^{i} = \mathrm{Re}\!\left(\eta\,\overline{\eth\hat x^{i}}\right).
    \label{eq:displacement}
\end{align}
The $\hat x^{i}$ are $\ell=1$ spin-weight-$0$ fields, so $\eth^{2}\hat x^{i}=0$
and $\eth\,\overline{\eth\hat x^{i}}=-2\hat x^{i}$; substituting
Eq.~\eqref{eq:displacement} into Eq.~\eqref{eq:chainrule} then gives
\begin{align}
    \delta\hat a = \eth\eta ,
    \qquad
    \overline{\delta\hat b} = \eth\bar\eta ,
    \label{eq:onepotential}
\end{align}
up to a term along $\hat x^{i}$, which is radial and is removed when the
displaced points are projected to the sphere.

The \emph{ConformalFactor} scheme in \texttt{SpECTRE} already uses the
displacement~\eqref{eq:displacement}, with the potential written as
$f=\bar\eth^{-1}\delta\hat b$ and
\begin{align}
    \delta\hat b = \frac{\hat\omega\,(\hat\omega_{\star}-\hat\omega)}{\hat b} ,
    \qquad
    \delta\hat a = \eth f ,
    \label{eq:cf_step}
\end{align}
which is Eq.~\eqref{eq:onepotential} with $\eta=f$. However,
\emph{ConformalFactor} targets $\hat\omega_{\star}=e^{2\beta}|_{\Gamma}$,
Eq.~\eqref{eq:CF_minimization}, by linearizing in $\delta\hat\omega$; it does
not solve the Beltrami problem. Our target is instead the
Jacobian~\eqref{eq:astar}, which solves the gauge condition analytically.

At each pass, the difference from the target Jacobian~\eqref{eq:astar}
determines the potential $\eta$ through
\begin{align}
    \eta = \eth^{-1}\!\left(\hat a_{\star} - \hat a\right),
    \label{eq:series_step}
\end{align}
and the map is updated using the displacement in Eq.~\eqref{eq:displacement}.
Each pass first computes $\hat a$, $\hat b$ and $\hat\omega$ from the current
$\hat x^{i}$ and interpolates $\tilde J^{(0)}$ onto those points before
evaluating Eq.~\eqref{eq:series_step}. Starting from the identity, where $\hat a=0$ and
$\hat b=2$, the first pass gives the first-order map in closed form,
$\eta=\eth^{-1}(2\mu)$. The kernel of $\eth^{-1}$ on spin weight $2$ is
$\ell<2$ and corresponds exactly to the conformal freedom left undetermined
in Sec.~\ref{sec:angular_coord_transformation}; we set it to zero.

For small $\lVert\mu\rVert_{\infty}$, each first-order
correction is expected to reduce the residual by a factor of order
$\lVert\mu\rVert_{\infty}$. Starting from the identity, where the residual is
of order $\lVert\mu\rVert_{\infty}$, we therefore estimate the number of passes
needed to reach a tolerance $\epsilon_{\mathrm{tol}}$ as

\begin{align}
    n_{\mathrm{pass}} \;=\; \left\lceil
        \frac{\ln\epsilon_{\mathrm{tol}}}{\ln\lVert\mu\rVert_{\infty}}
    \right\rceil - 1 ,
    \label{eq:passcount}
\end{align}

For the configurations studied here, Eq.~\eqref{eq:passcount} predicts two or
three passes, matching the number needed to bring $\breve J^{(0)}$ below
$10^{-14}$\ at $\ell_{\max}=20$. \emph{NoIncomingRadiation} also drives
$\breve J^{(0)}$ to zero, but uses a Newton step on the gauge condition rather
than its closed-form root. It converges much more slowly, requiring of order
$10^{2}$ passes for the same configurations: from $81$ for the $q=1$ binary at
the largest extraction radius to $217$ for the head-on collision at the
smallest.

\section{Existence and uniqueness of the fixed-point solution}
\label{app:fixed_point}
We establish existence and uniqueness of the fixed point in
Eq.~\eqref{eq:fixedpoint}, together with convergence of the iteration from
$x_{[N]}=0$, using the Banach fixed-point theorem~\cite{Banach:1922, Zeidler:1986}. At each angular point, the theorem
requires that $\Phi_{[N]}$ maps a closed disc into itself and is a contraction on
that disc. Both properties follow from a bound on the worldtube data supplied.

We choose the disc $|x_{[N]}|\le\varrho_{[N]}$, with radius
\begin{align}
    \varrho_{[N]} = \max\left(\big\lVert \tilde J_{[N]}^{(0)}(0)\big\rVert_{\infty} ,\;
        \big\lVert\tilde J_{[N]}^{(1)}(0)\big\rVert_{\infty}\right) \text{.}
    \label{eq:epsdef}
\end{align}
The coefficients evaluated at $x_{[N]}=0$ are determined entirely by the worldtube
data. We derive a sufficient
bound on $\varrho_{[N]}$ that ensures both conditions for every matching order
$N\ge1$.

The starting point is the linear dependence of the
coefficients on $x_{[N]}$. Collecting the $x_{[N]}$-independent terms of
Eq.~\eqref{eq:linearsolution} gives
\begin{align*}
    \tilde J_{[N]}^{(n)}(x_{[N]})
        &=\tilde J_{[N]}^{(n)}(0)+\alpha_{[N]}^{(n)}x_{[N]} .
\end{align*}
Equation~\eqref{eq:alphak} shows that $|\alpha_{[N]}^{(0)}|$ and
$|\alpha_{[N]}^{(1)}|$ are largest at $N=1$, where both equal $4$. Thus,
throughout the disc,
\begin{align*}
    \big|\tilde J_{[N]}^{(0)}\big|,\;
        \big|\tilde J_{[N]}^{(1)}\big|
        &\le\varrho_{[N]}+4\varrho_{[N]}=5\varrho_{[N]} .
\end{align*}

First, we show that the disc maps into itself. The inequality
$|\Re(\tilde J_{[N]}^{(0)}\bar{\tilde J}_{[N]}^{(1)})|\le|\tilde J_{[N]}^{(0)}||\tilde J_{[N]}^{(1)}|$,
together with Eq.~\eqref{eq:K1}, makes the bracket in Eq.~\eqref{eq:j2}
nonnegative and no larger than $|\tilde J_{[N]}^{(1)}|^2$. The $5\varrho_{[N]}$ coefficient bounds
therefore give
\begin{align}
    |\Phi_{[N]}| \;\le\; \tfrac12\big|\tilde J_{[N]}^{(0)}\big|\,\big|\tilde J_{[N]}^{(1)}\big|^{2}
        \;\le\; \tfrac{125}{2}\,\varrho_{[N]}^{3} \;\le\; \varrho_{[N]} ,
    \label{eq:selfmapbound}
\end{align}
where the last inequality holds for $\varrho_{[N]}\le\sqrt{2/125}$.

The second requirement is contraction on the disc. The map $\Phi_{[N]}$ is the right-hand side of
Eq.~\eqref{eq:j2}, with its dependence on $x_{[N]}$ and $\bar x_{[N]}$ entering through
$\tilde J_{[N]}^{(0)}$ and $\tilde J_{[N]}^{(1)}$. Let $L_{[N]}$ be its Lipschitz constant on the disc:
the smallest number such that
\begin{align*}
    \big|\Phi_{[N]}(x_{[N]})-\Phi_{[N]}(x_{[N]}')\big|
        &\le L_{[N]}\big|x_{[N]}-x_{[N]}'\big|
\end{align*}
for any two points in the disc. Contraction requires $L_{[N]}<1$.
Since $\Phi_{[N]}$ is not holomorphic, we have
\begin{align}
    d\Phi_{[N]}
        &=\partial_{x_{[N]}}\Phi_{[N]}\,d x_{[N]}
        +\partial_{\bar x_{[N]}}\Phi_{[N]}\,d\bar x_{[N]} , \notag\\
    L_{[N]} &= \sup\left(\big|\partial_{x_{[N]}}\Phi_{[N]}\big|
        + \big|\partial_{\bar x_{[N]}}\Phi_{[N]}\big|\right) \text{,}
    \label{eq:Ldef}
\end{align}
where the supremum is over the disc $|x_{[N]}|\le\varrho_{[N]}$. With
$\tilde K_{[N]}^{(0)}=\sqrt{1+|\tilde J_{[N]}^{(0)}|^2}$, we have
\begin{align*}
    \frac{|\tilde J_{[N]}^{(0)}|}{(\tilde K_{[N]}^{(0)})^2}
        &\le\frac12 , \qquad
    \frac{|\tilde J_{[N]}^{(0)}|^3}{(\tilde K_{[N]}^{(0)})^4}
        \le\frac{3\sqrt3}{16} .
\end{align*}
Differentiating Eq.~\eqref{eq:j2} and applying these inequalities, together
with the bounds on the coefficients and
$|\Re(\tilde J_{[N]}^{(0)}\bar{\tilde J}_{[N]}^{(1)})|$ above, gives
\begin{align}
    L_{[N]} \le 150\,\varrho_{[N]}^{2} + 665\,\varrho_{[N]}^{3}
     .
    \label{eq:Lbound}
\end{align}
The bound $\varrho_{[N]}\le7\times10^{-2}$ ensures $L_{[N]}<1$
in Eq.~\eqref{eq:Lbound}, whose right-hand side is just below unity at the
upper limit. It also satisfies the condition in
Eq.~\eqref{eq:selfmapbound}.
Both hypotheses of the Banach theorem are therefore satisfied. At every point
of the sphere, a fixed point exists, is unique in the disc $|x_{[N]}|\le\varrho_{[N]}$,
and is reached by iteration from $x_{[N]}=0$.

This bound is sufficient but not sharp. Physical worldtube data in the radiation zone comfortably
satisfy it; their small amplitudes also make the contraction factor in
Eq.~\eqref{eq:Lbound} small, explaining the rapid convergence of the
iteration. Data that do not satisfy this bound might still admit a unique solution,
but lie outside the reach of this proof.

The same estimates also describe the size and effect of the nonlinear
correction. Equation~\eqref{eq:selfmapbound} bounds the fixed-point value of
$x_{[N]}$. To estimate its effect on the angular map, consider the induced change
\begin{align*}
    \Delta\tilde J_{[N]}^{(0)}
        &=\tilde J_{[N]}^{(0)}(x_{[N]})-\tilde J_{[N]}^{(0)}(0) ,
\end{align*}
with $x_{[N]}$ evaluated at the fixed point.
Using the angular norm~\eqref{eq:angular_norm}, we estimate its relative size as
\begin{align}
    \frac{\lVert\Delta \tilde J_{[N]}^{(0)}\rVert}{\lVert \tilde J_{[N]}^{(0)}\rVert}
        &\;\simeq\; \frac{4}{N(N+1)}\,
        \big\lVert\tilde J_{[N]}^{(1)}\big\rVert^{2} ,
    \label{eq:xImpact}
\end{align}
which falls off as $N^{-2}$: increasing the matching order reduces the effect
of $x_{[N]}$ on the coordinates. For the binary of Fig.~\ref{fig:RadialCoeffs},
$\lVert\tilde J^{(1)}\rVert\sim10^{-4}$, so Eq.~\eqref{eq:xImpact} gives a
relative difference of $\sim10^{-8}$ for small values of $N$. This is far below
the precision to which the worldtube data are currently known.

\bibliography{CCEID}

@article{Moxon:2020gha,
    author        = {Moxon, Jordan and Scheel, Mark A. and Teukolsky, Saul A.},
    title         = {{Improved Cauchy-characteristic evolution system for high-precision numerical relativity waveforms}},
    journal       = {Phys. Rev. D},
    volume        = {102},
    pages         = {044052},
    year          = {2020},
    doi           = {10.1103/PhysRevD.102.044052},
    eprint        = {2007.01339},
    archivePrefix = {arXiv},
    primaryClass  = {gr-qc},
}

@article{Moxon:2021gbm,
    author        = {Moxon, Jordan and Scheel, Mark A. and Teukolsky, Saul A. and Deppe, Nils and Fischer, Nils and H{\'e}bert, Fran{\c{c}}ois and Kidder, Lawrence E. and Throwe, William},
    title         = {{The SpECTRE Cauchy-characteristic evolution system for rapid, precise waveform extraction}},
    journal       = {Phys. Rev. D},
    volume        = {107},
    pages         = {064013},
    year          = {2023},
    doi           = {10.1103/PhysRevD.107.064013},
    eprint        = {2110.08635},
    archivePrefix = {arXiv},
    primaryClass  = {gr-qc},
}

@article{Ma:2023qjn,
    author        = {Ma, Sizheng and Moxon, Jordan and Scheel, Mark A. and Nelli, Kyle C. and Deppe, Nils and Bonilla, Marceline S. and Kidder, Lawrence E. and Kumar, Prayush and Lovelace, Geoffrey and Throwe, William and Vu, Nils L.},
    title         = {{Fully relativistic three-dimensional Cauchy-characteristic matching for physical degrees of freedom}},
    journal       = {Phys. Rev. D},
    volume        = {109},
    pages         = {124027},
    year          = {2024},
    doi           = {10.1103/PhysRevD.109.124027},
    eprint        = {2308.10361},
    archivePrefix = {arXiv},
    primaryClass  = {gr-qc},
}

@article{Ma:2024qcv,
    author        = {Ma, Sizheng and Scheel, Mark A. and Moxon, Jordan and Nelli, Kyle C. and Deppe, Nils and Kidder, Lawrence E. and Throwe, William and Vu, Nils L.},
    title         = {{Merging black holes with Cauchy-characteristic matching: Computation of late-time tails}},
    journal       = {Phys. Rev. D},
    volume        = {112},
    pages         = {024003},
    year          = {2025},
    doi           = {10.1103/jd26-8q5w},
    eprint        = {2412.06906},
    archivePrefix = {arXiv},
    primaryClass  = {gr-qc},
}

@article{Mitman:2024uss,
    author        = {Mitman, Keefe and Boyle, Michael and Stein, Leo C. and Deppe, Nils and Kidder, Lawrence E. and Moxon, Jordan and Pfeiffer, Harald P. and Scheel, Mark A. and Teukolsky, Saul A. and Throwe, William and Vu, Nils L.},
    title         = {{A Review of Gravitational Memory and BMS Frame Fixing in Numerical Relativity}},
    journal       = {Class. Quant. Grav.},
    volume        = {41},
    number        = {22},
    pages         = {223001},
    year          = {2024},
    doi           = {10.1088/1361-6382/ad83c2},
    eprint        = {2405.08868},
    archivePrefix = {arXiv},
    primaryClass  = {gr-qc},
}

@article{Bondi:1962px,
    author        = {Bondi, H. and van der Burg, M. G. J. and Metzner, A. W. K.},
    title         = {{Gravitational waves in general relativity. 7. Waves from axisymmetric isolated systems}},
    journal       = {Proc. Roy. Soc. Lond. A},
    volume        = {269},
    pages         = {21--52},
    year          = {1962},
    doi           = {10.1098/rspa.1962.0161},
}

@article{Sachs:1962wk,
    author        = {Sachs, R. K.},
    title         = {{Gravitational waves in general relativity. 8. Waves in asymptotically flat space-times}},
    journal       = {Proc. Roy. Soc. Lond. A},
    volume        = {270},
    pages         = {103--126},
    year          = {1962},
    doi           = {10.1098/rspa.1962.0206},
}

@article{Bishop:1996gt,
    author        = {Bishop, Nigel T. and Gomez, Roberto and Lehner, Luis and Winicour, Jeffrey},
    title         = {{Cauchy characteristic extraction in numerical relativity}},
    journal       = {Phys. Rev. D},
    volume        = {54},
    pages         = {6153--6165},
    year          = {1996},
    doi           = {10.1103/PhysRevD.54.6153},
    eprint        = {gr-qc/9705033},
    archivePrefix = {arXiv},
}

@article{Babiuc:2011qi,
    author        = {Babiuc, M. C. and Winicour, J. and Zlochower, Y.},
    title         = {{Binary black hole waveform extraction at null infinity}},
    journal       = {Class. Quant. Grav.},
    volume        = {28},
    pages         = {134006},
    year          = {2011},
    doi           = {10.1088/0264-9381/28/13/134006},
    eprint        = {1106.4841},
    archivePrefix = {arXiv},
    primaryClass  = {gr-qc},
}

@article{Reisswig:2009us,
    author        = {Reisswig, C. and Bishop, N. T. and Pollney, D. and Szilagyi, B.},
    title         = {{Unambiguous determination of gravitational waveforms from binary black hole mergers}},
    journal       = {Phys. Rev. Lett.},
    volume        = {103},
    pages         = {221101},
    year          = {2009},
    doi           = {10.1103/PhysRevLett.103.221101},
    eprint        = {0907.2637},
    archivePrefix = {arXiv},
    primaryClass  = {gr-qc},
}

@article{Handmer:2014qha,
    author        = {Handmer, Casey J. and Szilagyi, Bela},
    title         = {{Spectral characteristic evolution: a new algorithm for gravitational wave propagation}},
    journal       = {Class. Quant. Grav.},
    volume        = {32},
    number        = {2},
    pages         = {025008},
    year          = {2015},
    doi           = {10.1088/0264-9381/32/2/025008},
    eprint        = {1406.7029},
    archivePrefix = {arXiv},
    primaryClass  = {gr-qc},
}

@article{Handmer:2016mls,
    author        = {Handmer, Casey J. and Szilagyi, Bela and Winicour, Jeffrey},
    title         = {{Gauge invariant spectral Cauchy characteristic extraction}},
    journal       = {Class. Quant. Grav.},
    volume        = {32},
    number        = {23},
    pages         = {235018},
    year          = {2015},
    doi           = {10.1088/0264-9381/32/23/235018},
    eprint        = {1502.06987},
    archivePrefix = {arXiv},
    primaryClass  = {gr-qc},
}

@article{Bishop:2016lgv,
    author        = {Bishop, Nigel T. and Rezzolla, Luciano},
    title         = {{Extraction of gravitational waves in numerical relativity}},
    journal       = {Living Rev. Rel.},
    volume        = {19},
    pages         = {2},
    year          = {2016},
    doi           = {10.1007/s41114-016-0001-9},
    eprint        = {1606.02532},
    archivePrefix = {arXiv},
    primaryClass  = {gr-qc},
}

@article{Winicour:2005ge,
    author        = {Winicour, Jeffrey},
    title         = {{Characteristic evolution and matching}},
    journal       = {Living Rev. Rel.},
    volume        = {15},
    pages         = {2},
    year          = {2012},
    doi           = {10.12942/lrr-2012-2},
}

@article{Boyle:2009vi,
    author        = {Boyle, Michael and Mroue, Abdul H.},
    title         = {{Extrapolating gravitational-wave data from numerical simulations}},
    journal       = {Phys. Rev. D},
    volume        = {80},
    pages         = {124045},
    year          = {2009},
    doi           = {10.1103/PhysRevD.80.124045},
    eprint        = {0905.3177},
    archivePrefix = {arXiv},
    primaryClass  = {gr-qc},
}

@article{Iozzo:2020jcu,
    author        = {Iozzo, Dante A. B. and Boyle, Michael and Deppe, Nils and Moxon, Jordan and Scheel, Mark A. and Teukolsky, Saul A. and Throwe, William and Vu, Nils L.},
    title         = {{Extending gravitational wave extrapolation to higher orders}},
    journal       = {Phys. Rev. D},
    volume        = {103},
    number        = {2},
    pages         = {024039},
    year          = {2021},
    doi           = {10.1103/PhysRevD.103.024039},
    eprint        = {2010.15200},
    archivePrefix = {arXiv},
    primaryClass  = {gr-qc},
}

@article{LIGOScientific:2018mvr,
    author        = {{LIGO Scientific and Virgo Collaborations}, B. P. Abbott and others},
    title         = {{GWTC-1: A Gravitational-Wave Transient Catalog of Compact Binary Mergers Observed by LIGO and Virgo during the First and Second Observing Runs}},
    journal       = {Phys. Rev. X},
    volume        = {9},
    number        = {3},
    pages         = {031040},
    year          = {2019},
    doi           = {10.1103/PhysRevX.9.031040},
    eprint        = {1811.12907},
    archivePrefix = {arXiv},
    primaryClass  = {astro-ph.HE},
}

@article{LIGOScientific:2021djp,
    author        = {{LIGO Scientific, Virgo and KAGRA Collaborations}, R. Abbott and others},
    title         = {{GWTC-3: Compact Binary Coalescences Observed by LIGO and Virgo During the Second Part of the Third Observing Run}},
    journal       = {Phys. Rev. X},
    volume        = {13},
    number        = {4},
    pages         = {041039},
    year          = {2023},
    doi           = {10.1103/PhysRevX.13.041039},
    eprint        = {2111.03606},
    archivePrefix = {arXiv},
    primaryClass  = {gr-qc},
}

@article{LIGOScientific:2014pky,
    author        = {{LIGO Scientific Collaboration}, J. Aasi and others},
    title         = {{Advanced LIGO}},
    journal       = {Class. Quant. Grav.},
    volume        = {32},
    pages         = {074001},
    year          = {2015},
    doi           = {10.1088/0264-9381/32/7/074001},
    eprint        = {1411.4547},
    archivePrefix = {arXiv},
    primaryClass  = {gr-qc},
}

@article{VIRGO:2014yos,
    author        = {{Virgo Collaboration}, F. Acernese and others},
    title         = {{Advanced Virgo: a second-generation interferometric gravitational wave detector}},
    journal       = {Class. Quant. Grav.},
    volume        = {32},
    number        = {2},
    pages         = {024001},
    year          = {2015},
    doi           = {10.1088/0264-9381/32/2/024001},
    eprint        = {1408.3978},
    archivePrefix = {arXiv},
    primaryClass  = {gr-qc},
}

@article{KAGRA:2020tym,
    author        = {{KAGRA Collaboration}, T. Akutsu and others},
    title         = {{Overview of KAGRA: Detector design and construction history}},
    journal       = {PTEP},
    volume        = {2021},
    number        = {5},
    pages         = {05A101},
    year          = {2021},
    doi           = {10.1093/ptep/ptaa125},
    eprint        = {2005.05574},
    archivePrefix = {arXiv},
    primaryClass  = {gr-qc},
}

@article{Reitze:2019iox,
    author        = {Reitze, David and others},
    title         = {{Cosmic Explorer: The U.S. Contribution to Gravitational-Wave Astronomy beyond LIGO}},
    journal       = {Bull. Am. Astron. Soc.},
    volume        = {51},
    number        = {7},
    pages         = {035},
    year          = {2019},
    eprint        = {1907.04833},
    archivePrefix = {arXiv},
    primaryClass  = {astro-ph.IM},
}

@misc{Evans:2021gyd,
    author        = {Evans, Matthew and others},
    title         = {{A Horizon Study for Cosmic Explorer: Science, Observatories, and Community}},
    year          = {2021},
    eprint        = {2109.09882},
    archivePrefix = {arXiv},
    primaryClass  = {astro-ph.IM},
}

@article{Punturo:2010zz,
    author        = {Punturo, M. and others},
    title         = {{The Einstein Telescope: A third-generation gravitational wave observatory}},
    journal       = {Class. Quant. Grav.},
    volume        = {27},
    pages         = {194002},
    year          = {2010},
    doi           = {10.1088/0264-9381/27/19/194002},
}

@article{Maggiore:2019uih,
    author        = {Maggiore, Michele and others},
    title         = {{Science Case for the Einstein Telescope}},
    journal       = {JCAP},
    volume        = {2020},
    number        = {03},
    pages         = {050},
    year          = {2020},
    doi           = {10.1088/1475-7516/2020/03/050},
    eprint        = {1912.02622},
    archivePrefix = {arXiv},
    primaryClass  = {astro-ph.CO},
}

@misc{LISA:2017pwj,
    author        = {{LISA Collaboration}, P. Amaro-Seoane and others},
    title         = {{Laser Interferometer Space Antenna}},
    year          = {2017},
    eprint        = {1702.00786},
    archivePrefix = {arXiv},
    primaryClass  = {astro-ph.IM},
}

@misc{Colpi:2024xhw,
    author        = {Colpi, Monica and others},
    title         = {{LISA Definition Study Report}},
    year          = {2024},
    eprint        = {2402.07571},
    archivePrefix = {arXiv},
    primaryClass  = {astro-ph.CO},
}

@article{Purrer:2019jcp,
    author        = {P{\"u}rrer, Michael and Haster, Carl-Johan},
    title         = {{Gravitational waveform accuracy requirements for future ground-based detectors}},
    journal       = {Phys. Rev. Res.},
    volume        = {2},
    number        = {2},
    pages         = {023151},
    year          = {2020},
    doi           = {10.1103/PhysRevResearch.2.023151},
    eprint        = {1912.10055},
    archivePrefix = {arXiv},
    primaryClass  = {gr-qc},
}

@article{Ferguson:2020xnm,
    author        = {Ferguson, Deborah and Jani, Karan and Laguna, Pablo and Shoemaker, Deirdre},
    title         = {{Assessing the readiness of numerical relativity for LISA and 3G detectors}},
    journal       = {Phys. Rev. D},
    volume        = {104},
    number        = {4},
    pages         = {044037},
    year          = {2021},
    doi           = {10.1103/PhysRevD.104.044037},
    eprint        = {2006.04272},
    archivePrefix = {arXiv},
    primaryClass  = {gr-qc},
}

@misc{Pompili:2024yec,
    author        = {Pompili, Lorenzo and others},
    title         = {{Accuracy of post-Newtonian and numerical-relativity hybrids in the LISA era}},
    year          = {2024},
    eprint        = {2410.02989},
    archivePrefix = {arXiv},
    primaryClass  = {gr-qc},
}

@article{Pretorius:2005gq,
    author        = {Pretorius, Frans},
    title         = {{Evolution of binary black hole spacetimes}},
    journal       = {Phys. Rev. Lett.},
    volume        = {95},
    pages         = {121101},
    year          = {2005},
    doi           = {10.1103/PhysRevLett.95.121101},
    eprint        = {gr-qc/0507014},
    archivePrefix = {arXiv},
}

@article{Campanelli:2005dd,
    author        = {Campanelli, M. and Lousto, C. O. and Marronetti, P. and Zlochower, Y.},
    title         = {{Accurate evolutions of orbiting black-hole binaries without excision}},
    journal       = {Phys. Rev. Lett.},
    volume        = {96},
    pages         = {111101},
    year          = {2006},
    doi           = {10.1103/PhysRevLett.96.111101},
    eprint        = {gr-qc/0511048},
    archivePrefix = {arXiv},
}

@article{Baker:2005vv,
    author        = {Baker, John G. and Centrella, Joan and Choi, Dae-Il and Koppitz, Michael and van Meter, James},
    title         = {{Gravitational wave extraction from an inspiraling configuration of merging black holes}},
    journal       = {Phys. Rev. Lett.},
    volume        = {96},
    pages         = {111102},
    year          = {2006},
    doi           = {10.1103/PhysRevLett.96.111102},
    eprint        = {gr-qc/0511103},
    archivePrefix = {arXiv},
}

@article{Boyle:2019kee,
    author        = {Boyle, Michael and others},
    title         = {{The SXS Collaboration catalog of binary black hole simulations}},
    journal       = {Class. Quant. Grav.},
    volume        = {36},
    number        = {19},
    pages         = {195006},
    year          = {2019},
    doi           = {10.1088/1361-6382/ab34e2},
    eprint        = {1904.04831},
    archivePrefix = {arXiv},
    primaryClass  = {gr-qc},
}

@misc{Scheel:2025jct,
    author        = {Scheel, Mark A. and others},
    title         = {{The SXS collaboration's third catalog of binary black hole simulations}},
    year          = {2025},
    eprint        = {2505.13378},
    archivePrefix = {arXiv},
    primaryClass  = {gr-qc},
}

@article{Mitman:2021xkq,
    author        = {Mitman, Keefe and Iozzo, Dante A. B. and Khera, Neev and Boyle, Michael and Hinderer, Tanja and Moxon, Jordan and Pfeiffer, Harald P. and Scheel, Mark A. and Teukolsky, Saul A. and Throwe, William},
    title         = {{Fixing the BMS frame of numerical relativity waveforms with BMS charges}},
    journal       = {Phys. Rev. D},
    volume        = {106},
    number        = {8},
    pages         = {084029},
    year          = {2022},
    doi           = {10.1103/PhysRevD.106.084029},
    eprint        = {2208.04356},
    archivePrefix = {arXiv},
    primaryClass  = {gr-qc},
}

@article{Mitman:2022kwt,
    author        = {Mitman, Keefe and Lagos, Macarena and Stein, Leo C. and Ma, Sizheng and Hui, Lam and Chen, Yanbei and Deppe, Nils and H{\'e}bert, Fran{\c{c}}ois and Kidder, Lawrence E. and Moxon, Jordan and Pfeiffer, Harald P. and Scheel, Mark A. and Teukolsky, Saul A. and Throwe, William and Vu, Nils L.},
    title         = {{Nonlinearities in black hole ringdowns}},
    journal       = {Phys. Rev. Lett.},
    volume        = {130},
    number        = {8},
    pages         = {081402},
    year          = {2023},
    doi           = {10.1103/PhysRevLett.130.081402},
    eprint        = {2208.07380},
    archivePrefix = {arXiv},
    primaryClass  = {gr-qc},
}

@article{Berti:2009kk,
    author        = {Berti, Emanuele and Cardoso, Vitor and Starinets, Andrei O.},
    title         = {{Quasinormal modes of black holes and black branes}},
    journal       = {Class. Quant. Grav.},
    volume        = {26},
    pages         = {163001},
    year          = {2009},
    doi           = {10.1088/0264-9381/26/16/163001},
    eprint        = {0905.2975},
    archivePrefix = {arXiv},
    primaryClass  = {gr-qc},
}

@misc{Berti:2025hly,
    author        = {Berti, Emanuele and others},
    title         = {{Black hole spectroscopy: status report}},
    year          = {2025},
    eprint        = {2505.23895},
    archivePrefix = {arXiv},
    primaryClass  = {gr-qc},
}

@article{Favata:2010zu,
    author        = {Favata, Marc},
    title         = {{The gravitational-wave memory effect}},
    journal       = {Class. Quant. Grav.},
    volume        = {27},
    pages         = {084036},
    year          = {2010},
    doi           = {10.1088/0264-9381/27/8/084036},
    eprint        = {1003.3486},
    archivePrefix = {arXiv},
    primaryClass  = {gr-qc},
}

@misc{Strominger:2017zoo,
    author        = {Strominger, Andrew},
    title         = {{Lectures on the Infrared Structure of Gravity and Gauge Theory}},
    year          = {2017},
    eprint        = {1703.05448},
    archivePrefix = {arXiv},
    primaryClass  = {hep-th},
}

@article{Madler:2016xju,
    author        = {M{\"a}dler, Thomas and Winicour, Jeffrey},
    title         = {{Bondi-Sachs Formalism}},
    journal       = {Scholarpedia},
    volume        = {11},
    pages         = {33528},
    year          = {2016},
    doi           = {10.4249/scholarpedia.33528},
    eprint        = {1609.01731},
    archivePrefix = {arXiv},
    primaryClass  = {gr-qc},
}

@misc{spectreCode,
    author        = {Deppe, Nils and Throwe, William and Kidder, Lawrence E. and Vu, Nils L. and Nelli, Kyle C. and Boyle, Michael and others},
    title         = {{\textsc{SpECTRE} v2025.07.30}},
    year          = {2025},
    doi           = {10.5281/zenodo.4290404},
    howpublished  = {Zenodo},
    note          = {\url{https://github.com/sxs-collaboration/spectre}},
}

@article{Bini:2017wfr,
    author        = {Bini, Donato and Damour, Thibault and Geralico, Andrea},
    title         = {{Novel approach to binary dynamics: application to the fifth post-Newtonian level}},
    journal       = {Phys. Rev. Lett.},
    volume        = {123},
    number        = {23},
    pages         = {231104},
    year          = {2019},
    doi           = {10.1103/PhysRevLett.123.231104},
    eprint        = {1909.02375},
    archivePrefix = {arXiv},
    primaryClass  = {gr-qc},
}

@article{Bern:2019nnu,
    author        = {Bern, Zvi and Cheung, Clifford and Roiban, Radu and Shen, Chia-Hsien and Solon, Mikhail P. and Zeng, Mao},
    title         = {{Black Hole Binary Dynamics from the Double Copy and Effective Theory}},
    journal       = {JHEP},
    volume        = {2019},
    number        = {10},
    pages         = {206},
    year          = {2019},
    doi           = {10.1007/JHEP10(2019)206},
    eprint        = {1908.01493},
    archivePrefix = {arXiv},
    primaryClass  = {hep-th},
}

@article{Newman:1961qr,
    author        = {Newman, Ezra and Penrose, Roger},
    title         = {{An Approach to gravitational radiation by a method of spin coefficients}},
    journal       = {J. Math. Phys.},
    volume        = {3},
    pages         = {566--578},
    year          = {1962},
    doi           = {10.1063/1.1724257},
}

@article{Teukolsky:1982nz,
    author        = {Teukolsky, Saul A.},
    title         = {{Linearized quadrupole waves in general relativity and the motion of test particles}},
    journal       = {Phys. Rev. D},
    volume        = {26},
    pages         = {745},
    year          = {1982},
    doi           = {10.1103/PhysRevD.26.745},
}

@article{Flanagan:2015pxa,
    author        = {Flanagan, {\'E}anna {\'E}. and Nichols, David A.},
    title         = {{Conserved charges of the extended Bondi-Metzner-Sachs algebra}},
    journal       = {Phys. Rev. D},
    volume        = {95},
    number        = {4},
    pages         = {044002},
    year          = {2017},
    doi           = {10.1103/PhysRevD.95.044002},
    eprint        = {1510.03386},
    archivePrefix = {arXiv},
    primaryClass  = {hep-th},
}

@article{DeAmicis:2024eoy,
    author        = {De Amicis, Marina and R{\"u}ter, Hannes and Carullo, Gregorio and Albanesi, Simone and Ferrus, C. Melize and Mitman, Keefe and Stein, Leo C. and Cardoso, Vitor and Bernuzzi, Sebastiano and Boyle, Michael and Deppe, Nils and Kidder, Lawrence E. and Moxon, Jordan and Nagar, Alessandro and Nelli, Kyle C. and Pfeiffer, Harald P. and Scheel, Mark A. and Throwe, William and Vu, Nils L. and Zengino{\u{g}}lu, An{\i}l},
    title         = {{Late-time tails in nonlinear evolutions of merging black holes}},
    journal       = {Phys. Rev. Lett.},
    volume        = {135},
    number        = {17},
    pages         = {171401},
    year          = {2025},
    doi           = {10.1103/2brx-xnyr},
    eprint        = {2412.06887},
    archivePrefix = {arXiv},
    primaryClass  = {gr-qc},
}

@article{DaRe:2025glj,
	author = "Da Re, Guido and others",
	title = "{Modeling the BMS transformation induced by a binary black hole merger}",
	eprint = "2503.09569",
	archivePrefix = "arXiv",
	primaryClass = "gr-qc",
	doi = "10.1103/PhysRevD.111.124019",
	journal = "Phys. Rev. D",
	volume = "111",
	number = "12",
	pages = "124019",
	year = "2025"
}

@article{LIGOScientific:2025rid,
	author = "Abac, A. G. and others",
	collaboration = "LIGO Scientific, Virgo, KAGRA",
	title = "{GW250114: Testing Hawking{\textquoteright}s Area Law and the Kerr Nature of Black Holes}",
	eprint = "2509.08054",
	archivePrefix = "arXiv",
	primaryClass = "gr-qc",
	reportNumber = "LIGO-P2500421",
	doi = "10.1103/kw5g-d732",
	journal = "Phys. Rev. Lett.",
	volume = "135",
	number = "11",
	pages = "111403",
	year = "2025"
}

@article{LIGOScientific:2025wao,
	author = "Abac, A. G. and others",
	collaboration = "LIGO Scientific, Virgo, KAGRA",
	title = "{Black Hole Spectroscopy and Tests of General Relativity with GW250114}",
	eprint = "2509.08099",
	archivePrefix = "arXiv",
	primaryClass = "gr-qc",
	reportNumber = "LIGO P2500461",
	doi = "10.1103/6c61-fm1n",
	journal = "Phys. Rev. Lett.",
	volume = "136",
	number = "4",
	pages = "041403",
	year = "2026"
}

@article{Mitman:2020pbt,
	author = "Mitman, Keefe and Moxon, Jordan and Scheel, Mark A. and Teukolsky, Saul A. and Boyle, Michael and Deppe, Nils and Kidder, Lawrence E. and Throwe, William",
	title = "{Computation of displacement and spin gravitational memory in numerical relativity}",
	eprint = "2007.11562",
	archivePrefix = "arXiv",
	primaryClass = "gr-qc",
	doi = "10.1103/PhysRevD.102.104007",
	journal = "Phys. Rev. D",
	volume = "102",
	number = "10",
	pages = "104007",
	year = "2020"
}

@article{Pollney:2010hs,
	author = "Pollney, Denis and Reisswig, Christian",
	title = "{Gravitational memory in binary black hole mergers}",
	eprint = "1004.4209",
	archivePrefix = "arXiv",
	primaryClass = "gr-qc",
	doi = "10.1088/2041-8205/732/1/L13",
	journal = "Astrophys. J. Lett.",
	volume = "732",
	pages = "L13",
	year = "2011"
}

@misc{companion,
  author  = {Da Re, Guido and Mitman, Keefe and Stein, Leo C. and Deppe, Nils
             and Chaudhary, Himanshu and Sun, Dongze and Boyle, Michael
             and Nelli, Kyle C. and Scheel, Mark A. and Teukolsky, Saul A.},
  title   = {{Calculating gravitational waves at infinity: the missing link}},
  howpublished = {Companion Letter, in preparation},
  year    = {2026}
}

@article{Barkett:2019uae,
	author = "Barkett, Kevin and Moxon, Jordan and Scheel, Mark A. and Szil{\'a}gyi, B{\'e}la",
	title = "{Spectral Cauchy-Characteristic Extraction of the Gravitational Wave News Function}",
	eprint = "1910.09677",
	archivePrefix = "arXiv",
	primaryClass = "gr-qc",
	doi = "10.1103/PhysRevD.102.024004",
	journal = "Phys. Rev. D",
	volume = "102",
	number = "2",
	pages = "024004",
	year = "2020"
}

@article{AhlforsBers:1960,
    author  = {Ahlfors, Lars and Bers, Lipman},
    title   = {{Riemann's mapping theorem for variable metrics}},
    journal = {Ann. Math.},
    volume  = {72},
    number  = {2},
    pages   = {385--404},
    year    = {1960},
    doi     = {10.2307/1970141}
}

@book{Ahlfors:1966,
    author    = {Ahlfors, Lars V.},
    title     = {{Lectures on Quasiconformal Mappings}},
    series    = {University Lecture Series},
    volume    = {38},
    edition   = {2},
    publisher = {American Mathematical Society},
    address   = {Providence, RI},
    year      = {2006},
    note      = {First edition: Van Nostrand, 1966},
    isbn      = {978-0-8218-3644-6}
}

@misc{LIGOScientific:2026wfs,
    author = "Abac, A. G. and others",
    collaboration = "LIGO Scientific, VIRGO, KAGRA",
    title = "{GWTC-5.0: Observations from the Second Part of the Fourth LIGO-Virgo-KAGRA Observing Run and Updates to the Gravitational-Wave Transient Catalog}",
    eprint = "2605.27225",
    archivePrefix = "arXiv",
    primaryClass = "gr-qc",
    reportNumber = "LIGO-P2600152",
    month = "5",
    year = "2026"
}

@misc{LIGOScientific:2026ctl,
    author = "Abac, A. G. and others",
    collaboration = "LIGO Scientific, VIRGO, KAGRA",
    title = "{GWTC-5.0: Population Properties of Merging Compact Binaries}",
    eprint = "2605.27226",
    archivePrefix = "arXiv",
    primaryClass = "astro-ph.HE",
    reportNumber = "LIGO-P2600045",
    month = "5",
    year = "2026"
}

@misc{LIGOScientific:2026oim,
    author = "Abac, A. G. and others",
    collaboration = "LIGO Scientific, VIRGO, KAGRA",
    title = "{GWTC-5.0: Tests of General Relativity}",
    eprint = "2607.19293",
    archivePrefix = "arXiv",
    primaryClass = "gr-qc",
    reportNumber = "LIGO-P2500781",
    month = "7",
    year = "2026"
}

@misc{LIGOScientific:2026uyd,
    author = "Abac, A. G. and others",
    collaboration = "LIGO Scientific, VIRGO, KAGRA",
    title = "{GWTC-5.0: Constraints on the Cosmic Expansion Rate and Modified Gravitational-wave Propagation}",
    eprint = "2605.27227",
    archivePrefix = "arXiv",
    primaryClass = "astro-ph.CO",
    reportNumber = "LIGO-P2600018",
    month = "5",
    year = "2026"
}

@article{Lindblom:2008cm,
    author = "Lindblom, Lee and Owen, Benjamin J. and Brown, Duncan A.",
    title = "{Model Waveform Accuracy Standards for Gravitational Wave Data Analysis}",
    eprint = "0809.3844",
    archivePrefix = "arXiv",
    primaryClass = "gr-qc",
    doi = "10.1103/PhysRevD.78.124020",
    journal = "Phys. Rev. D",
    volume = "78",
    pages = "124020",
    year = "2008"
}

@misc{Toubiana:2024car,
    author = "Toubiana, Alexandre and Gair, Jonathan R.",
    title = "{Indistinguishability criterion and estimating the presence of biases}",
    eprint = "2401.06845",
    archivePrefix = "arXiv",
    primaryClass = "gr-qc",
    month = "1",
    year = "2024"
}

@article{Knapp:2025ecw,
    author = "Knapp, Taylor and Chatziioannou, Katerina and Mitman, Keefe and Scheel, Mark A. and Boyle, Michael and Kidder, Lawrence E. and Pfeiffer, Harald",
    title = "{Comprehensive look into the accuracy of spectral Einstein code binary black hole waveforms}",
    eprint = "2510.06393",
    archivePrefix = "arXiv",
    primaryClass = "gr-qc",
    doi = "10.1103/dmpn-q3fk",
    journal = "Phys. Rev. D",
    volume = "113",
    number = "12",
    pages = "124001",
    year = "2026"
}

@article{Healy:2017psd,
    author = "Healy, James and Lousto, Carlos O. and Zlochower, Yosef and Campanelli, Manuela",
    title = "{The RIT binary black hole simulations catalog}",
    eprint = "1703.03423",
    archivePrefix = "arXiv",
    primaryClass = "gr-qc",
    doi = "10.1088/1361-6382/aa91b1",
    journal = "Class. Quant. Grav.",
    volume = "34",
    number = "22",
    pages = "224001",
    year = "2017"
}

@article{Healy:2019jyf,
    author = "Healy, James and Lousto, Carlos O. and Lange, Jacob and O'Shaughnessy, Richard and Zlochower, Yosef and Campanelli, Manuela",
    title = "{Second RIT binary black hole simulations catalog and its application to gravitational waves parameter estimation}",
    eprint = "1901.02553",
    archivePrefix = "arXiv",
    primaryClass = "gr-qc",
    doi = "10.1103/PhysRevD.100.024021",
    journal = "Phys. Rev. D",
    volume = "100",
    number = "2",
    pages = "024021",
    year = "2019"
}

@article{Healy:2020vre,
    author = "Healy, James and Lousto, Carlos O.",
    title = "{Third RIT binary black hole simulations catalog}",
    eprint = "2007.07910",
    archivePrefix = "arXiv",
    primaryClass = "gr-qc",
    doi = "10.1103/PhysRevD.102.104018",
    journal = "Phys. Rev. D",
    volume = "102",
    number = "10",
    pages = "104018",
    year = "2020"
}

@article{Healy:2022wdn,
    author = "Healy, James and Lousto, Carlos O.",
    title = "{Fourth RIT binary black hole simulations catalog: Extension to eccentric orbits}",
    eprint = "2202.00018",
    archivePrefix = "arXiv",
    primaryClass = "gr-qc",
    doi = "10.1103/PhysRevD.105.124010",
    journal = "Phys. Rev. D",
    volume = "105",
    number = "12",
    pages = "124010",
    year = "2022"
}

@article{Ficarra:2026cej,
    author = "Ficarra, Giuseppe and Healy, James and Lousto, Carlos O.",
    title = "{Fifth RIT catalog of binary black hole simulations: Multiple-resolution studies of eccentric orbits}",
    eprint = "2606.10104",
    archivePrefix = "arXiv",
    primaryClass = "gr-qc",
    doi = "10.1103/p77n-6fxw",
    journal = "Phys. Rev. D",
    volume = "114",
    number = "6",
    pages = "064095",
    year = "2026"
}

@article{Jani:2016wkt,
    author = "Jani, Karan and Healy, James and Clark, James A. and London, Lionel and Laguna, Pablo and Shoemaker, Deirdre",
    title = "{Georgia Tech Catalog of Gravitational Waveforms}",
    eprint = "1605.03204",
    archivePrefix = "arXiv",
    primaryClass = "gr-qc",
    doi = "10.1088/0264-9381/33/20/204001",
    journal = "Class. Quant. Grav.",
    volume = "33",
    number = "20",
    pages = "204001",
    year = "2016"
}

@article{Ferguson:2023vta,
    author = "Ferguson, Deborah and others",
    title = "{Second MAYA catalog of binary black hole numerical relativity waveforms}",
    eprint = "2309.00262",
    archivePrefix = "arXiv",
    primaryClass = "gr-qc",
    reportNumber = "UTWI-32-2023",
    doi = "10.1103/gk7x-9cds",
    journal = "Phys. Rev. D",
    volume = "112",
    number = "4",
    pages = "044043",
    year = "2025"
}

@article{Hamilton:2023qkv,
    author = "Hamilton, Eleanor and others",
    title = "{Catalog of precessing black-hole-binary numerical-relativity simulations}",
    eprint = "2303.05419",
    archivePrefix = "arXiv",
    primaryClass = "gr-qc",
    doi = "10.1103/PhysRevD.109.044032",
    journal = "Phys. Rev. D",
    volume = "109",
    number = "4",
    pages = "044032",
    year = "2024"
}

@article{Friedrich:1985afv,
    author = "Friedrich, Helmut",
    title = "{On the hyperbolicity of Einstein's and other gauge field equations}",
    doi = "10.1007/BF01217728",
    journal = "Commun. Math. Phys.",
    volume = "100",
    number = "4",
    pages = "525--543",
    year = "1985"
}

@article{Lindblom:2005qh,
    author = "Lindblom, Lee and Scheel, Mark A. and Kidder, Lawrence E. and Owen, Robert and Rinne, Oliver",
    title = "{A New generalized harmonic evolution system}",
    eprint = "gr-qc/0512093",
    archivePrefix = "arXiv",
    doi = "10.1088/0264-9381/23/16/S09",
    journal = "Class. Quant. Grav.",
    volume = "23",
    pages = "S447--S462",
    year = "2006"
}

@article{Shibata:1995we,
    author = "Shibata, Masaru and Nakamura, Takashi",
    title = "{Evolution of three-dimensional gravitational waves: Harmonic slicing case}",
    doi = "10.1103/PhysRevD.52.5428",
    journal = "Phys. Rev. D",
    volume = "52",
    pages = "5428--5444",
    year = "1995"
}

@article{Baumgarte:1998te,
    author = "Baumgarte, Thomas W. and Shapiro, Stuart L.",
    title = "{On the numerical integration of Einstein's field equations}",
    eprint = "gr-qc/9810065",
    archivePrefix = "arXiv",
    doi = "10.1103/PhysRevD.59.024007",
    journal = "Phys. Rev. D",
    volume = "59",
    pages = "024007",
    year = "1998"
}

@article{Damour:2014afa,
    author = "Damour, Thibault and Guercilena, Federico and Hinder, Ian and Hopper, Seth and Nagar, Alessandro and Rezzolla, Luciano",
    title = "{Strong-Field Scattering of Two Black Holes: Numerics Versus Analytics}",
    eprint = "1402.7307",
    archivePrefix = "arXiv",
    primaryClass = "gr-qc",
    doi = "10.1103/PhysRevD.89.081503",
    journal = "Phys. Rev. D",
    volume = "89",
    number = "8",
    pages = "081503",
    year = "2014"
}

@article{Damour:2022ybd,
    author = "Damour, Thibault and Rettegno, Piero",
    title = "{Strong-field scattering of two black holes: Numerical relativity meets post-Minkowskian gravity}",
    eprint = "2211.01399",
    archivePrefix = "arXiv",
    primaryClass = "gr-qc",
    doi = "10.1103/PhysRevD.107.064051",
    journal = "Phys. Rev. D",
    volume = "107",
    number = "6",
    pages = "064051",
    year = "2023"
}

@article{Hopper:2022rwo,
    author = "Hopper, Seth and Nagar, Alessandro and Rettegno, Piero",
    title = "{Strong-field scattering of two spinning black holes: Numerics versus analytics}",
    eprint = "2204.10299",
    archivePrefix = "arXiv",
    primaryClass = "gr-qc",
    doi = "10.1103/PhysRevD.107.124034",
    journal = "Phys. Rev. D",
    volume = "107",
    number = "12",
    pages = "124034",
    year = "2023"
}

@article{Rettegno:2023ghr,
    author = "Rettegno, Piero and Pratten, Geraint and Thomas, Lucy M. and Schmidt, Patricia and Damour, Thibault",
    title = "{Strong-field scattering of two spinning black holes: Numerical relativity versus post-Minkowskian gravity}",
    eprint = "2307.06999",
    archivePrefix = "arXiv",
    primaryClass = "gr-qc",
    doi = "10.1103/PhysRevD.108.124016",
    journal = "Phys. Rev. D",
    volume = "108",
    number = "12",
    pages = "124016",
    year = "2023"
}

@article{Long:2025tvk,
    author = "Long, Oliver and Pfeiffer, Harald P. and Kidder, Lawrence E. and Scheel, Mark A.",
    title = "{Black-hole scattering with numerical relativity: Self-force extraction and post-Minkowskian validation}",
    eprint = "2511.10196",
    archivePrefix = "arXiv",
    primaryClass = "gr-qc",
    doi = "10.1103/bdsb-sp9c",
    journal = "Phys. Rev. D",
    volume = "112",
    number = "12",
    pages = "124038",
    year = "2025"
}

@article{Bern:2021yeh,
    author = "Bern, Zvi and Parra-Martinez, Julio and Roiban, Radu and Ruf, Michael S. and Shen, Chia-Hsien and Solon, Mikhail P. and Zeng, Mao",
    title = "{Scattering Amplitudes, the Tail Effect, and Conservative Binary Dynamics at O(G4)}",
    eprint = "2112.10750",
    archivePrefix = "arXiv",
    primaryClass = "hep-th",
    doi = "10.1103/PhysRevLett.128.161103",
    journal = "Phys. Rev. Lett.",
    volume = "128",
    number = "16",
    pages = "161103",
    year = "2022"
}

@article{Dlapa:2022lmu,
    author = {Dlapa, Christoph and K{\"a}lin, Gregor and Liu, Zhengwen and Neef, Jakob and Porto, Rafael A.},
    title = "{Radiation Reaction and Gravitational Waves at Fourth Post-Minkowskian Order}",
    eprint = "2210.05541",
    archivePrefix = "arXiv",
    primaryClass = "hep-th",
    doi = "10.1103/PhysRevLett.130.101401",
    journal = "Phys. Rev. Lett.",
    volume = "130",
    number = "10",
    pages = "101401",
    year = "2023"
}

@article{Driesse:2024xad,
    author = "Driesse, Mathias and Jakobsen, Gustav Uhre and Mogull, Gustav and Plefka, Jan and Sauer, Benjamin and Usovitsch, Johann",
    title = "{Conservative Black Hole Scattering at Fifth Post-Minkowskian and First Self-Force Order}",
    eprint = "2403.07781",
    archivePrefix = "arXiv",
    primaryClass = "hep-th",
    reportNumber = "HU-EP-24/08-RTG, CERN-TH-2024-032",
    doi = "10.1103/PhysRevLett.132.241402",
    journal = "Phys. Rev. Lett.",
    volume = "132",
    number = "24",
    pages = "241402",
    year = "2024"
}

@article{Chern:1955,
    author  = {Chern, Shiing-Shen},
    title   = {{An elementary proof of the existence of isothermal parameters on a surface}},
    journal = {Proc. Amer. Math. Soc.},
    volume  = {6},
    pages   = {771--782},
    year    = {1955}
}

@book{LehtoVirtanen:1973,
    author    = {Lehto, Olli and Virtanen, Kaarlo I.},
    title     = {{Quasiconformal Mappings in the Plane}},
    series    = {Grundlehren der mathematischen Wissenschaften},
    volume    = {126},
    edition   = {2},
    publisher = {Springer},
    address   = {Berlin},
    year      = {1973}
}

@article{Banach:1922,
    author  = {Banach, Stefan},
    title   = {{Sur les op\'erations dans les ensembles abstraits et leur application aux \'equations int\'egrales}},
    journal = {Fund. Math.},
    volume  = {3},
    pages   = {133--181},
    year    = {1922}
}

@book{Zeidler:1986,
    author    = {Zeidler, Eberhard},
    title     = {{Nonlinear Functional Analysis and its Applications I: Fixed-Point Theorems}},
    publisher = {Springer},
    address   = {New York},
    year      = {1986}
}

@article{Rashti:2024yoc,
	author = "Rashti, Alireza and Gamba, Rossella and Chandra, Koustav and Radice, David and Daszuta, Boris and Cook, William and Bernuzzi, Sebastiano",
	title = "{Binary black hole waveforms from high-resolution gr-athena++ simulations}",
	eprint = "2411.11989",
	archivePrefix = "arXiv",
	primaryClass = "gr-qc",
	doi = "10.1103/n5pz-qv3x",
	journal = "Phys. Rev. D",
	volume = "111",
	number = "10",
	pages = "104078",
	year = "2025"
}

@article{Huerta:2019oxn,
	author = "Huerta, E. A. and others",
	title = "{Physics of eccentric binary black hole mergers: A numerical relativity perspective}",
	eprint = "1901.07038",
	archivePrefix = "arXiv",
	primaryClass = "gr-qc",
	doi = "10.1103/PhysRevD.100.064003",
	journal = "Phys. Rev. D",
	volume = "100",
	number = "6",
	pages = "064003",
	year = "2019"
}

@article{Nakano:2015pta,
    author = "Nakano, Hiroyuki and Healy, James and Lousto, Carlos O. and Zlochower, Yosef",
    title = "{Perturbative extraction of gravitational waveforms generated with Numerical Relativity}",
    eprint = "1503.00718",
    archivePrefix = "arXiv",
    primaryClass = "gr-qc",
    doi = "10.1103/PhysRevD.91.104022",
    journal = "Phys. Rev. D",
    volume = "91",
    number = "10",
    pages = "104022",
    year = "2015"
}

@article{Rinne:2009Hyperboloidal,
    author        = {Rinne, Oliver},
    title         = {{An axisymmetric evolution code for the Einstein equations on hyperboloidal slices}},
    journal       = {Class. Quant. Grav.},
    volume        = {27},
    pages         = {035014},
    year          = {2010},
    doi           = {10.1088/0264-9381/27/3/035014},
    eprint        = {0910.0139},
    archivePrefix = {arXiv},
    primaryClass  = {gr-qc},
}

@article{Peterson:2024Hyperboloidal,
    author        = {Peterson, Christian and Gautam, Shalabh and Va{\~n}{\'o}-Vi{\~n}uales, Alex and Hilditch, David},
    title         = {{Spherical evolution of the generalized harmonic gauge formulation of general relativity on compactified hyperboloidal slices}},
    journal       = {Phys. Rev. D},
    volume        = {110},
    pages         = {124033},
    year          = {2024},
    doi           = {10.1103/PhysRevD.110.124033},
    eprint        = {2409.02994},
    archivePrefix = {arXiv},
    primaryClass  = {gr-qc},
}

\end{document}